# Unveiling High Selectivity Origin of Pt-Bi Catalysts for Alkaline Methanol Electrooxidation via CO-free pathway

Lecheng Liang,[1,5] Hengyu Li,[2,5] Peng Li,[4] Jinhui Liang,[1] Shao Ye,[1] Binwen Zeng,[1] Yanhong Xie,[3] Yucheng Wang,[3] Taisuke Ozaki, [2] Shengli Chen,[4] and Zhiming Cui [1,*]

[1]Guangdong Provincial Key Laboratory of Fuel Cell Technology, School of Chemistry and Chemical Engineering, South China University of Technology, Guangzhou 510641, China.

[2]Institute for Solid State Physics, The University of Tokyo, Kashiwa 277-8581, Japan.

[3]State Key Laboratory of Physical Chemistry of Solid Surfaces, College of Chemistry and Chemical Engineering, Xiamen University, Xiamen 361005, China.

[4]College of Chemistry and Molecular Sciences, Wuhan University, Wuhan, China.

[5]These authors contributed equally

*Correspondence: zmcui@scut.edu.cn

**Abstract**: A long-standing puzzle for methanol electrooxidation is how to achieve a CO-free pathway and accurately understand the origin of electrocatalytic selectivity. Herein, we unequivocally demonstrate that the Bi-modified Pt/C follows a CO-free dominated pathway during alkaline methanol electrooxidation, and unveil the formaldehyde (HCHO) intermediate as a critical factor influencing pathway selectivity. These findings are substantiated by kinetic isotope effects, formate Faradaic efficiency, in situ spectroscopy, ab initio molecular dynamic simulations, and density functional theory calculations. Bi modification significantly increases the HCHO dehydrogenation barrier, which facilitates its desorption and subsequent conversion to the H2COOH- anion at the alkaline interface, intrinsically avoiding CO formation. More specifically, the formation of ensemble sites featuring V-shaped Bi-Pt-Bi configuration inhibits the cleavage of C-H bond, and the weak OH binding energy at Bi adatoms effectively prevents blockage of oxygenated species, allowing such ensemble sites to fulfill their functional role. Our study opens up a novel dimension for designing advanced CO-free catalysts.

## Introduction

Methanol electrooxidation, as a classical C1 molecular model reaction, is one of the cornerstones of important energy conversion technology, such as fuel cells and electrolyzers.[1,2] The dual-pathway mechanism for methanol electrooxidation, proposed by Breiter, has been well-established with the community. [3,4] This mechanism presumes that there are two parallel pathways in the reaction scheme: CO pathway and CO-free pathway. However, a standing puzzle is that Pt-based catalysts typically follow a CO pathway rather than a CO-free pathway, whether in acidic or alkaline electrolytes, leading to serious CO-poisoning issue. Such dilemma has spurred numerous research to rationally design the advanced catalysts with high CO-tolerance, but it is still unknown how to achieve CO-free dominant pathway.[5-8] A comprehensive understanding of the origin of pathway selectivity is urgently needed by the community.

So far, in contrast to the fruitful progress in catalyst exploration, fundamental insights into the mechanism at atomic level are still mainly originated from early pioneering works that were based on single-crystal electrodes.[9] In 2006, Cuesta. emphasized the significant influence of atomic ensemble effect on the selectivity of reaction pathways.[10] This conclusion coincides with the density functional theory (DFT) results by Neurock et al. who further connected the selectivity of reaction pathways and first proton-coupled electron transfer (PCET) step, that is, initial O–H bond cleavage favors a CO-free pathway while initial C–H bond cleavage prefers a CO pathway.[11] By using ab initio molecular dynamics (AIMD), Herron et al. proposed that cleavage of methanol O−H bond is slightly more facile than that of C−H bond on solvated positively biased Pt (111).[12] Such assumption can well explain the experimental results that CO-free pathway happens at high potentials.[13-16] In addition, based on the results from in situ spectroscopy, Katayama et al. suggested that adsorbed methoxy ($CH_3O$*) is only observed at high potentials and serves as the key intermediate to improve the alkaline methanol electrooxidation activity.[17] These findings imply that initial O–H bond activation of methanol to form $CH_3O$* seems an important factor to achieve a CO-free dominant pathway. However, Mekazni et al. found that $CH_3O$* can be easily formed in the presence of OH* and is gradually dehydrogenated to form CO*.[18] Apparently, the intrinsic influencing factors in determining selectivity of reaction pathway remain an unknown "black box". It is noteworthy that Pt-Bi alloys have been recently reported to have extraordinary catalytic properties toward methanol electrooxidation in alkaline electrolytes and are speculated to follow CO-free dominated pathway according to the results from spectral evidence.[19-22] These intriguing findings inspire us to leverage Bi-Pt system as an appropriate model to delve into the underlying mechanism, especially to uncover the inherent origin that determine the selectivity of methanol electrooxidation.

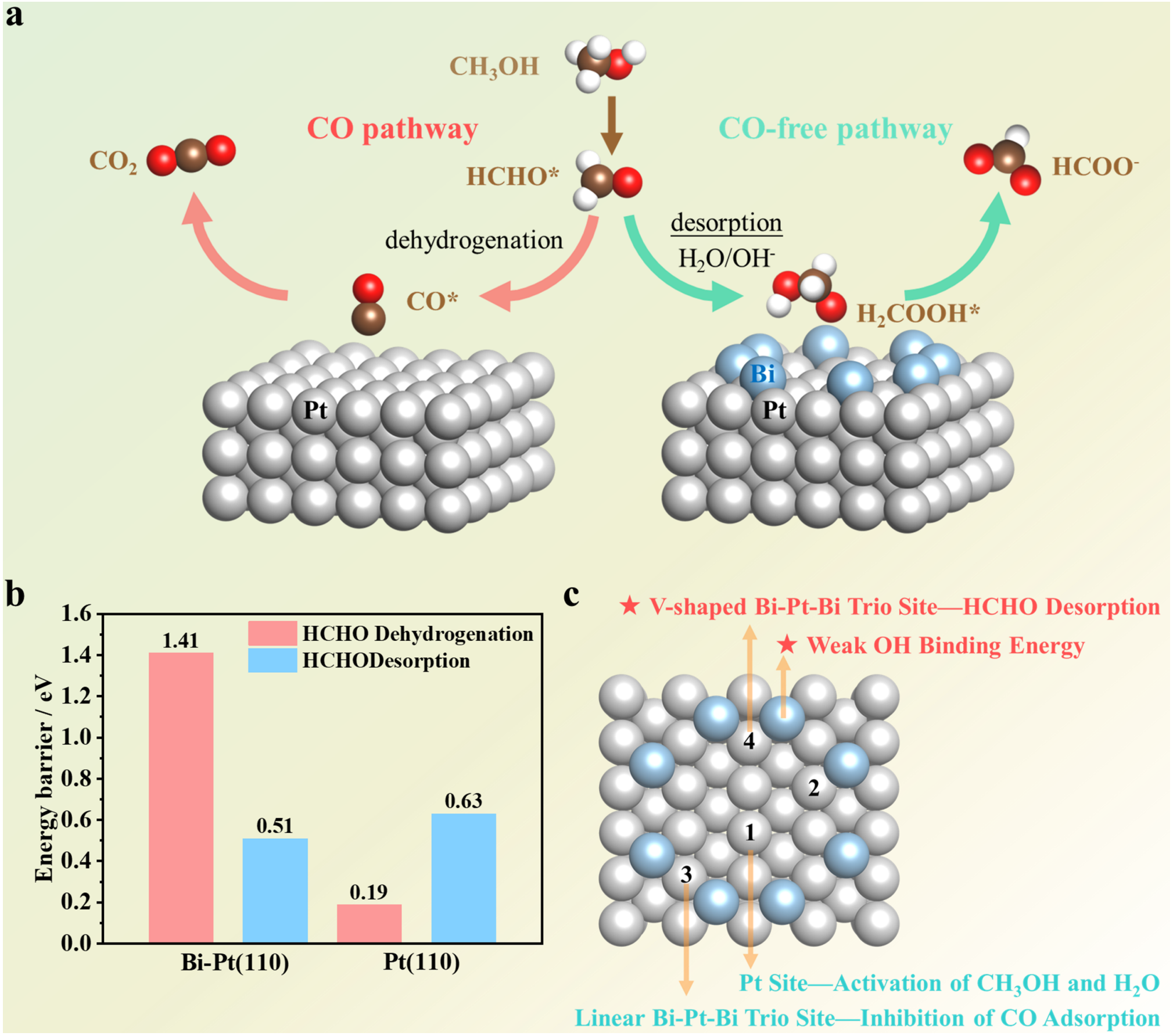

**Scheme 1. Origin of High Selectivity for CO-free Pathway on Bi-Pt Surfaces.** (a) Schematic illustration

of the alkaline methanol electrooxidation mechanism on Pt and Bi-modified Pt surfaces. (b) A comparison of HCHO* desorption barrier and its dehydrogenation barrier on Bi-Pt(110) and Pt(110) at 0.5 $V_{RHE}$. (c) Multi-sites effect of Bi-Pt surface and two intrinsic factors through which Bi modification enhances selectivity for the CO-free pathway.

In this study, we constructed a Bi-modified Pt/C model catalyst using underpotential deposition method and employed a combination of various electrochemical characterization, in situ spectroscopy, DFT calculations, and AIMD simulations to investigate the underlying alkaline methanol electrooxidation mechanism as well as the selectivity origin. Concretely, kinetic isotope effects tests show that the activation barrier of the O-H bond in methanol is negligible, whereas the cleavage of the C-H bond on Bi-modified Pt/C is the rate-determining step. Additionally, Bi modification not only inhibits the formation of CO across the entire potential range as observed by in situ surface-enhanced infrared absorption spectroscopy but also enables a formate Faradaic efficiency of nearly 90%, indicating that the reaction follows a CO-free dominated pathway. Further assisted by AIMD simulations combined with slow-growth method and DFT calculations, we clarify the comprehensive mechanism and reveal that the dynamic behavior of HCHO intermediate is the pivotal factor influencing pathway selectivity. As illustrated in Scheme 1a-1b, Bi modification significantly enhance the HCHO dehydrogenation barrier compared to its desorption barrier, enabling the conversion of HCHO into $H_2COOH^-$ anion through hydration and deprotonation at alkaline interfaces, thereby inherently preventing CO formation. More importantly, Scheme 1c provides a comprehensive understanding of multi-sites effect of Bi-Pt surface and identify two intrinsic factors by which Bi modification significantly enhances selectivity towards the CO-free pathway: (i) The formation of ensemble sites featuring V-shaped Bi-Pt-Bi ($BPB_V$) configuration inhibits the cleavage of the C-H bond, thereby promoting HCHO desorption; (ii) The weak OH binding energy at Bi adatoms effectively prevents OH species from blocking the BPBV sites, thus guaranteeing the fulfillment of their functional effect. Our study not only sheds light on the fundamental factor behind the selectivity of methanol electrooxidation from both interfacial and kinetic perspectives but also establishes an investigative framework to promote an in-depth understanding of C1 molecular electrocatalytic behavior.

## Results

### Exceptional Inherent Activity of Bi-modified Pt/C for Alkaline Methanol Electrooxidation

We adopted a underpotential deposition method to prepare Bi-modified Pt/C model catalyst (denoted Bi-Pt/C), which was electrodeposited in 0.5 M $H_2SO4$ containing 10 mM $Bi^{3+}$ ions at -0.16 V vs. Ag/AgCl (saturated KCl) for a duration of 120 s, with detailed procedures outlined in the Experimental Section and Figure S1. Both high-angle annular dark-field scanning transmission electron microscopy (HAADF-STEM) and TEM images confirm that the size and morphology of Pt nanoparticles remain unchanged upon the introduction of Bi adatoms (Figures S2-S4). Energy-dispersive spectroscopy (EDS) elemental mapping and line scans across a single particle show that Pt surfaces are uniformly covered with a sub-monolayer of Bi adatoms. Such Bi adatoms significantly altered the electrochemical behavior of Pt surfaces. In Figure 1A and S5, the cyclic voltammograms (CV) curve of Bi-Pt/C displays a distinct reduction in current within the hydrogen region (0.05-0.45 $V_{RHE}$) compared to Pt/C and Bi/C. Particularly, a marked decrease in the sharp peak intensity around 0.3 $V_{RHE}$ indicates that Bi adatoms inhibit hydrogen and hydroxide adsorption at the Pt (110) step.[23-27] In the hydroxyl region, the peak position of Bi-Pt/C shifts to higher potential values, suggesting weakened binding strength of oxygen-related species on the Bi-Pt surface.

Prior to conducting the methanol oxidation reaction (MOR) test, these catalysts were activated and cleaned in $N_2$-saturated 1M KOH until stable cyclic voltammograms curves were obtained (Figure S6). As shown in Figure 1B, Bi-Pt/C exhibits an ultrahigh mass activity (MA) of 13.7 A $mg_{Pt}^{-1}$ at peak potential, surpassing the MA values found for PtRu/C (3.8 A $mg_{Pt}^{-1}$) and Pt/C (3.2 A $mg_{Pt}^{-1}$) by a factor of 3.57 and 4.25, respectively, which ranks among the state-of-the-art alkaline MOR electrocatalysts reported in the literature (Figure S7 and Table S1). It is noteworthy that the MA of Bi-Pt/C even outperforms that of Pt/C by 12 times at 0.5 $V_{RHE}$, indicating its extremely low onset potential and fast intrinsic reaction kinetics (Figure 1C). In addition, chronoamperometry tests in Figure S8 show that Bi-Pt/C maintains higher MA compared to those of commercial catalysts after 3600 s, suggesting its excellent durability toward MOR.

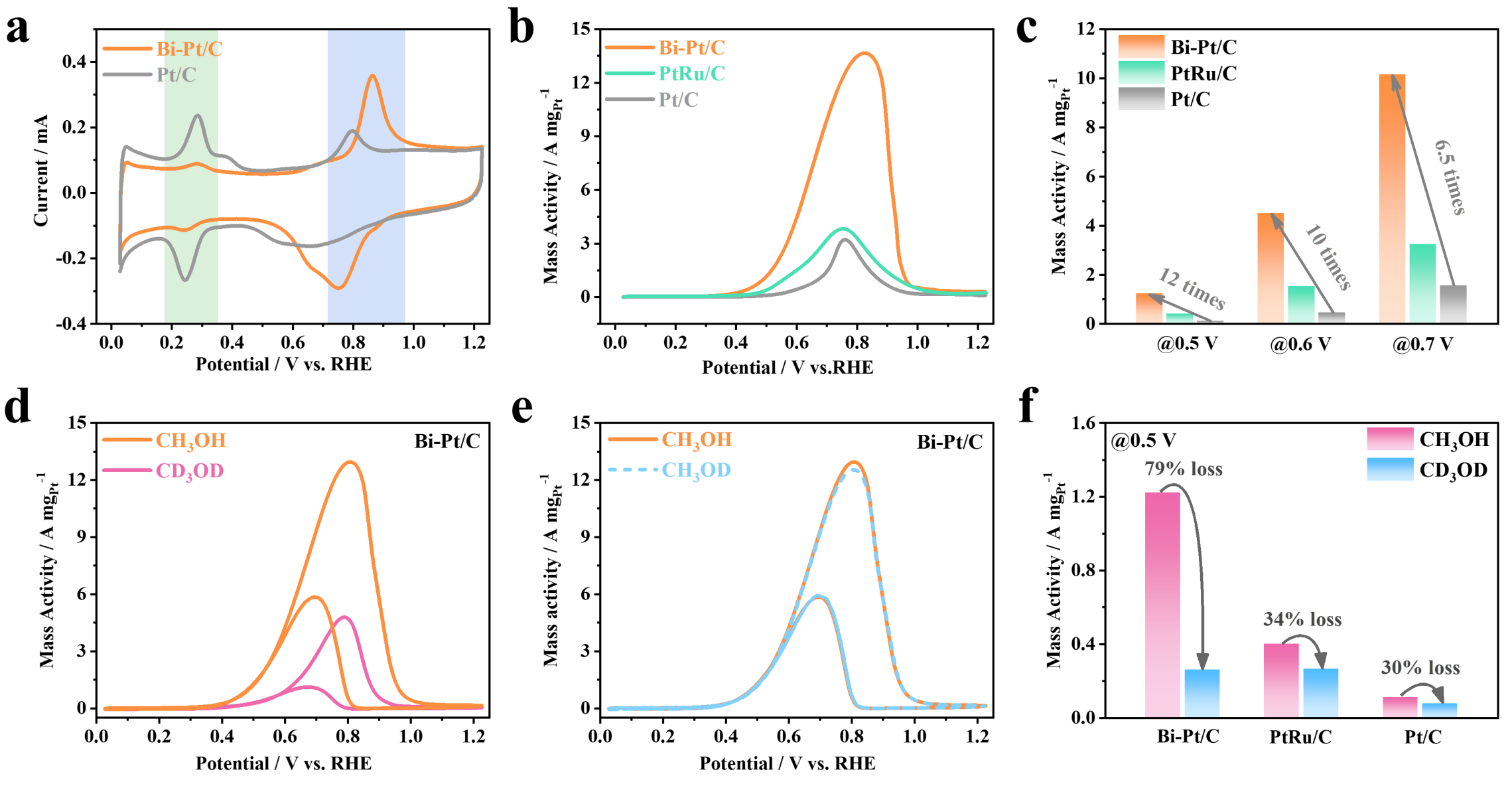

**Figure 1.** Alkaline methanol electrooxidation and kinetic isotope effect studies on model catalysts. (a) CV curves of Bi-Pt/C and Pt/C in $N_2$-saturated 1M KOH. (b) Positive-going polarization curves of different catalysts recorded at a scan rate of 50 mV $s^{-1}$ in 1M $CH_3OH$/1M KOH. (c) Histograms of summarized mass activity at different potentials. (d) A comparison of CV curves of Bi-Pt/C between 1M $CH_3OH$/1M KOH and 1M $CD_3OD$/ 1M KOH. (e) A comparison of CV curves of Bi-Pt/C between 1M $CH_3OH$/1M KOH and 1M $CH_3OD$/1M KOH. (f) A comparison of the mass activity of different catalysts under H/D exchange at 0.5 $V_{RHE}$.

**Kinetic Isotope Effect Analysis.**

The intrinsic reaction kinetics of Bi-Pt/C stem from its distinctive mechanism, and the measurement of kinetic isotope effects (KIEs) serves as one of the key tools to probe mechanistic information.[28] Figures 1D, 1e and S9, show that the deuterium KIEs tests were conducted in 1M $CD_3OD$/1M KOH and 1M $CH_3OD$/1M KOH solutions to aid in identifying the rate-determining step (RDS) of the electrocatalytic processes. After substituting $CH_3OH$ with $CD_3OD$ in the electrolyte, Bi-Pt/C and commercial catalysts exhibit noticeable drops in mass activities at peak potential. Electrochemical impedance spectroscopy (EIS) further indicates that Bi-Pt/C exhibits the fastest charge transfer kinetics and H/D exchange can significantly influence alkaline MOR kinetics (refer to Figure S10 for details). Comparatively, in $CH_3OD$ solutions, their mass activities are nearly unchanged. Such results demonstrate that C−H bond cleavage significantly influences the reaction kinetics, while the influence of O-H bond cleavage can be negligible. Furthermore, it was observed that the activation of the C−H bond has a different impact on Bi-Pt/C and commercial catalysts. As depicted in Figures 1F and S9E, the loss of MA of Bi-Pt/C at 0.5 $V_{RHE}$ increased by up to 79% relative to its value at peak potential, whereas commercial catalysts only exhibited around a 30% reduction in mass activity loss, suggesting that Bi modification renders cleavage of C-H bond as RDS. Tafel slope analysis further supports this consequence (Figure S11). The insights gained from these KIEs results promote to precisely understand the unique reaction mechanism of Bi-Pt/C, complemented by subsequent theoretical calculations.

**Determination of a CO-free Pathway for Bi-modified Pt/C.**

To further understand the origin of the high MOR performance of Bi-Pt/C, we conducted CO-stripping experiments, in situ attenuated total reflection surface-enhanced infrared absorption spectra (ATR-SEIRA), and ion chromatography (IC) analysis. The CO stripping experiments in Figures 2A and 2B show a significant reduction in current on Bi-Pt/C compared to Pt/C, indicating that Bi adatoms greatly inhibit CO poisoning, in agreement with the previous report.[19,21,29] However, the comparable onset potential and CO oxidation peak potential of Pt/C and Bi-Pt/C catalysts, measured at 0.48 $V_{RHE}$ and 0.75 $V_{RHE}$ respectively, suggest that some unmodified Pt sites persist on the Bi-Pt surfaces, making them susceptible to CO poisoning. In other words, this evidence indicates that the unique arrangement of Bi adatoms on Pt surface leads to the coexistence of both CO-poisoned and CO-resistant Pt sites.

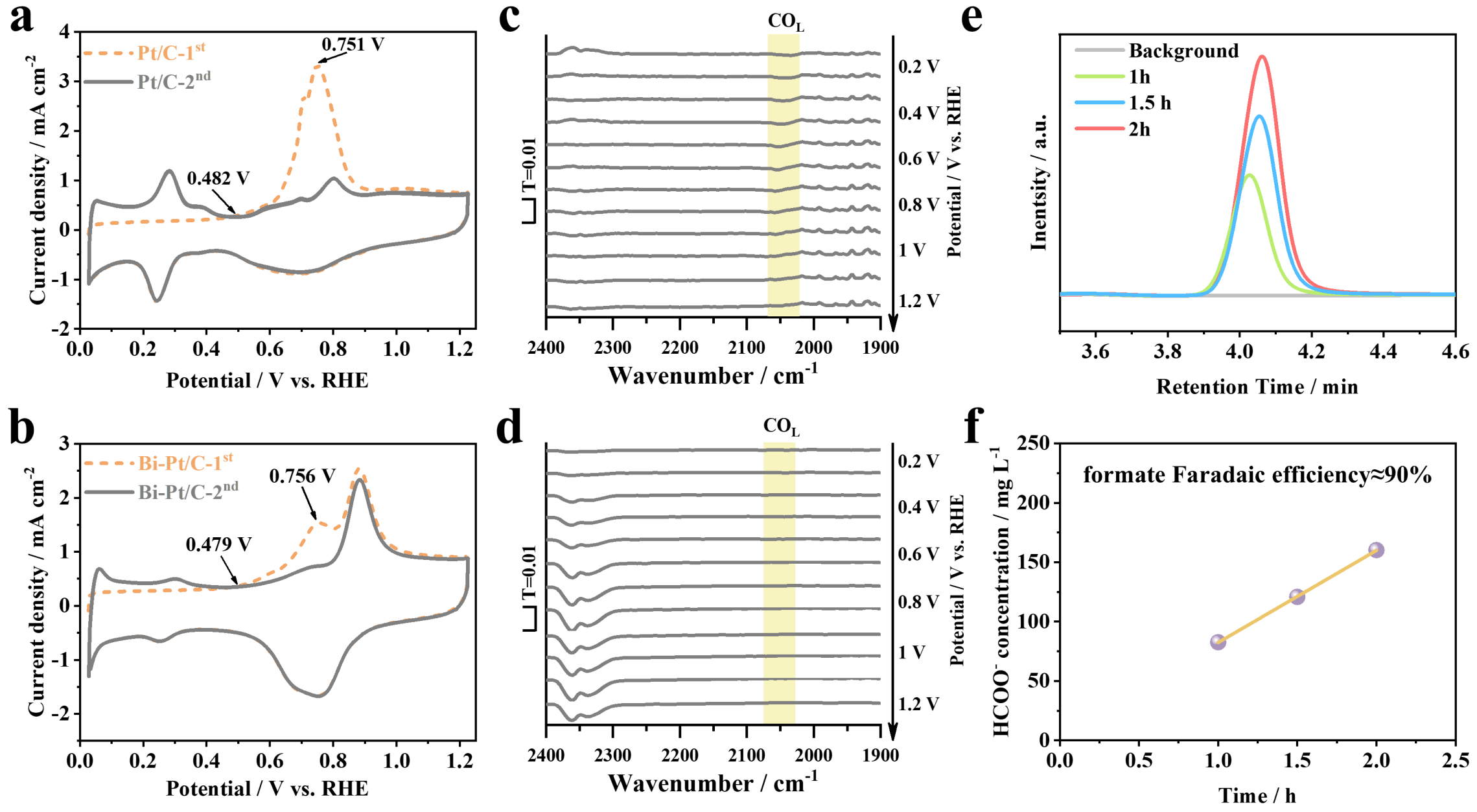

**Figure 2. Determination of a CO-free pathway for Bi-modified Pt/C.** (a and b) CO-stripping curves of (a) Pt/C and (b) Bi-Pt/C in 1M KOH solution. (c and d) In situ ATR-SEIRA measurements on (c) Pt/C and (d) Bi-Pt/C in 1M $CH_3OH$/1 M KOH over the potential range from 0.1 to 1.2 $V_{RHE}$.(e) IC spectra of HCOO- for the electrolyte after different electrolysis time at 50 mA $cm^{-2}$ by Bi-Pt/C. (f) The $HCOO^-$ concentration after different electrolysis time derived from the IC spectra. Calculations of corresponding formate Faradaic efficiency (%) are listed in Table S2.

Figures 2C and 2D exhibit the ATR-SEIRA spectra of Pt/C and Bi-Pt/C, respectively, which were acquired with the potential ranging from 0.1 to 1.2 $V_{RHE}$, and the reference potential was selected at -0.25 VRHE. The absorption band at ca. 2050 $cm^{-1}$ is assigned to the signal from adsorbed linearly bonded CO ($CO_L$) species, which was detected on Pt/C. Notably, despite the presence of CO-poisoned Pt sites on the Bi-Pt surfaces, no $CO_L$ signal was observed covering the entire potential range, confirming that MOR process on Bi-Pt/C does not involve the formation of CO. In addition, we performed a chronopotentiometry test on Bi-Pt/C at 50 mA $cm^{-2}$ and anodically produced formate was quantified by analyzing electrolyte using IC measurements to investigate the selectivity of formate (Figure 2E). Figure 2F summarizes the concentration of formate observed under varying electrolysis durations, revealing a formate Faradaic efficiency of approximately 90% throughout the continuous electrolysis process (Table S2). The aforementioned experimental results clearly demonstrate that the Bi modification not only inhibits the CO adsorption, but fundamentally shifts the MOR process from a dual-pathway mechanism to a CO-free dominated pathway (formate pathway).

**Bi-Modified Pt(110) Surface Model and Thermodynamic Calculations.**

To gain an in-depth understanding of the origins of selectivity towards CO-free pathway at the molecular-atomic level, we constructed the theoretical models of Bi-Pt surface. Considering the electrochemical characteristics of Bi-Pt/C in Figure 1A, we chose Pt(110) as initial surface models. To obtain a reasonable coverage of Bi atoms on the (110) surface, a truncated octahedral model was used to calculate the surface-to-core atomic ratio (0.43) of a 2.4 nm Pt nanoparticle. By incorporating inductively coupled plasma optical emission spectrometry (ICP-OES) data on the Pt to Bi atomic ratio (10:1), the surface coverage of Bi atoms was estimated to be approximately 0.33 monolayers (ML). Bader charge analysis reveals significant charge transfer from Bi to Pt which is in agreement with the X-ray photoelectron spectroscopy (XPS) observations (Figure S12). Figures 3A and S13 display four distinct types of Pt adsorption sites on the Bi-Pt surface, labeled as sites 1, 2, 3, and 4, based on the local coordination environments of the Pt and Bi atoms. We further characterize these active sites by the configurations and combinations of Pt and Bi atom: Site 1, composed solely of Pt atoms, is designated as the Pt (P) site; Site 2 features a Bi atom bridging two Pt atoms in a V-shaped configuration, earning it the name Pt-Bi-$Pt_V$ trio ($PBP_V$) site; Site 3 consists of a Pt atom bridging two Bi atoms in a linear arrangement, which we refer to as the Bi-Pt-BiL trio ($BPB_L$) site; Finally, site 4 has a configuration similar to that of site 2 and is therefore named the Bi-Pt-$Bi_V$ trio ($BPB_V$) site.

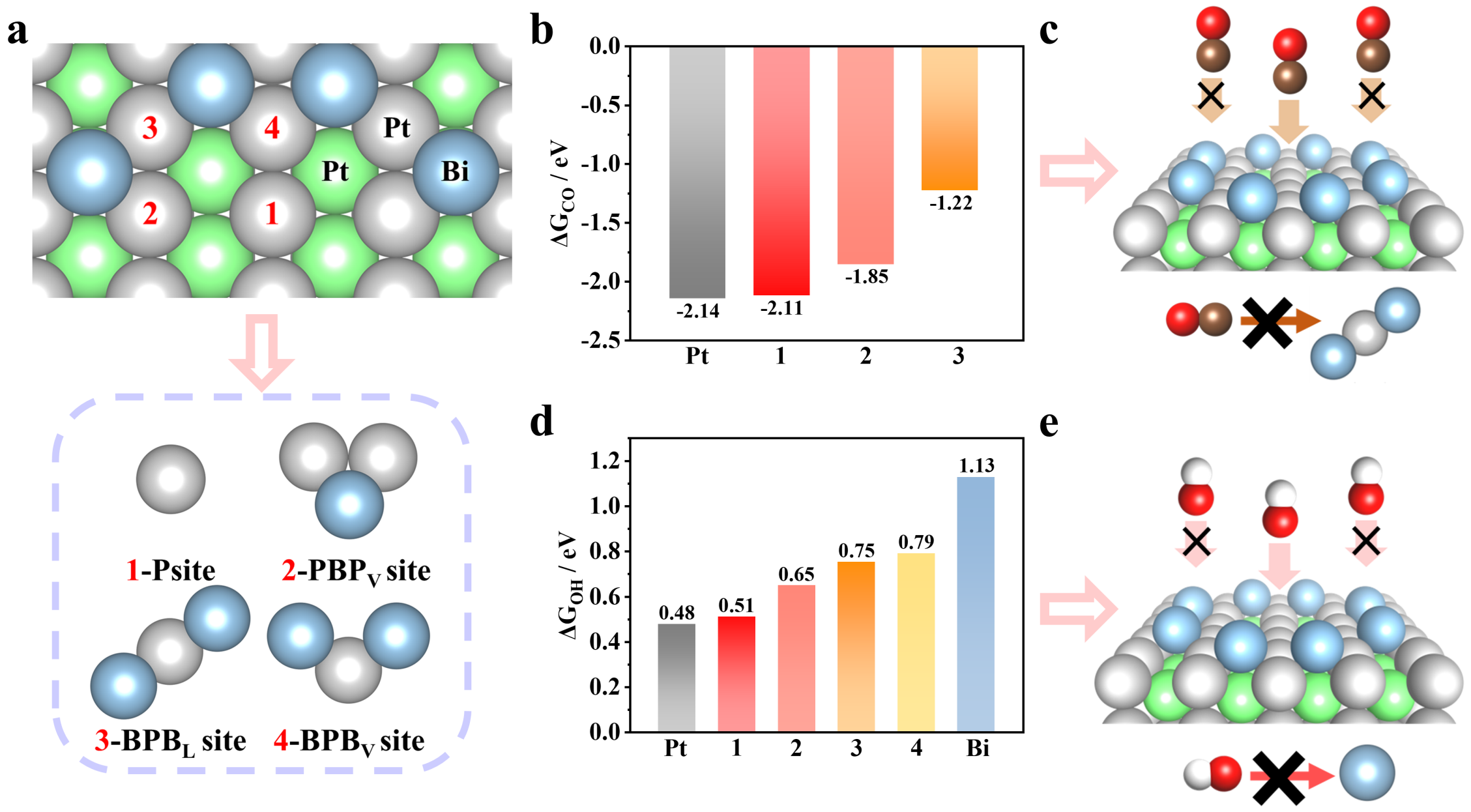

**Figure 3.** Bi-modified Pt(110) surface models and thermodynamic calculations. (a) surface structure of Bi-Pt model with multiple sites and the definition of active sites depends on the local coordination environment

of Pt and Bi atoms. (b and c) The binding energy comparison of (b) CO and (c) OH at top sites. "Pt", denoted as the top site of Pt(110). (d and e) Schematic illustration of the influence of multi-site effects on adsorption strength of (d) CO and (e) OH. Color code: outermost layer Pt, gray; subsurface Pt, green; Bi, blue; O, red; and C, brown.

The CO binding energies (COBEs) at all potential active sites were calculated to identify the interactions between different Pt sites and CO. Figures 3C and S14 display CO* models and COBE at different top sites. The value of COBE at site 1 is -2.11 eV, differing by only 0.03 eV from the top site of Pt(110). Besides, the calculated COBE at different bridge sites are close to that of bridge site of Pt(110) and their maximum differences do not exceed 0.15 eV i.e, both of site 1, site 2, and site 4 can be poisoned by CO (Figure S15). Only site 3 has the weakest COBE with -1.22 eV, indicating relatively unfavorable binding with CO. The COBE results elucidate that the multi-site effect on Bi-Pt surface leads to the coexistence of both CO-poisoned and CO-resistant Pt sites, which is consistent with the experimental phenomena observed from CO-stripping tests. Moreover, the results of the calculated HBE in Figure S16 show that hydrogen atoms cannot adsorb at site 3, which aligns with the findings of the electrochemical characterization (Figure 1A), preliminarily validating the rationality of constructed Bi-Pt surface model. As illuminated in Figure 3C, Bi modification results in the formation of multiple sites on the Bi-Pt surface, among which only site 3 can be resistant to CO poisoning, implying that resistance to CO poisoning may not be a critical factor influencing selectivity.

We also calculated the OH binding energies (OHBEs) at different sites on Bi-Pt surface. As shown in Figures 3D and S17, the OHBE at site 1 is also similar to the Pt surface with difference only 0.03 eV. However, the Pt sites adjacent to Bi atoms show significantly weaker OHBE, especially those at sites 3 and 4, which are adjacent to two Bi atoms, with values of 0.75 eV and 0.79 eV, respectively. This result can be explained by the d-band center theory (Figure S18). Theoretically, oxophilic metal Bi would have a stronger OHBE than Pt. However, OHBE at the Bi site is 1.13 eV, much weaker than that at other sites. Despite Bi atoms being ca. +0.7 e by Bader charge analysis, the reason for the weak OHBE is the local charge accumulation on the top of Bi adatoms, as revealed by differential charge density (Figure S19). In other words, contrary to the conclusions in previous studies, as shown in Figure 3E, our study suggests that the Bi modification weakens OHBE, which is consistent with the positive reoxidation position shift under hydroxyl region (Figure 1A).[19,20,30] Furthermore, the weak OHBE at Bi adatoms indicates that Bi modification does not play a role in facilitating water dissociation and thereby enabling the subsequent coupling of OH* with carbonaceous intermediates.

Further, we constructed the Gibbs free energy diagrams to investigate the thermodynamics of MOR at different sites. Due to the multi-sites effect on Bi-Pt surface, the arrangement of Bi atoms was adjusted to obtain surface structural models with different site combinations (Figure S20). Figures S21-S24 display the optimized structures of reaction intermediates and the relevant free energy diagrams which were calculated by using the computational hydrogen electrode (CHE) method.[31] The free energy diagrams in Figure S21 indicate that despite site 3 exhibiting the weakest CO binding energy, it is thermodynamically more conducive for HCO* to undergo dehydrogenation to form CO*. Consequently, the MOR process predominantly favors the CO pathway, which diverges from experimental observations. Additionally, stable HCO* cannot be obtained at some sites after structural optimization, making it unclear whether the CO-free pathway occurs at these sites (Figures S23 and S24). Apparently, the results from Gibbs free energy diagrams obtained under vacuum conditions are insufficient to thoroughly explain the fundamental reasons behind the enhancement of the CO-free pathway selectivity due to Bi modification. This is primarily because the thermodynamic metal-vacuum models used in DFT calculations overlook the crucial role of the electrolyte and the associated dynamic characteristics.

**AIMD Simulations for the Bi-Pt(110)/Electrolyte Interface Structure.**

In order to further explore the origin of selectivity and to more accurately reproduce the real electrochemical experimental conditions, we performed AIMD simulations on the Bi-Pt(110) electrolyte interface system incorporating a water layer. The credibility of electrode potential calculation of the Bi-Pt/electrolyte interfaces can be verified by calculating potential of zero charge (PZC). Figure S25 shows that the PZC of the Pt(110)/water interface was calculated to ca 0.08 $V_{SHE}$, which aligns with the experimental value. After Bi modification of the Pt(110) surface, the PZC is increased up to approximately 0.33 $V_{SHE}$, which weakens the interaction between the surface and water molecules, thereby reducing the charge redistribution at the interface.[32] Meanwhile, K cations were introduced into the water film to simulate the alkaline conditions and to modulate the electrode potential to approximately 0.5 $V_{RHE}$, corresponding to pH = 14 (Figure 4A). The required number of K cations for Bi-Pt(110) is less than that of Pt(110) to reach the applied electrode potential, suggesting its better connectivity of hydrogen-bond networks in the electric double layer for rapid hydrogen transfer process (Figure S26).[33] Based on preliminary validation and observations of the computational model, the dynamic behavior of $CH_3OH$ molecule was simulated on the Bi-Pt(110)/electrolyte interface. Different from the DFT results (Figure S27), it can be observed that $CH_3OH$ cannot steadily adsorb on Bi-Pt surface during AIMD simulations. Even the electrode potentials of systems are driven to higher values by adjusting the number of K cations or introducing an $OH^-$ anion, $CH_3OH$ remains distant from the surfaces (Figure S28). This suggests that the initial proton-coupled electron transfer step (PCET) may encounter high kinetic free energy barriers due to the weak interaction between $CH_3OH$ and the surface.

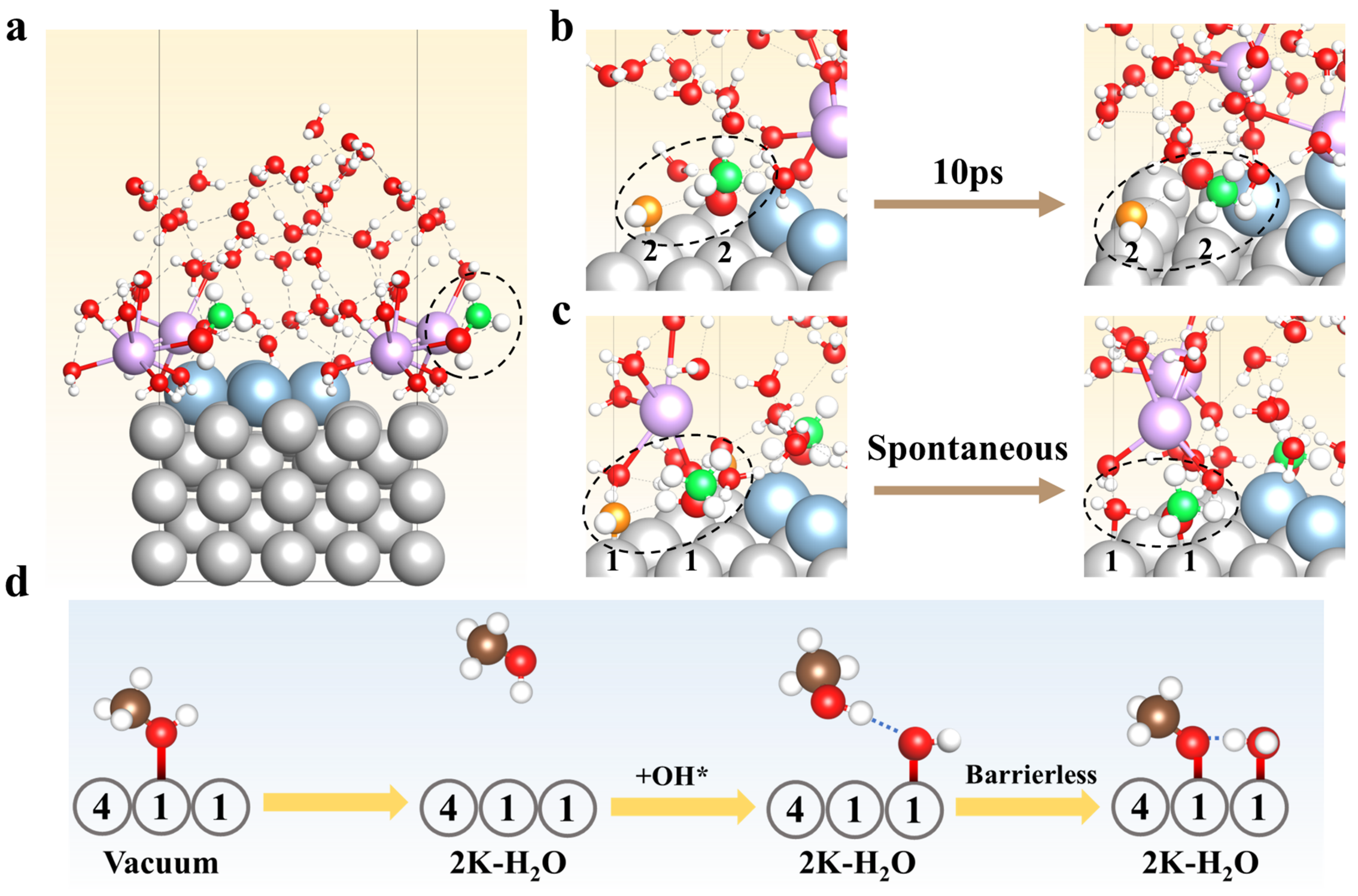

**Figure 4. The first proton transfer step and the identification of active site.** (a) Representative snapshot of the interfacial structure on Bi-Pt (110) at alkaline MOR potential. (b) Representative snapshots from AIMD

trajectory when OH adsorbs at site 2 and $CH_3OH$ is close to top of adjacent site 2. (c) Representative snapshots of from AIMD trajectory when OH adsorbs at site 1 and $CH_3OH$ is close to top of adjacent site 1. (d) Illustration of influence of alkaline interface and OH* on methanol activation. $CH_3OH$ and OH* are highlighted in black ellipses. Color code: Pt, gray; K, purple; Bi, blue; O, red; O in OH*, orange; C, green; and H, white.

Inspired by the findings reported by Herrero and his co-workers, which indicate that adsorbed OH (OH*) may play a crucial role in the initial PECT step, we further introduced an OH* onto the Bi-Pt surface.[18,34] Considering the OHBE results, the OH* was placed at site 1 or site 2, with the $CH_3OH$ above it. As shown in Figure 4B, during 10 ps AIMD simulation, $CH_3OH$ can only remain stably connected to the surface via hydrogen bond when OH* is located at site 2 and $CH_3OH$ is suspended above the adjacent site,2. In contrast, when OH and $CH_3OH$ are relocated to site 1, the O−H bond in $CH_3OH$ aligns with OH* via hydrogen bond connection and undergoes spontaneous dehydrogenation (Figure 4C). In addition, when $CH_3OH$ suspended above site 1, OH* located at site 2 also can promote the activation of $CH_3OH$ (Figure S29). The $CH_3OH$ activation will only take place in the presence of OH* at site 1, corresponding to favorable barrierless hydrogen transference from the hydroxy group of $CH_3OH$ to OH*. Consequently, as illustrated in Figure 4D, only site 1 can sever as the active site of methanol dissociation on Bi-Pt surface. The initial dissociation step occurs when H in the hydroxyl group of $CH_3OH$ is attracted to the adsorbed OH, resulting in the formation of $CH_3O$* and adsorbed $H_2O$. Such dynamic behavior is consistent with the KIE results (Figures 1D-E), where replacing OH with OD has a negligible impact on the activation of $CH_3OH$. Therefore, the updated model which introduced an OH* on the site 1 better reconciles experimental and computational results.

**Fundamental Factors Affecting Selectivity and Ture Role of Bi Modification**

According to the classical reaction network displayed in Figure S30, we proceeded to simulate models of HCO* and HCHO*, which are potential intermediates for determining reaction selectivity. Previous study suggested that the key determinant of selectivity depends on whether HCO intermediate dehydrogenates to CO* or reacts with OH* to form HCOOH*.[21,35-37] However, our AIMD simulation results indicate that HCO* is a thermodynamically unfavorable intermediate species. Figure 5A shows the initial structure of HCO* and OH* adsorbed on Pt(110). It can be observed that HCO* undergoes spontaneous dehydrogenation to produce CO* and H*during AIMD stimulation. Such phenomenon also occurs on Bi-Pt(110), suggesting that the selectivity of reaction pathway is not determined by the HCO intermediate (Figure S31).

Drawing inspiration from the alkaline formaldehyde electrooxidation mechanism, the evaluation of the electrocatalytic activity of Pt/C and Bi-Pt/C for formaldehyde (HCHO) oxidation reaction (FOR) can provide valuable insights into the alkaline MOR mechanism.[38-40] Electrochemical tests in Figures S32A-B indicate that both Bi-Pt/C and Pt/C catalysts exhibit higher activity for the FOR under alkaline conditions than for the MOR, suggesting an important role of HCHO intermediate in pathway selectivity. Besides, Bi modification significantly narrow the activity gap between FOR and MOR, which can be ascribed to the significant impact of Bi adatoms on the dynamic behavior of the HCHO intermediate. As the FOR mechanism depicted in Figure S32D, if the MOR process follows a CO-free pathway, HCHO* should desorb from the catalyst surface and diffuse into the bulk solution, subsequently undergo hydration and deprotonation to form the $H_2COOH$- anion (Figure S33). Conversely, if HCHO* prefers to undergo the dehydrogenation, the CO pathway will predominate. Thus, the desorption and dehydrogenation barriers of HCHO* are critical determinants of reaction pathway selectivity.

To substantiate the hypothesis derived from the electrochemical results on HCHO regarding its impact on MOR selectivity, we utilized the slow-growth method to calculate the critical energy barriers of HCHO. Additionally, to accurately replicate the electrode potentials in electrochemical experiments, we implemented the "constant-potential hybrid-solvation dynamic model" (CP-HS-DM) developed by Liu et al.[41] Figures 5B and S34 show the representative CP-HS-DM simulation snapshots for OH* and HCHO* adsorbed on Bi-Pt(110) and Pt(110) models, respectively. As illustrated in Figure S35, the reaction collective variable (CV) during HCHO* desorption was defined as CV = d1+d2, while C−H bond length was used as the CV in HCHO* dehydrogenation. As shown in Figure S36, for Pt(110) model, the free energy profile exhibits a 0.21 eV barrier for C–H bond cleavage of HCHO* while HCHO* desorption requires higher barrier (0.63 eV). Notably, when HCHO adsorbs on the Pt-Bi(110), with its oxygen and carbon ends binding to site 1 and adjacent site 4 respectively, the desorption barrier is reduced to 0.51 eV while C-H bond activation requires overcoming a significantly higher reaction barrier of 1.41 eV (Figures 5C and S37). Due to the weak interaction between Bi-Pt surface and intermediates, once HCHO* desorbs from surface into the bulk solution, it can easily remain in aqueous states [HCHO(aq)] and then HCHO(aq)-to-$H_2COOH^-$ conversion occurs at interfacial region (Figure S38). As schematically illustrated in Figure 5D, the selectivity of the reaction pathway is primarily determined by HCHO intermediate, rather than HCO intermediate. Meanwhile, the true role of Bi modification is to facilitate the desorption of HCHO* from the surface, thereby achieving high selectivity for CO-free pathway.

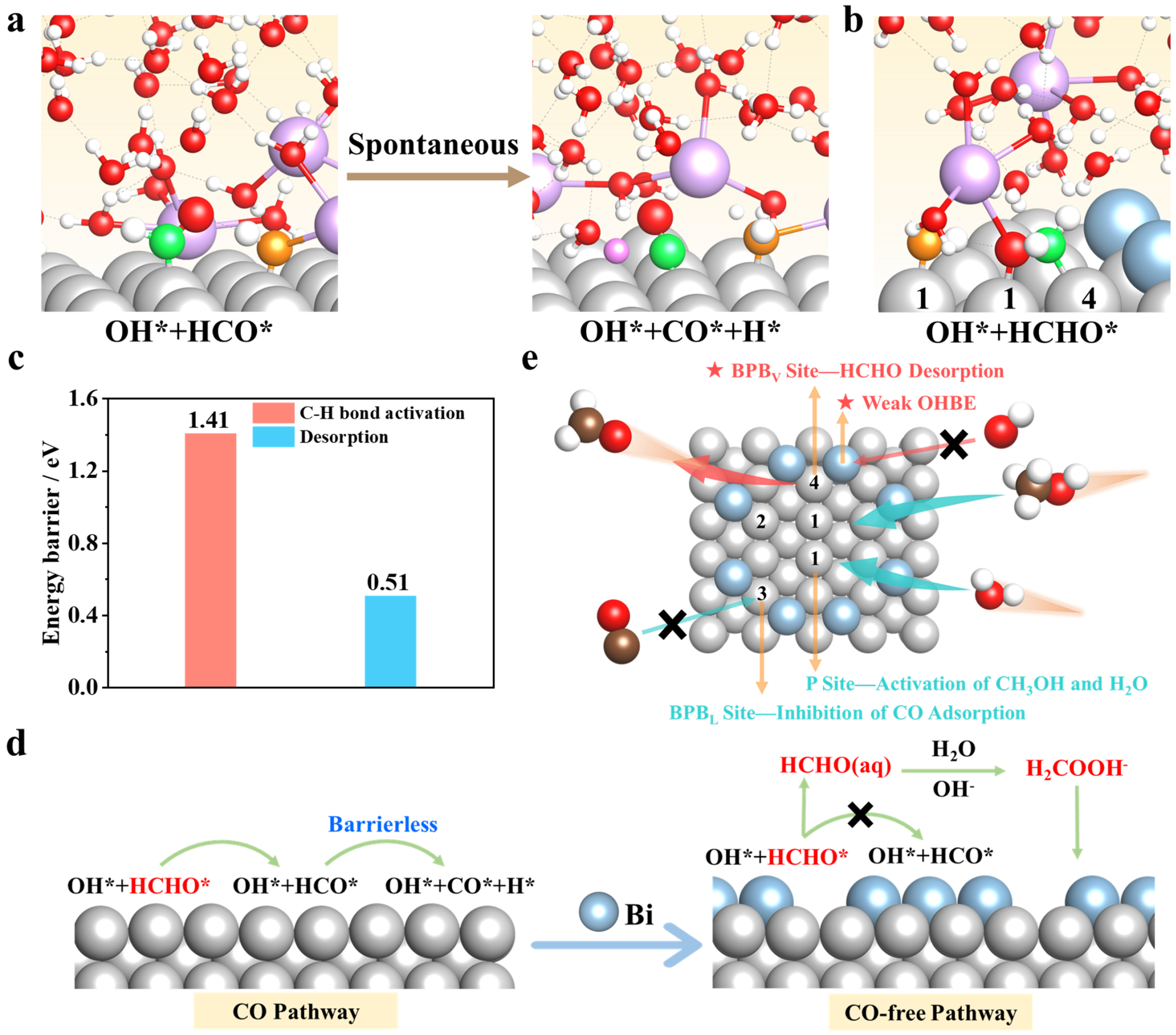

**Figure 5. The fundamental origin of pathway selectivity and the true role of Bi modification.** (a) Representative snapshots from AIMD trajectory when HCO and OH adsorb on Pt site of Pt(110). (b)

Representative snapshots of $OH^*$ and $HCHO^*$ on Bi-Pt (110). Color code: Pt, gray; Bi, blue; H; white; adsorbed H, pink; C, green; K, purple; O, red; and O in $OH^*$, orange. (c) A comparison of desorption barrier of $HCHO^*$ and its dehydrogenation barrier on Bi-Pt(110) at 0.5 $V_{RHE}$. (d) Illustration of the key role of HCHO intermediate and Bi adatoms in pathway selectivity. (e) Elucidation of the multi-sites effect on Bi-Pt Surface and identification of two Intrinsic factors pivotal for high selectivity towards the CO-free pathway.

Building on the previous analysis of thermodynamic calculations and AIMD simulations at the four sites, we summarize the multi-site effects on the Bi-Pt(110) surface in Figure 5E. Site 1 serves as the active site for activation of H2O and $CH_3OH$, subsequently leading to the formation of $OH^*$ and $CH_3O^*$. The weak COBE at site 3 protects against CO poisoning and helps suppress CO diffusion and migration, thereby preventing continuous CO generation. More importantly, both site 4 and the weak OHBE at Bi adatoms are crucial for achieving high selectivity towards CO-free pathway. Specifically, site 4, with its V-shaped Bi-Pt-Bi trio configuration, interrupts the continuity of Pt sites, which greatly impedes C-H bond activation. As a result, the formation of HCHO at this site promotes its desorption from Bi-Pt surface rather than further dehydrogenation. In addition, the weak OH binding energy at Bi adatoms effectively prevents OH species from blocking the $BPB_V$ sites, thus guaranteeing the fulfillment of their functional effect. In contrast, the dehydrogenation of $HCHO^*$ is facilitated on Pt surfaces due to large Pt atomic ensembles and the unimpeded diffusion and migration of CO result in consistently poisoning the active sites, thereby hindering the MOR process (Figure S39).

**Complete Alkaline MOR Pathway Landscape**

Subsequently, we further elucidated the complete alkaline MOR pathway mechanism on the Bi-Pt surface, along with the associated RDS. Figure 6A-B presents the representative CP-HS-DM simulation snapshots for the models of $CH_3O^*$ and $H_2COOH^*$ on Bi-Pt(110). In both models, the oxygen ends of the intermediate bind to site 1, and C−H bond length was used as the CV in dehydrogenation step, as illustrated in Figure S40. The calculation details were consistently maintained to evaluate the kinetic barriers. As shown in Figures 6C and S41, it can be observed that the C–H bond cleavage of $CH_3O^*$ requires overcoming a kinetic barrier of 0.99 eV, significantly higher than the kinetic barrier of $H_2COOH^*$ (0.70 eV), indicating that the conversion of $CH_3O^*$ to $HCHO^*$ is the RDS. The underlying reason is that the C–H bond in $CH_3O^*$ exhibits lower polarization compared to that in $H_2COOH^*$ (Figure S42), which impedes the ability of surface d-orbital electrons to fill the σ* antibonding orbital of the C-H bond, thereby resulting in a substantial energy barrier. Additionally, we also investigated the cation effect on electrocatalytic kinetics by changing the cation types. Figure S43 shows similar MOR activities with different types of cation conditions, indicating that the cation effect on RDS can be ignored. Eventually, the schematic diagrams of the alkaline MOR pathway on Bi-Pt surface are shown in Figure 6D. Overall, before the activation of $CH_3OH$, $OH^*$ from water dissociation first adsorbs on the active site (site 1), which can promote the immediate activation of $CH_3OH$. Subsequently, $HCHO^*$, the key intermediate, is generated via cleavage of C-H bond and then desorbs from surface. As the next step, HCHO(aq) can be converted to $H_2COOH_-$ anion at alkaline interface through a hydration and deprotonation process, which further adsorbed on surface to produce HCOOH.

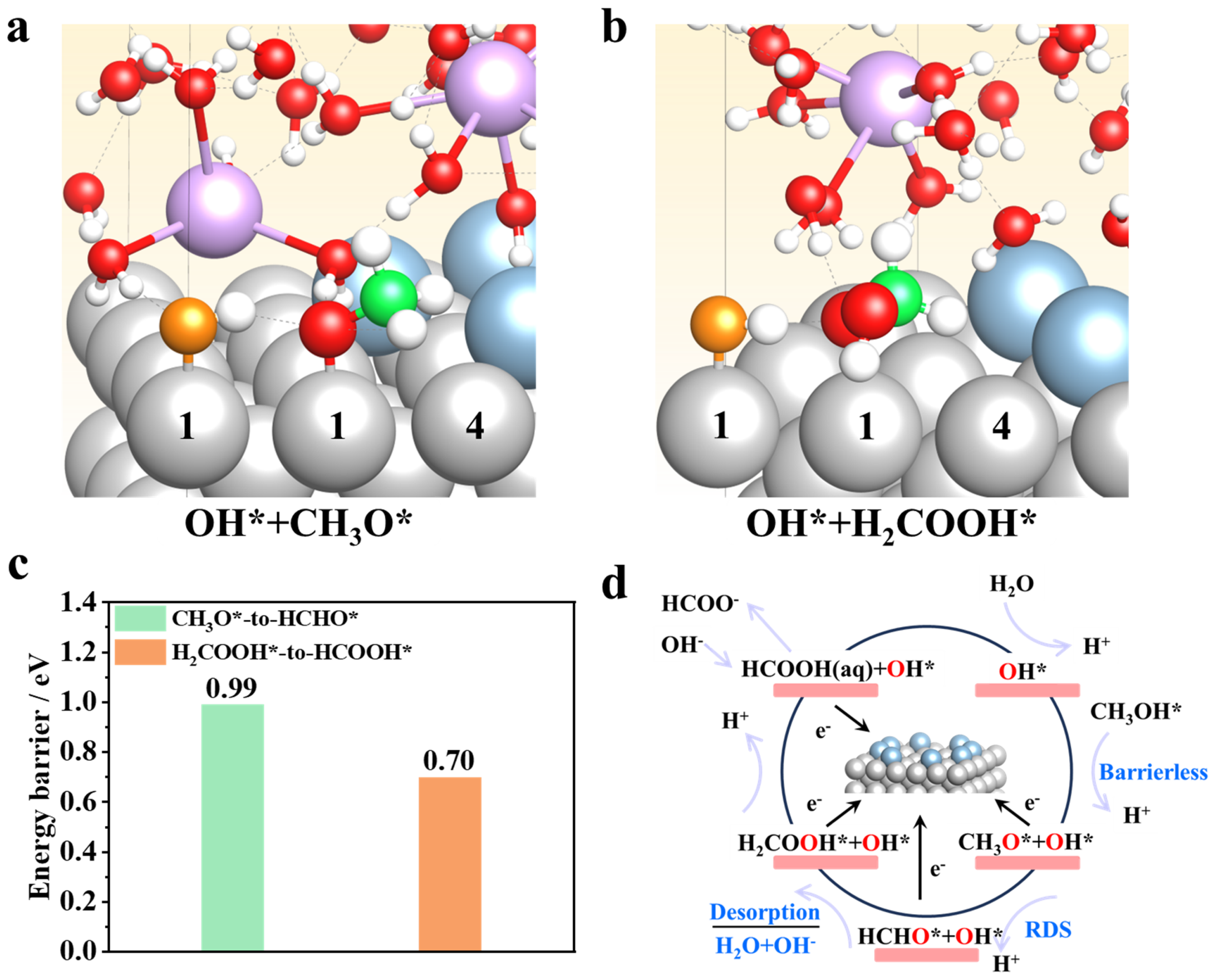

**Figure 6. The complete mechanism for alkaline MOR and corresponding RDS on Bi-modified Pt surface.** (a and b) Representative snapshots of (a) OH*+$CH_3O$* and (b) OH*+$H_2COOH$* on Bi-Pt (110). (c) A comparison of kinetic barriers for both $CH_3O$*-to-HCHO* conversion and $H_2COOH$*-to-HCOOH* conversion on Bi-Pt surface at 0.5 $V_{RHE}$. (d) Schematic diagrams of the alkaline MOR mechanism on Bi-Pt surface. Color code: Pt, gray; K, purple; Bi, blue; O, red; O in OH, orange; C, green; and H, white.

After confirming the alkaline MOR pathways on the Bi-Pt surface and the mechanisms of Bi modification, we also observed that the pH effect significantly influences the selectivity of the reaction pathways on the Bi-Pt surface.[22] In Figure S44, Bi-Pt/C exhibits similar MOR activity to Pt/C in acidic conditions at low over-potentials and significantly less than PtRu/C, suggesting that Bi-Pt/C follows a CO dominated pathway under acidic conditions. Meanwhile, as shown in Figure S45, CO-stripping experiments indicates that PtRu/C exhibits the lowest CO oxidation onset potentials and peak potentials under both acidic and alkaline conditions. A significant reduction in CO current on Bi-Pt/C compared to Pt/C is also observed under acidic condition. This finding further confirms that the CO tolerance is not the dominant factor influencing the selectivity of the MOR pathways.

The pH effect may modulate pathway selectivity by altering the charge state of the electrode surface or the structure of the electrical double layer.[33,41,42] To validate this hypothesis, we performed AIMD calculations to simulate Bi-Pt(110)/electrolyte interfaces under acidic MOR conditions. Different form alkaline condition, Bi-Pt surface is almost positively charged under acidic condition, cause its PZC is ca. 0.33 VSHE. As shown in Figure S46, we introduced a $CH_3OH$ and a $ClO_4^-$ anion into water film to simulate acidic conditions. It can be observed that when OH* is located at site 2, the adjacent $CH_3OH$ can spontaneously undergo hydrogen transfer through hydrogen bond, thereby forming $CH_3O$* at the neighboring site 2. In other words, site 2 can also serve as the active site for the initial activation of methanol at acidic interface. Furthermore, we utilized the CP-HS-DM to control potential at 0.5 $V_{RHE}$ (pH=1) and conducted a 10 ps simulation. As illustrated in Figure S47, after 10 ps, two $H_2O$ molecules adsorb at site 1, while $CH_3OH$ only remains suspended above site 2 by hydrogen bond connection. This indicates that under acidic conditions, $CH_3OH$ is more inclined to activate at site 2, and unable to promote methanol decomposition through site 1 and site 4, leading the MOR process towards the CO dominated pathway. Evidently, the pH effect alters the initial methanol dissociation active sites by influencing the charge state of the electrode surface, thereby affecting the selection of the MOR pathway. This can explain why Bi-Pt/C can only proceed with the CO-free pathway under alkaline conditions. Our simulation results effectively elucidate a series of experimental phenomena and validate the rationality of the metal/electrolyte interface model construction.

## Conclusion

In summary, a Bi-modified Pt/C model catalyst prepared by underpotential deposition method was used to investigate the selectivity origin of methanol electrooxidation. We demonstrate that Bi-modified Pt/C follow a CO-free dominated pathway under alkaline conditions, as evidenced by kinetic isotope effects, formate Faradaic efficiency, and in situ infrared spectroscopy. Furthermore, a detail mechanism study is provided by AIMD stimulations integrated with slow-growth method as well as DFT calculations. Our results reveal that the HCHO intermediate is the essential factor that affects the selectivity and Bi modification significantly increases the dehydrogenation barrier of HCHO* compared to its desorption barrier, thereby promoting its conversion to $H_2COOH^-$ anion at the alkaline interface and fundamentally eliminating the CO pathway. More importantly, we provide a comprehensive understanding of the multi-sites effect on the Bi-Pt surface, highlighting the formation of the $BPB_V$ sites and the weak OHBE at Bi adatoms as two key factors that facilitate the desorption of HCHO.

## Supplemental Information

This section should include the titles and (optional) legends of all supplemental items. Document S1 is the main supplemental PDF: Document S1. Figures S1–S47 and Table S1-S2.

## Acknowledgements

This work was financially supported by the National Natural Science Foundation of China (22072048 and 22372062) and Guangdong Provincial Department of Science and Technology (2022A0505050013). The computation in this work has been done using the facilities (supercomputer Ohtaka) of the Supercomputer Center, the Institute for Solid State Physics, the University of Tokyo. The authors acknowledge Beijing PARATERA Technology Co., LTD for providing high-performance and AI computing resources for contributing to the research results reported within this paper. URL: http://cloud.paratera.com. The authors would like to thank Prof. Yuanyue Liu for sharing the CP-VASP code, and to thank Dr. Xiaowan Bai for providing valuable insights on free energy calculations.

# Supporting Information

## Unveiling High Selectivity Origin of Pt-Bi Catalysts for Alkaline Methanol Electrooxidation via CO-free pathway

Lecheng Liang,[1, 5] Hengyu Li,[2, 5] Peng Li,[4] Jinhui Liang,[1] Shao Ye,[1] Binwen Zeng,[1] Mingjia Lu,[1] Yanhong Xie,[3] Yucheng Wang,[3] Taisuke Ozaki,[2] Shengli Chen,[4] and Zhiming Cui[1, *]

[1] Guangdong Provincial Key Laboratory of Fuel Cell Technology, School of Chemistry and Chemical Engineering, South China University of Technology, Guangzhou 510641, China.

[2] Institute for Solid State Physics, The University of Tokyo, Kashiwa 277-8581, Japan

[3] State Key Laboratory of Physical Chemistry of Solid Surfaces, College of Chemistry and Chemical Engineering, Xiamen University, Xiamen 361005, China.

[4] College of Chemistry and Molecular Sciences, Wuhan University, Wuhan, China.

[5] These authors contributed equally to this work.

**Corresponding Author:**

*E-mail: zmcui@scut.edu.cn

## Experimental Procedures

### Resource availability

*Lead contact*

Further information and requests for resources should be directed to and will be fulfilled by the lead contact, Chiming Cui (zmcui@scut.edu.cn).

*Materials availability*

All materials generated in this study are available from the lead contact without restriction.

*Data and code availability*

This study did not generate any datasets.

### Chemicals

Bismuth nitrate pentahydrate ($Bi(NO_3)_3 \cdot 5H_2O$), methanol ($CH_3OH$, 99.5 %), and potassium hydroxide (KOH, 95%) were all purchased from Aladdin. Formaldehyde (HCHO, 37 %) was purchased from MACKLIN. Deuterium-substituted methanol ($CD_3OD$ and $CH_3OD$; 99.8 atom % D) were purchased from Energy Chemical. Sulfuric acid ($H_2SO_4$, AR) and perchloric acid ($HClO_4$, GR) were purchased from Guangzhou Chemical Reagent Factory. Ketjen Black EC300J was purchased from Suzhou Sinero Technology Co., Ltd. Commercial Pt/C was provided by was provided by TANAKA (20 wt.%). Commercial PtRu/C was provided by Johnson Matthey (30 wt.%). All regents were used without further purification, and all solutions were freshly prepared with ultrapure water (18.2 MΩ $cm^{-1}$).

### Preparation of Bi-modified Pt/C

As shown in Figure S1, the Bi-modified Pt/C was synthesized simply using the room-temperature electrodeposition method. 2.0 mg of Pt/C was dispersed in a mixed solvent containing 950 μl isopropanol, 50 μl ultrapure water, and 20 μL Nafion by ultrasound for 1 hour to form a homogeneous ink. The 8 μL of ink was dropped onto a clean glassy-carbon electrode to prepare the working electrode (0.196 $cm^2$). The Bi-modified Pt/C was electrodeposited at -0.16 V vs. Ag/AgCl (saturated KCl) for 120 s from the 0.5M $H_2SO_4$ solutions of 10 mM $Bi^{3+}$ ions.

### Preparation of Bi/C

The Bi/C was synthesized simply using the room-temperature electrodeposition. 2.0 mg of Ketjen black was dispersed in a mixed solvent containing 950 μl isopropanol, 50 μl ultrapure water, and 20 μL Nafion by ultrasound for 1 hour to form a homogeneous ink. The 8 μL of ink was dropped onto a clean glassy-carbon electrode to prepare the working electrode (0.196 $cm^2$), and graphite rod was served as a counter electrode. The Bi/C was electrodeposited at -0.20 V vs. Ag/AgCl (saturated KCl) for 120 s from the 0.5M $H_2SO_4$ solutions of 10 mM $Bi^{3+}$ ions.

### Materials Characterization

A Talos F200x microscope was used to obtain transmission electron microscopy (TEM) images and energy dispersive X-ray spectroscopy (EDS) spectra, which were measured at 200 kV. STEM HighresolutionTEM (HRTEM), high-angle annular dark-field scanningTEM (HAADF-STEM) and HAADF-STEM energy dispersive X-rayspectroscopy (HAADF-STEM-EDS) were characterized by a Titan

Themis G2 at an accelerating voltage of 300 KV. The samples were analyzed by inductively coupled plasma optical emission spectrometry (ICP-OES) using an IRIS Intrepid II XSP instrument (Thermo Fisher). XPS was performed on an Axis Ultra DLD 1X-ray photoelectron spectrometer employing monochromated Al-Ka X-ray sources (hv=1486.6 eV).

**Electrochemical Measurement**

All the electrochemical measurements were performed on a PINE-WaveDriver 200 electrochemical workstation with a three-electrode cell. The counter, reference, and working electrodes consisted of a platinum mesh, standard Hg/HgO (1M KOH solution), glassy-carbon electrode (0.196 $cm^2$), respectively. The methanol oxidation reaction (MOR) polarization curves were obtained at the scan rate of 50 mV $s^{-1}$ in $N_2$-saturated 1M KOH/1M $CH_3OH$ and 0.1M $HClO_4$/0.5M $CH_3OH$ solutions, respectively. The chronoamperometry (CA) measurements of MOR were conducted at 0.45 V vs. RHE. The formaldehyde oxidation reaction (FOR) polarization curves were obtained at the scan rate of 50 mV $s^{-1}$ in $N_2$-saturated 1M KOH+ 1M HCHO solution.

For the CO-stripping measurements, the CO oxidation experiments were carried out in 1M KOH and 0.1M $HClO_4$ solutions, respectively. Graphite rod was served as a counter electrode. Before the test, the solution was bubbled with CO gas (99.9%) for 15 min to achieve the maximum coverage of CO at the Pt active centers. Afterward, the working electrode was quickly moved into a fresh $N_2$-saturated electrolyte and recorded at a scan rate of 50 $mVs^{-1}$.

**In situ SEIRAS Measurements**

The electrochemical in-situ attenuated total reflection surface-enhanced infrared absorption spectra (ATR-SEIRA) of methanol electrooxidation on Pt/C and Bi-Pt/C were collected using a Bruker Vertex 70V FTIR spectrometer equipped with a liquid-nitrogen-cooled MCT-A detector. Measurements were conducted in an IR cell with an Au film-coated silicon prism. The counter electrode was a platinum wire, and the reference electrode was a Hg/HgO electrode. The potential was converted to the reversible hydrogen electrode (RHE) scale using the following equation: E(RHE)=E(Hg/HgO)+0.059×pH+0.098

Prior to each experiment, Nitrogen was bubbled through the electrolyte for 20 minutes to remove dissolved air and the prism surface was cleaned by cycling the potential between 0 and 1.2 $V_{RHE}$ in a 1 M KOH solution at a scan rate of 100 mV $s^{-1}$. All spectral resolution was 4 $cm^{-1}$ and 64 interferograms were co-added for each spectrum in the 1M KOH/1M $CH_3OH$. In this work, single-beam spectra collected at the sample potential varied from 0.1 to 1.2 $V_{RHE}$ and the reference potential was selected at -0.25 $V_{RHE}$.

**Analysis of MOR Products**

The obtained liquid phase products after electrolysis were analyzed on an ion chromatography (Thermo Fisher Aquion). The faradic efficiency for the $HCOO^-$ product was calculated based on 4 electrons consumed for the formation of $HCOO^-$ and the total moles of electrons passed.

**Computations and Models**

All DFT and AIMD simulations are preformed using Vienna ab initio Simulation Package (VASP) with projector augmented wave (PAW) method.[43,44] The revised Perdew−Burke−Ernzerhof functional (RPBE) within the generalized gradient approximation (GGA) framework is employed to characterize electron exchange−correlation interactions. For DFT calculations, the kinetic energy cutoff is set to 520 eV and the Gaussian smearing width is set to 0.1 eV. The zero–damping method of Grimme (DFT-D3) is added to consider the dispersion corrections. For structure optimization, the convergence criteria are set to $10^{-6}$ eV and $10^{-2}$ eV $Å^{-1}$ for energy and force, respectively. A five-layer 3 × 4 Pt(110) slab were built for representing surface which the bottom three layers were fixed, and a vacuum of 15 Å is adopted along the z-axis. The Brillouin zone integration was sampled using a Monkhorst-Pack (3 × 3 × 1) k-point mesh. According to the computational hydrogen electrode (CHE) model proposed by Nørskov and co-workers, the free energy of the proton-electron pair is equal to that of 1/2 $H_2$ (g).[31] The free energy change for each fundamental step was determined by:

$$\Delta G = \Delta E + \Delta ZPE - T\Delta S$$

where $\Delta E$ is the difference of electronic energy directly obtained from DFT simulation. $\Delta ZPE$ is the contribution of variation of zero-point energy (ZPE), $\Delta S$ is the entropy (S) change, T is the temperature 300K. The ZPE and S of MOR intermediates were obtained by the vibrational frequencies.

For AIMD stimulations, the atomic core and valence electrons were expanded by PAW method and plane-wave basis functions, and the kinetic energy cutoff is set to 400 eV. The GGA with the RPBE exchange-correlation functional is employed to account for core–valence interaction. The Gaussian smearing with a width of 0.2 eV is employed. The DFT-D3 is added to consider the dispersion corrections. The convergence criteria is set to $10^{-5}$ eV for energy. The Pt(110)/water interface is modelled by adding 40 water molecules above a five-layer 3 × 4 Pt(110) slab with a vacuum of 15 Å, in which the bottom three layers were fixed. K cations are introduced into the Pt(110)/water interfacial model thus constructing the cation-containing interfaces. A time step of 1 fs and only the Γ k-point is sampled in the reciprocal space of the Brillouin zone throughout. The canonical ensemble condition (NVT) is imposed by a Nose-Hoover thermostat with a target temperature of 330 K. For each AIMD simulation, an initial 10 ps of MD trajectory is used to adequately equilibrate the system.

To replicate the electrode potential applied in experiments within AIMD simulations, we employed "constant-potential hybrid-solvation dynamic model" (CP-HS-DM) developed by Liu et al to conduct the constant potential MD.[41,45] Within the CP-HS-DM framework, the VASPsol implicit solvation model is employed to dynamically adjust the electron count during MD simulations, thereby ensuring that the potential of system aligns with the specified target value. Based on the well-equilibrated interface, the constrained AIMD (cAIMD) simulations with a slow-growth sampling approach were used to calculate the activation barrier by thermodynamic integration.[46] In this method, one suitable collective variable (CV, namely $\xi$) can be defined as the reaction coordinate, which is linearly changed from the initial state to the final state with a transformation velocity $\xi$. The work required to perform the transformation from initial to final states can be computed as:

$$W_{\text{initial-to-final}} = \int_{\xi(\text{ initial })}^{\xi(\text{ final })} \left(\frac{\partial F}{\partial \xi}\right) \cdot \dot{\xi} dt$$

where F is the computed free energy, which is evolving along with t, $\frac{\partial F}{\partial \xi}$ can be computed along cAIMD using the blue-moon ensemble with the SHAKE algorithm.[47] With the limit of infinitesimally small $\partial\xi$, the needed work (Winitial-to-final) corresponds to the free-energy difference between the final and initial states.

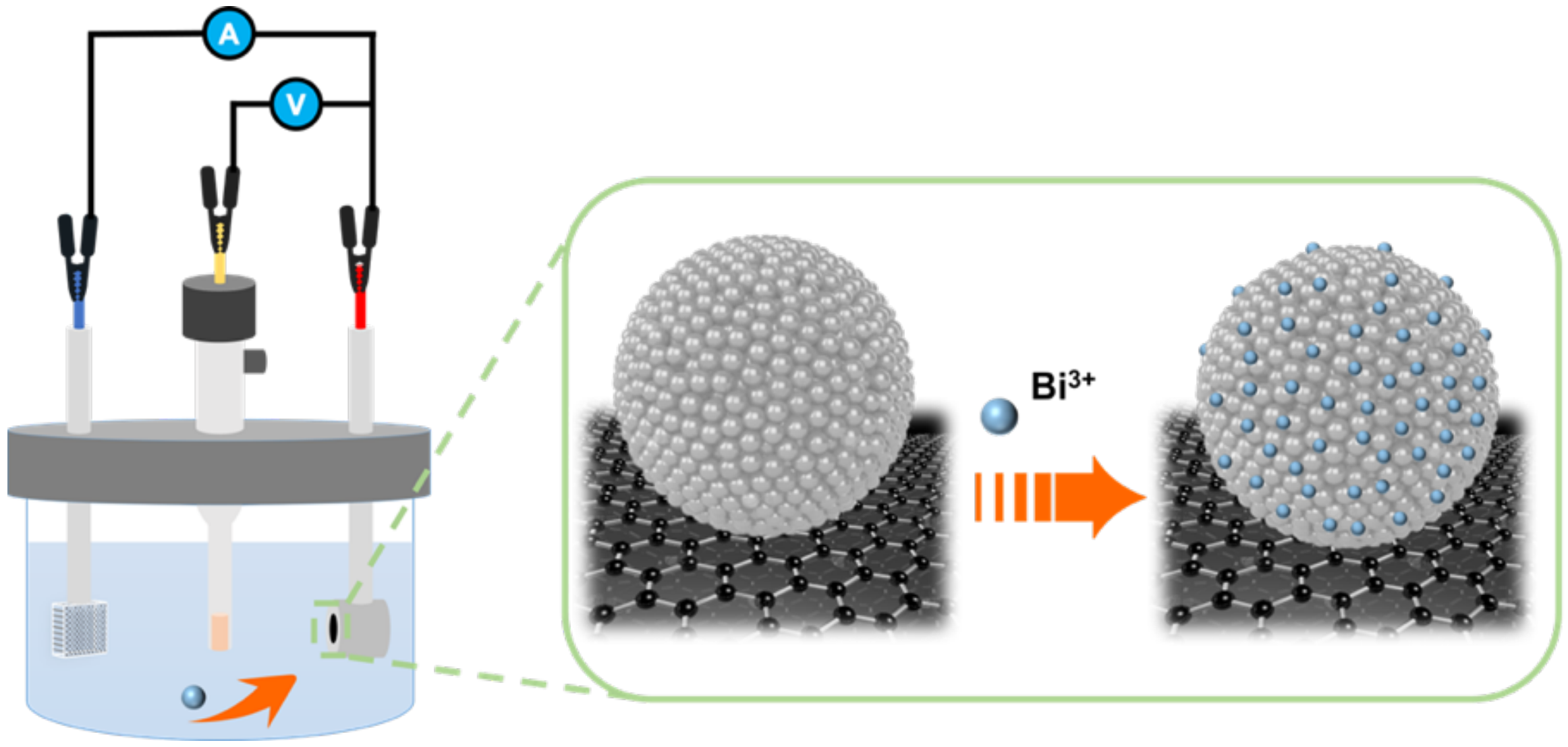

**Figure S1.** Schematic illustration of Bi-Pt/C preparation by electrodeposition method.

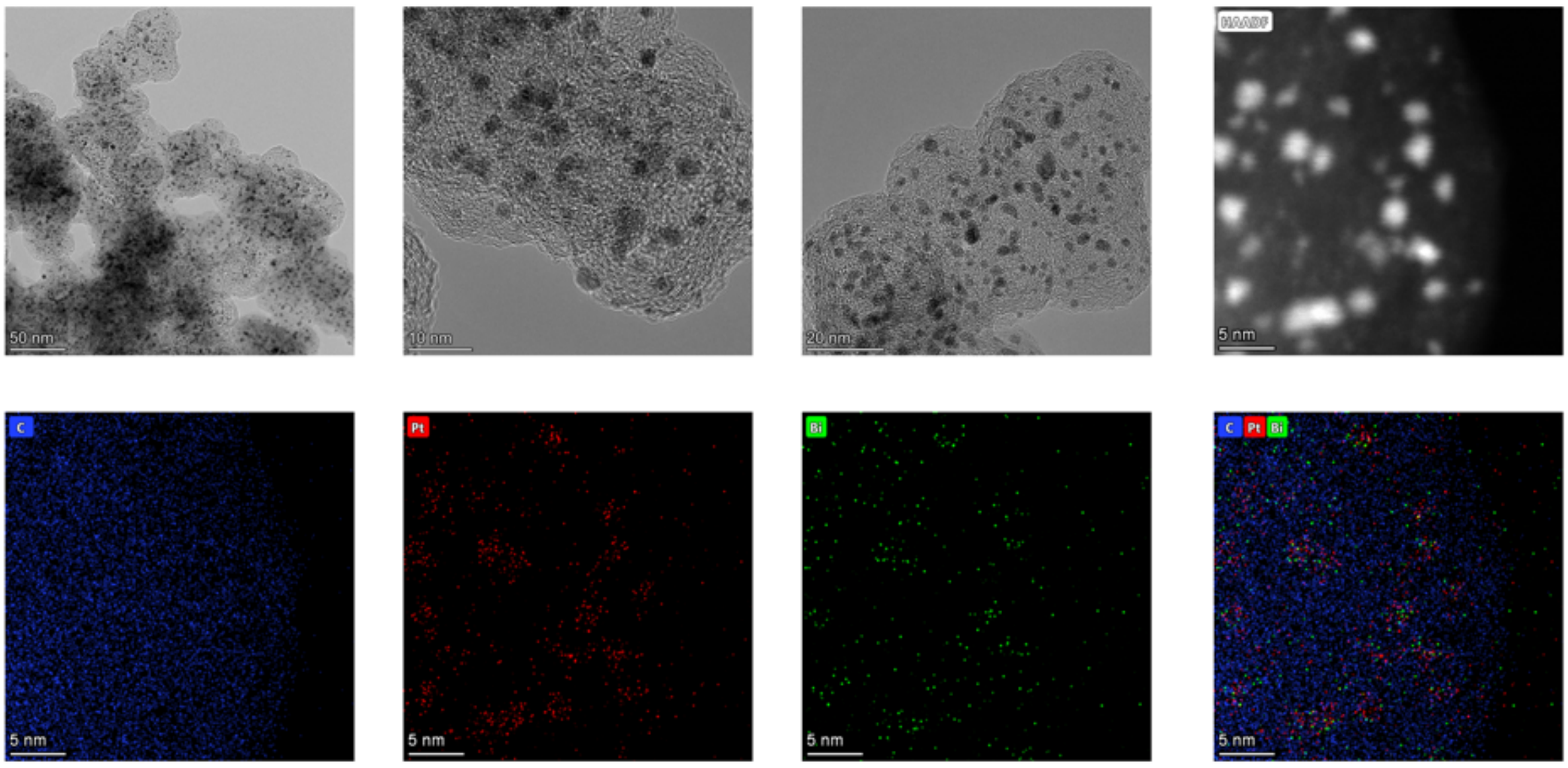

**Figure S2.** TEM and EDS mapping images of Bi-Pt/C.

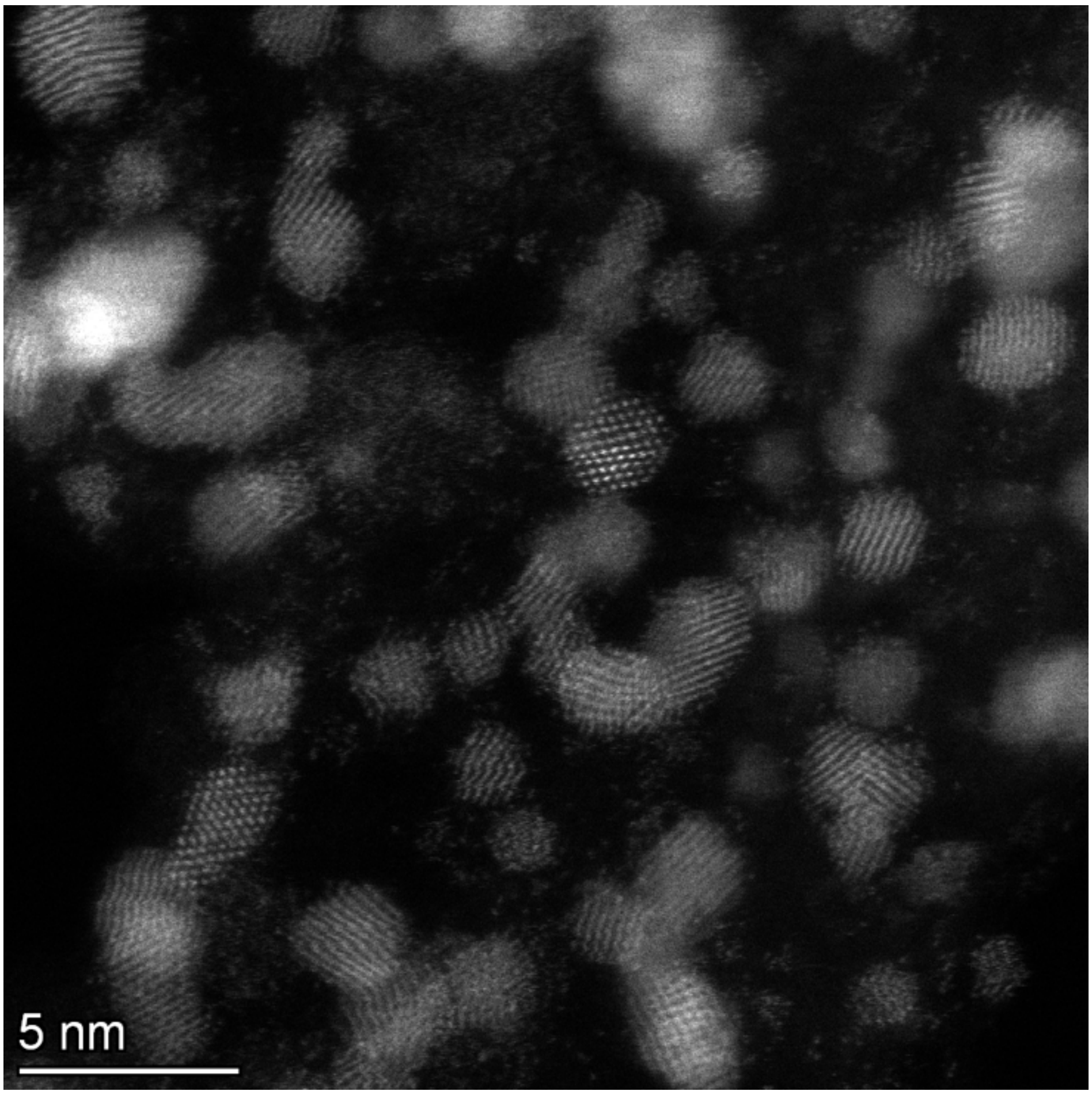

**Figure S3.** HAADF-STEM image of Bi-Pt NPs.

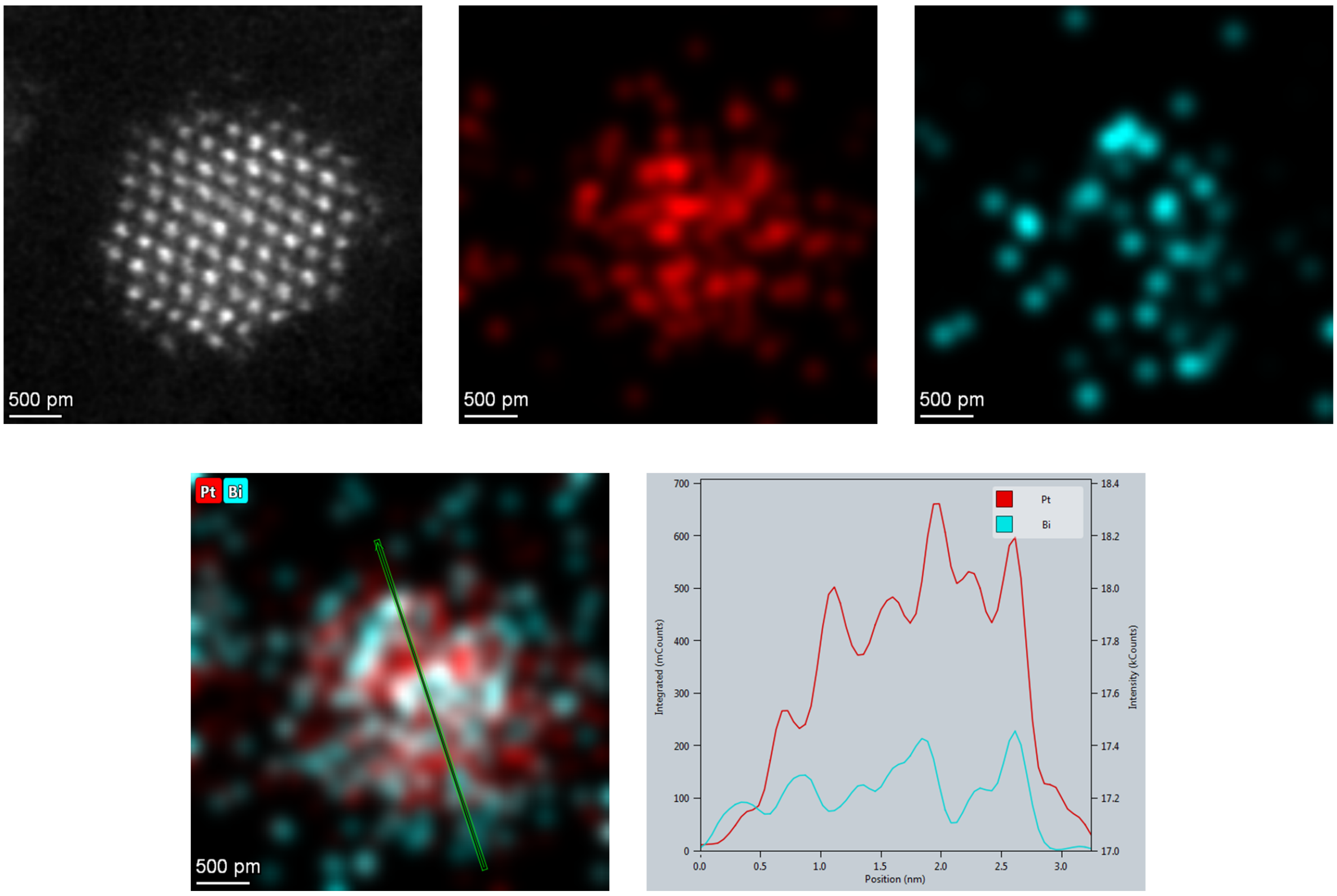

**Figure S4.** HAADF-STEM, EDS mapping, and EDS line images of a Bi-Pt NP.

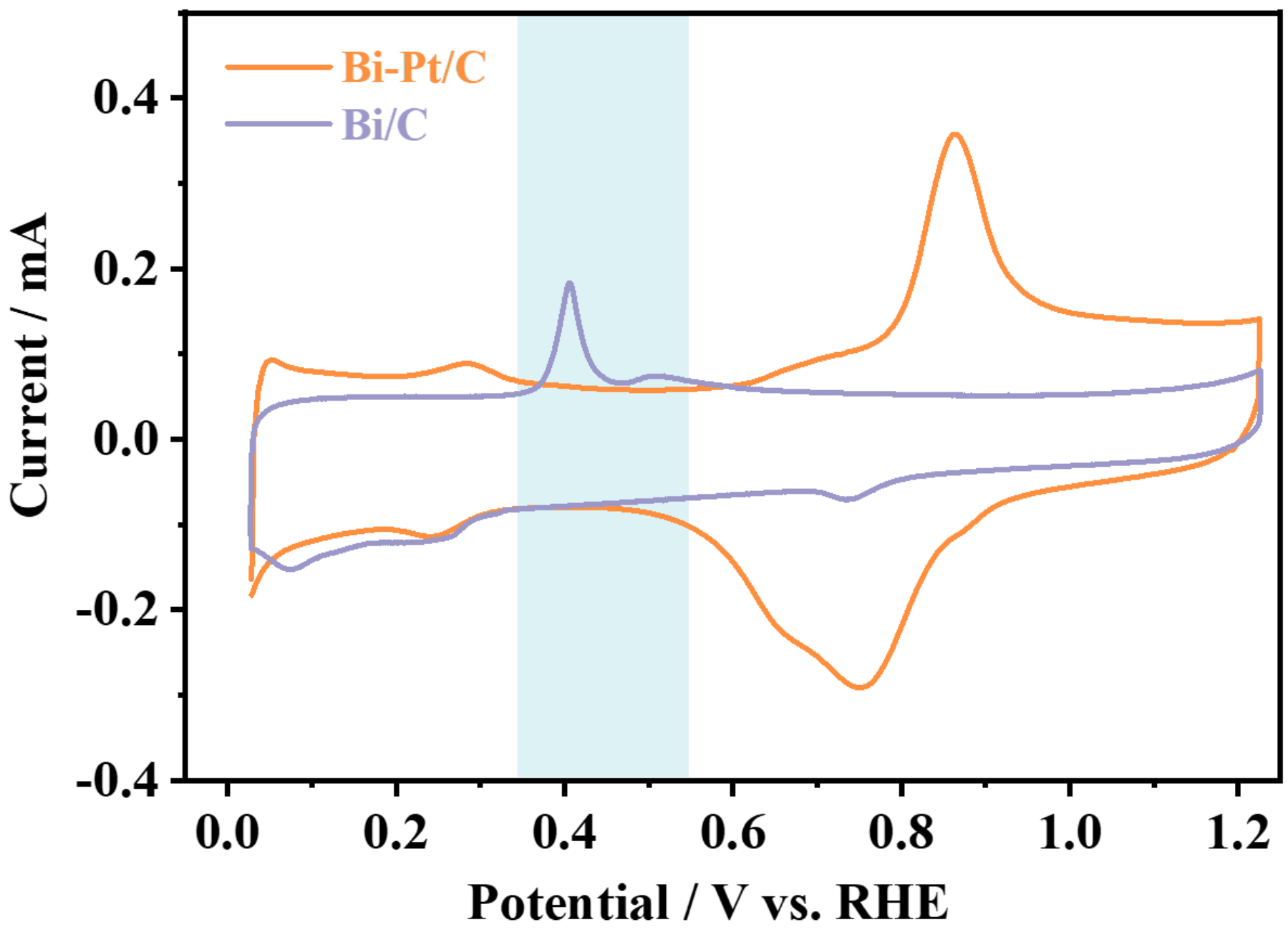

**Figure S5.** CV curves of Bi/C and Bi-Pt/C in 1M KOH solution.

Note for Figure S5: To deeper understand the property of Bi adatoms, we obtained Bi/C by a simple overpotential deposition method and compared its CV curve with that of Bi-Pt/C. It can be found that Bi/C shows an obvious oxidation current at approximately. 0.4 $V_{RHE}$, meaning the oxidation of the bulk phase while it does not happen at Bi-Pt/C. The result indicates that the electronic interactions between Pt atoms and Bi adatoms help to enhance oxidation resistance of Bi during electrochemical process, as supported by theoretical calculations.

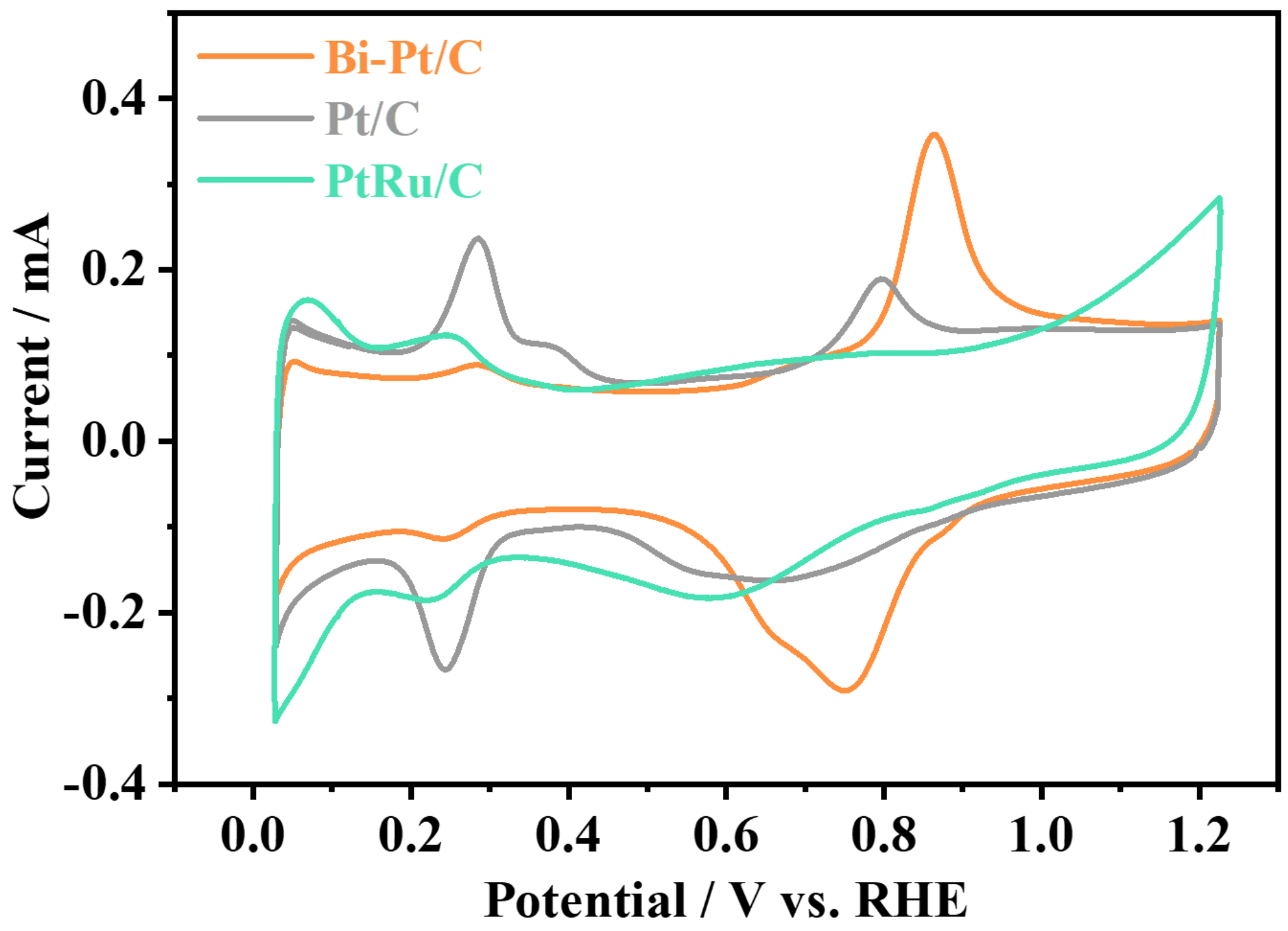

**Figure S6.** CV curves of different catalysts in 1M KOH solution.

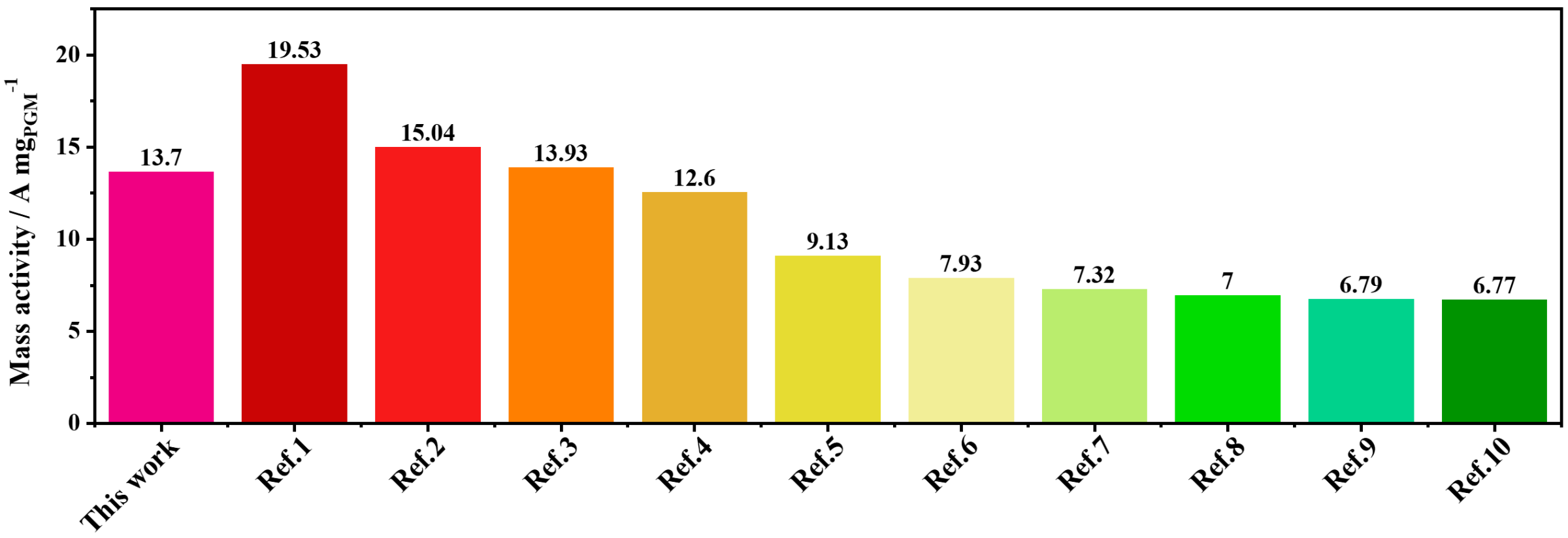

**Figure S7.** A comparison of mass activity between Bi-Pt/C and other reported catalysts. More details are listed in Table S1.

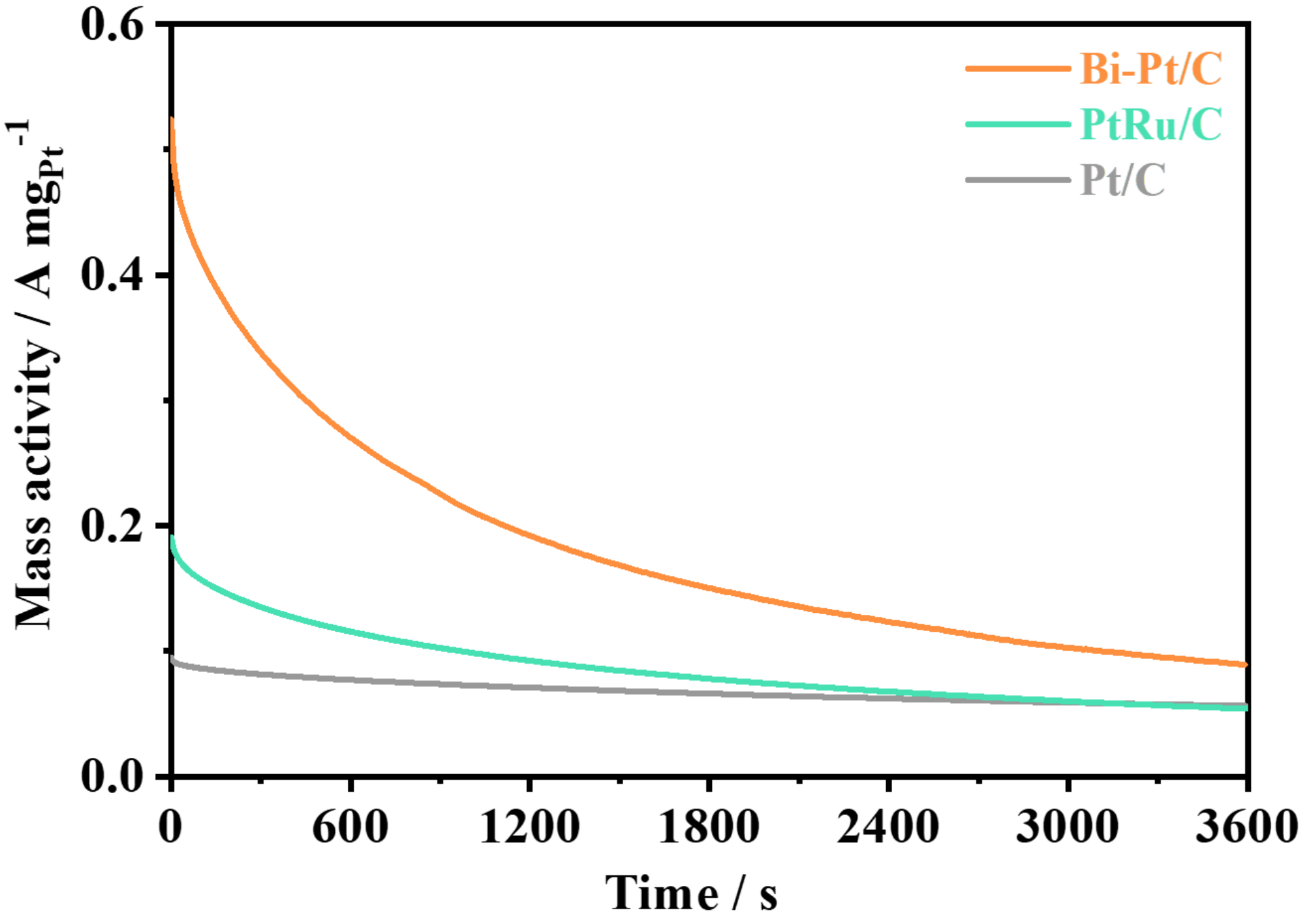

**Figure S8.** Mass-normalized i-t curves recorded at 0.45 V $_{RHE}$.

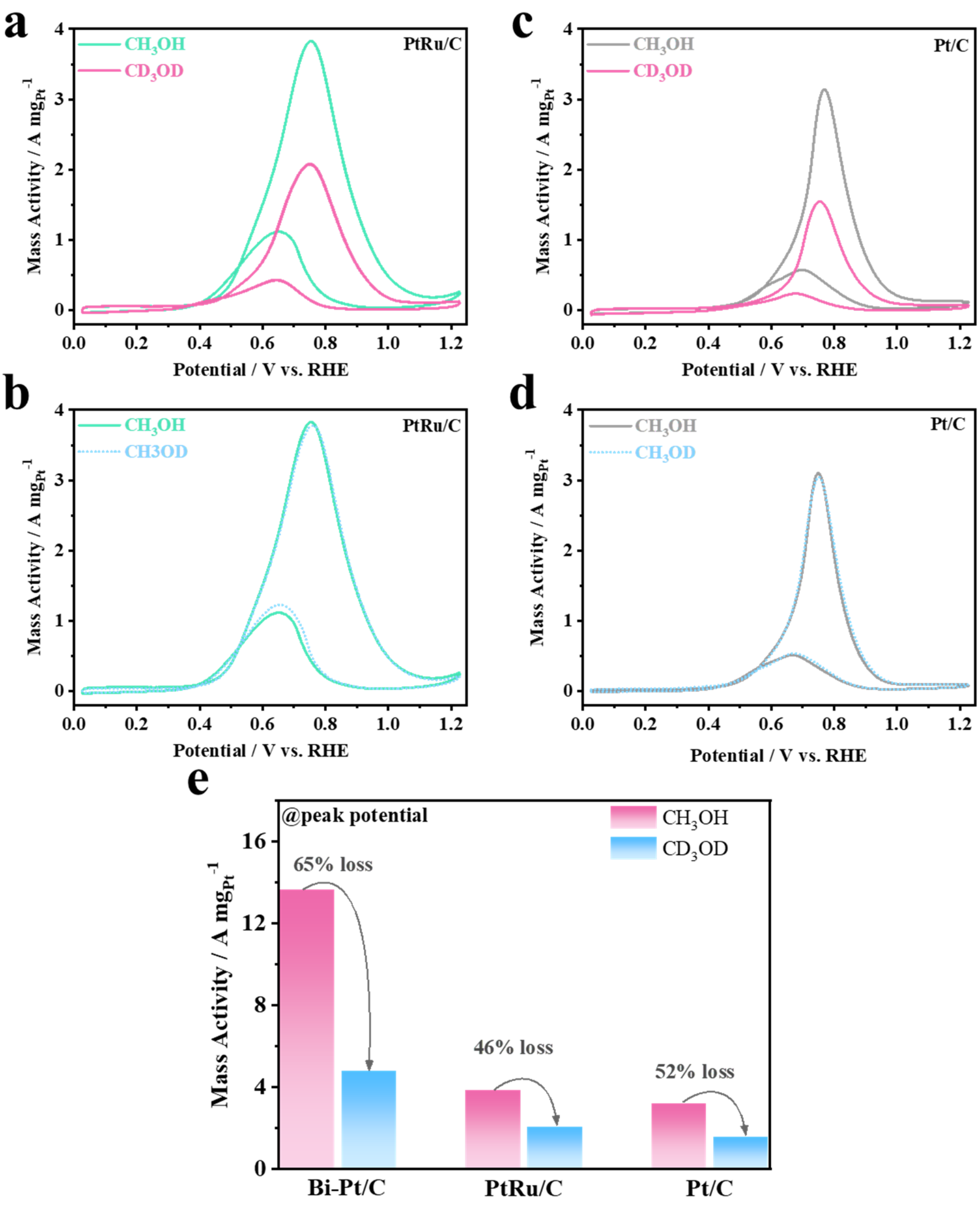

**Figure S9.** A comparison of CV curves of (A, B) PtRu/C and (C, DD) Pt/C under H/D exchange (1M $CH_3OH$ and 1M $CD_3OD$; 1M $CH_3OH$ and 1M $CH_3OD$). (E) A comparison of the mass activity between different catalysts under H/D exchange at peak potential.

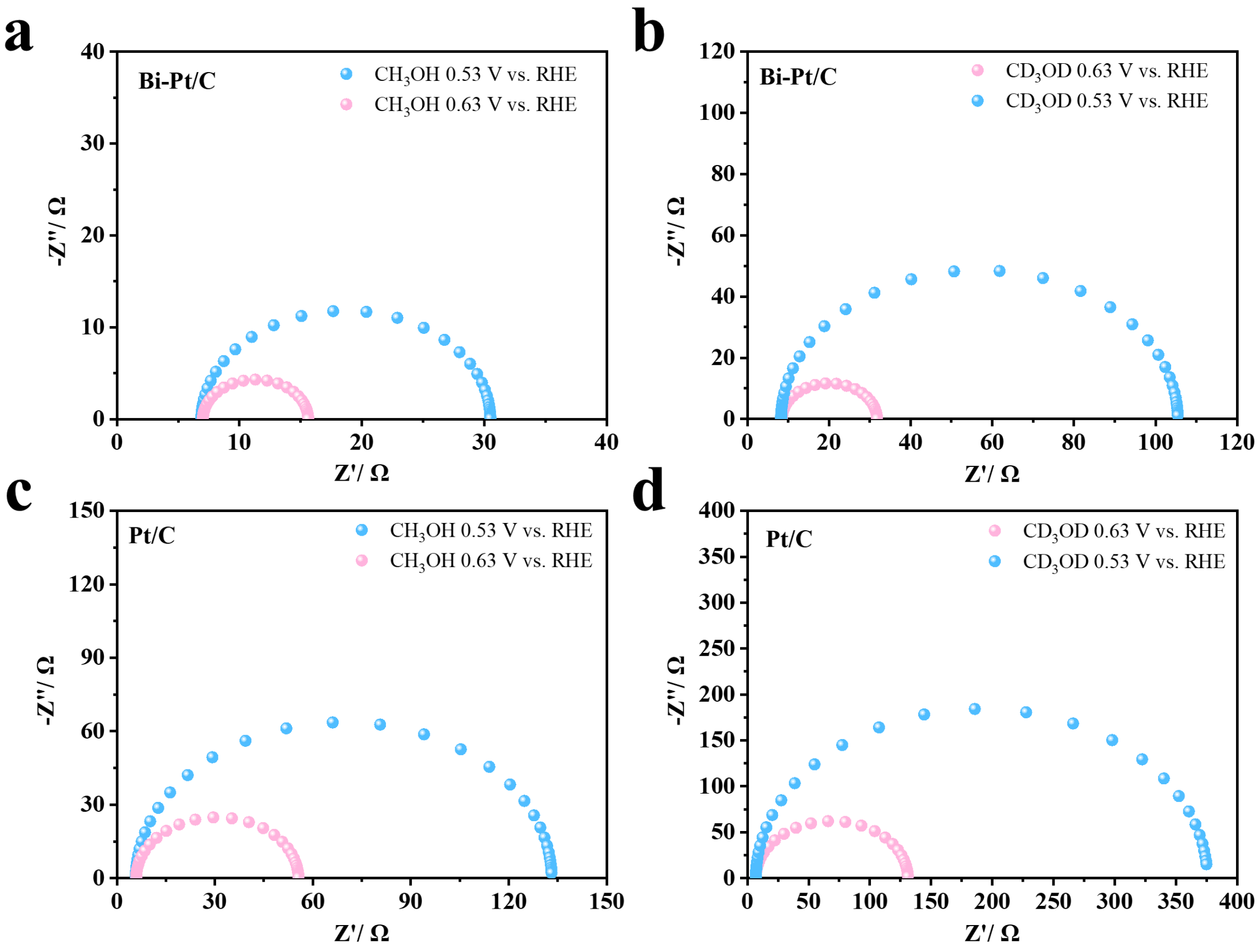

**Figure S10.** EIS plots of (A, B) Bi-Pt/C and (C, D) Pt/C under H/D exchange (1M $CH_3OH$ and 1M $CD_3OD$) at different potentials.

Note for Figure S10: When $CD_3OD$ was used to displace $CH_3OH$ in electrolyte, the obvious enhancement in electrochemical impedance can be detected for both Bi-Pt/C and Pt/C, indicating that H/D exchange can significantly affect alkaline MOR kinetics. In addition, Bi-Pt/C still represents a lower electrochemical impedance than Pt/C, whether in $CH_3OH$ or $CD_3OD$, indicating its superior charge-transfer rate and electronic conductivity.

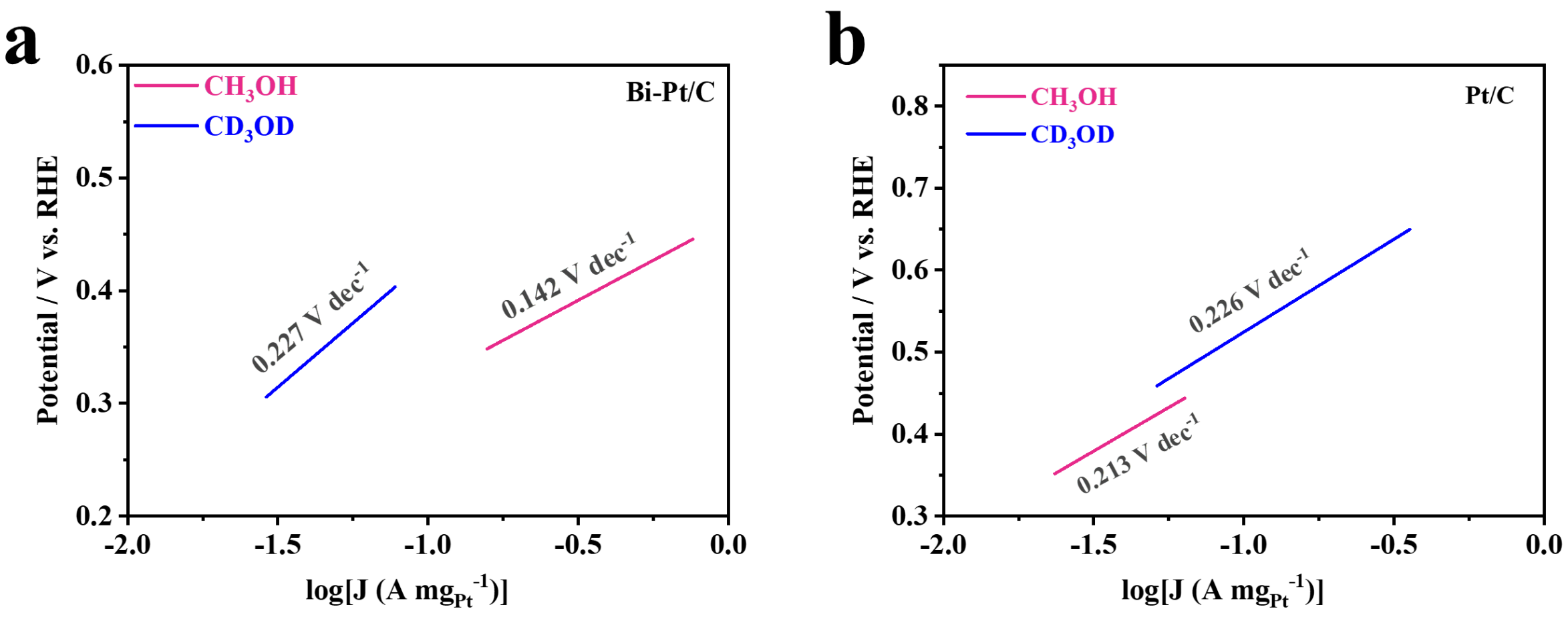

**$KIE_{H/D}$=1.60** **$KIE_{H/D}$=1.06**

**Figure S11.** Tafel plots of (A) Bi-Pt/C and (B) Pt/C under H/D exchange (1M $CH_3OH$ and 1M $CD_3OD$).

Note for Figure S11: The KIE value of Bi-Pt/C is >1.5, indicating a primary KIE where dehydrogenation steps are directly involved in the RDS. However, the KIE value of Pt/C is 1.06, indicating a secondary KIE.

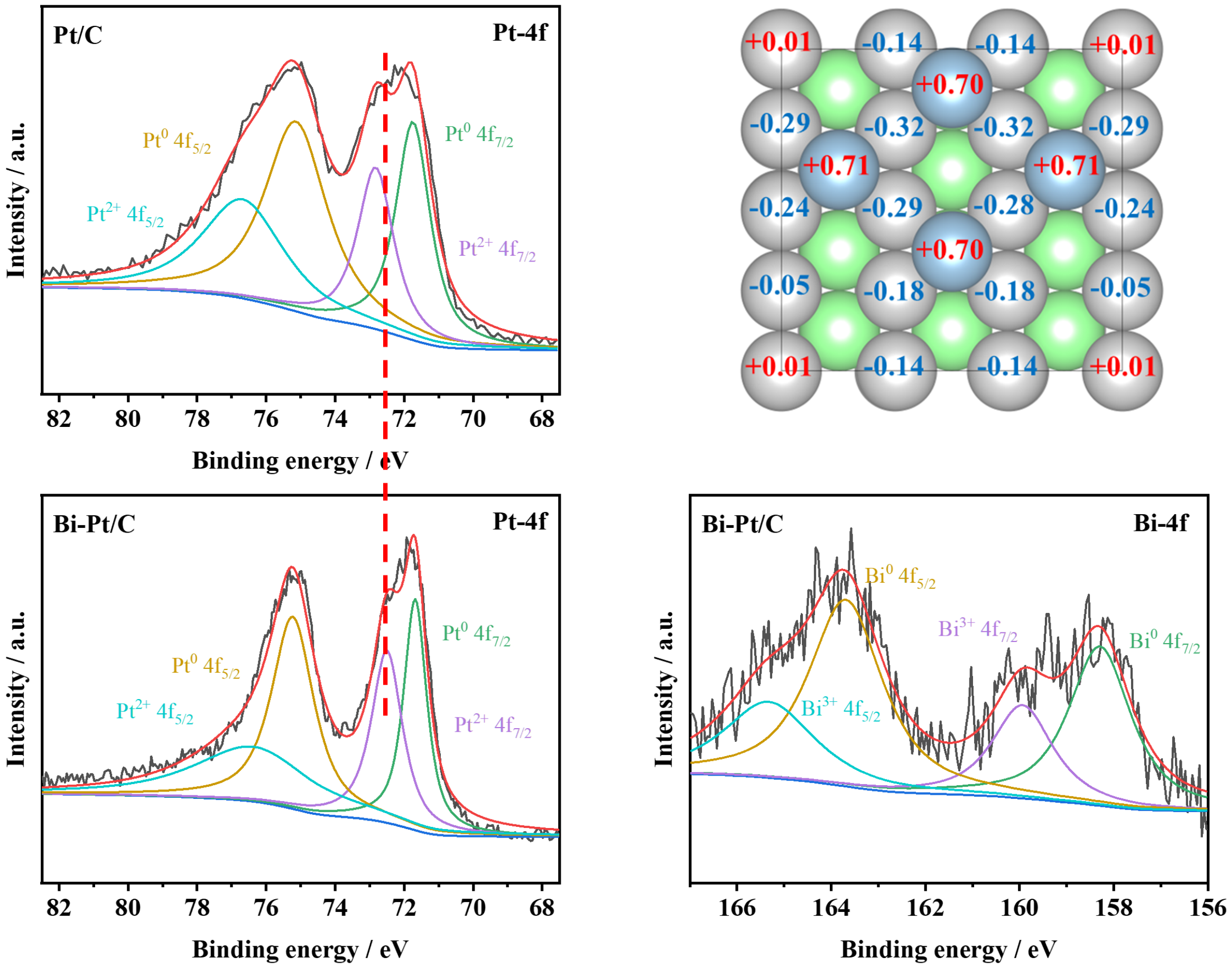

**Figure S12.** Pt 4f and Bi 4f XPS spectra of Pt/C and Bi-Pt/C as well as Bader charge analysis of Bi-Pt(110).

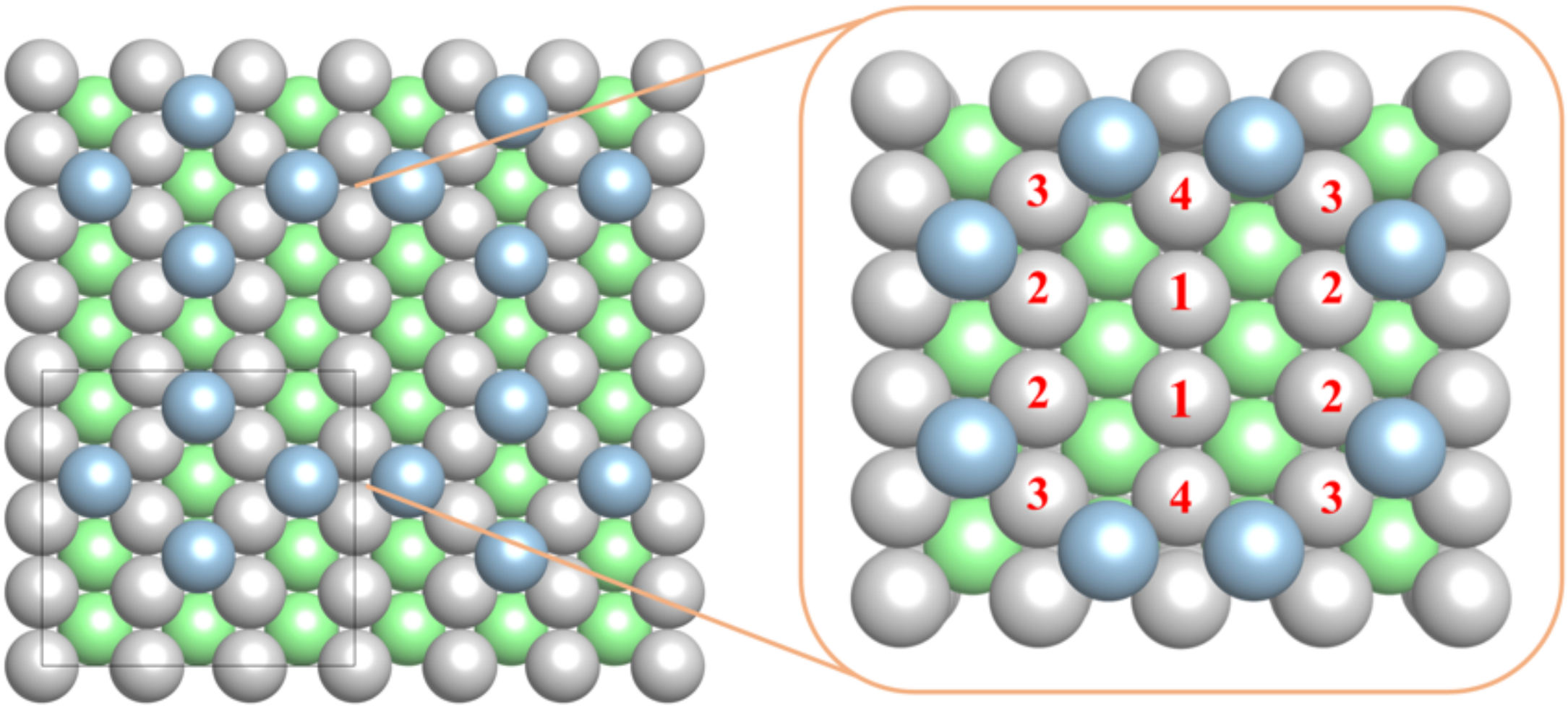

**Figure S13.** The structural model of Bi-Pt(110) surface. Color code: outermost layer Pt, gray; subsurface Pt, green; and Bi, blue.

Note for Figure S13: Based on the previous report, Bi adatoms tend to occupy surface sites where their coordination with Pt is maximized.[1] Consequently, they preferably occupy fourfold hollow sites on Pt(110), allowing interaction with four Pt atoms. For Bi adatoms, this affinity for high-coordination sites at low coverage is attributed to lateral repulsion between Bi adatoms, leading to a uniform distribution of Bi across the surface.

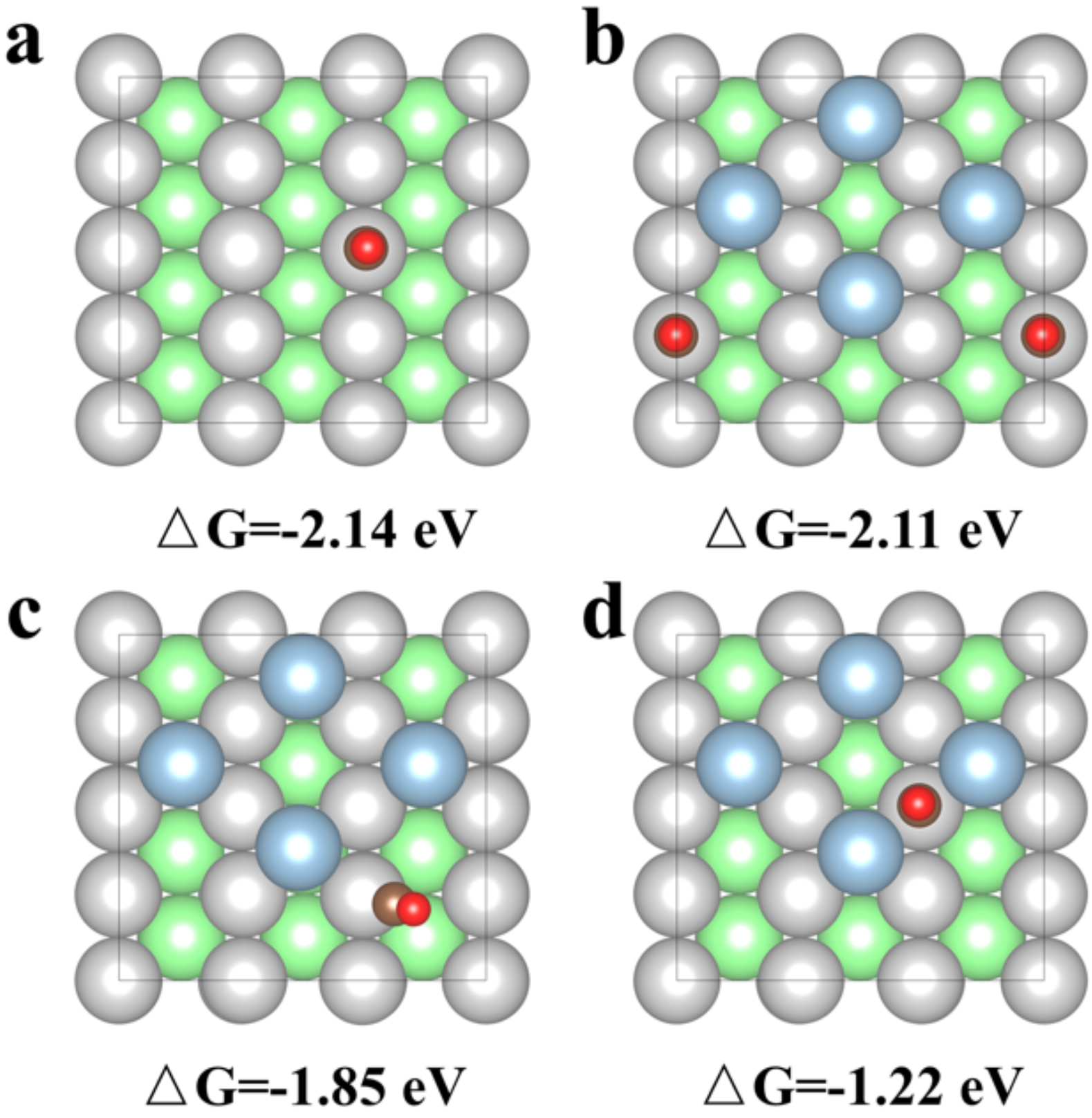

**Figure S14.** Adsorbed CO at (A) Pt top site of Pt(110), (B) site 1, (C) site 2, and (D) site 3 of Bi-Pt(110). Color code: outermost layer Pt, gray; subsurface Pt, green; Bi, blue; O, red; and C, brown.

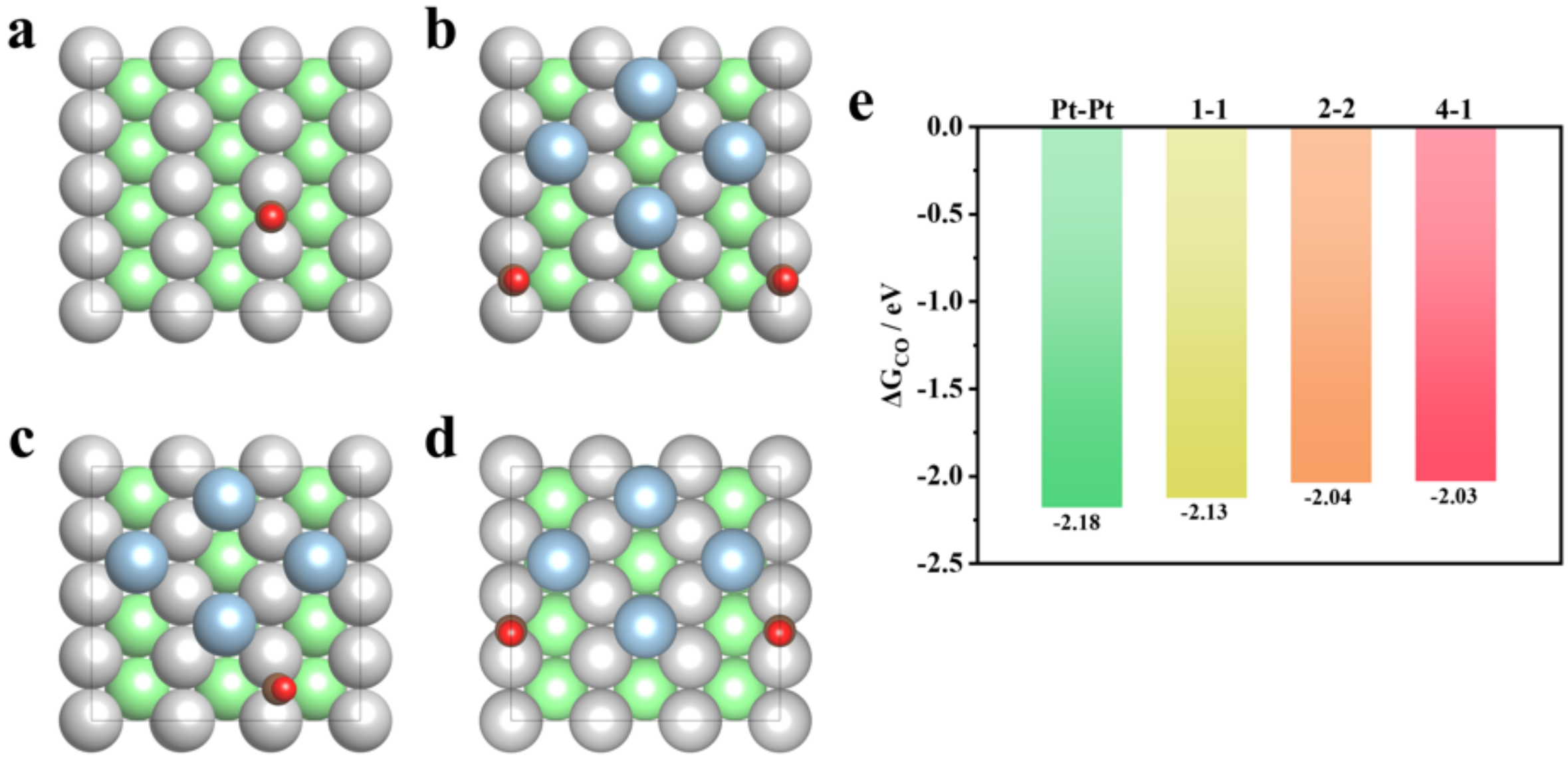

**Figure S15.** Adsorbed CO at (A) Pt-Pt bridge sites of Pt(110), (B) site 1-1, (C) site 2-2, and (D) site 4-1 of Bi-Pt(110). (E) The corresponding COBE comparison. Color code: outermost layer Pt, gray; subsurface Pt, green; Bi, blue; O, red; and C, brown.

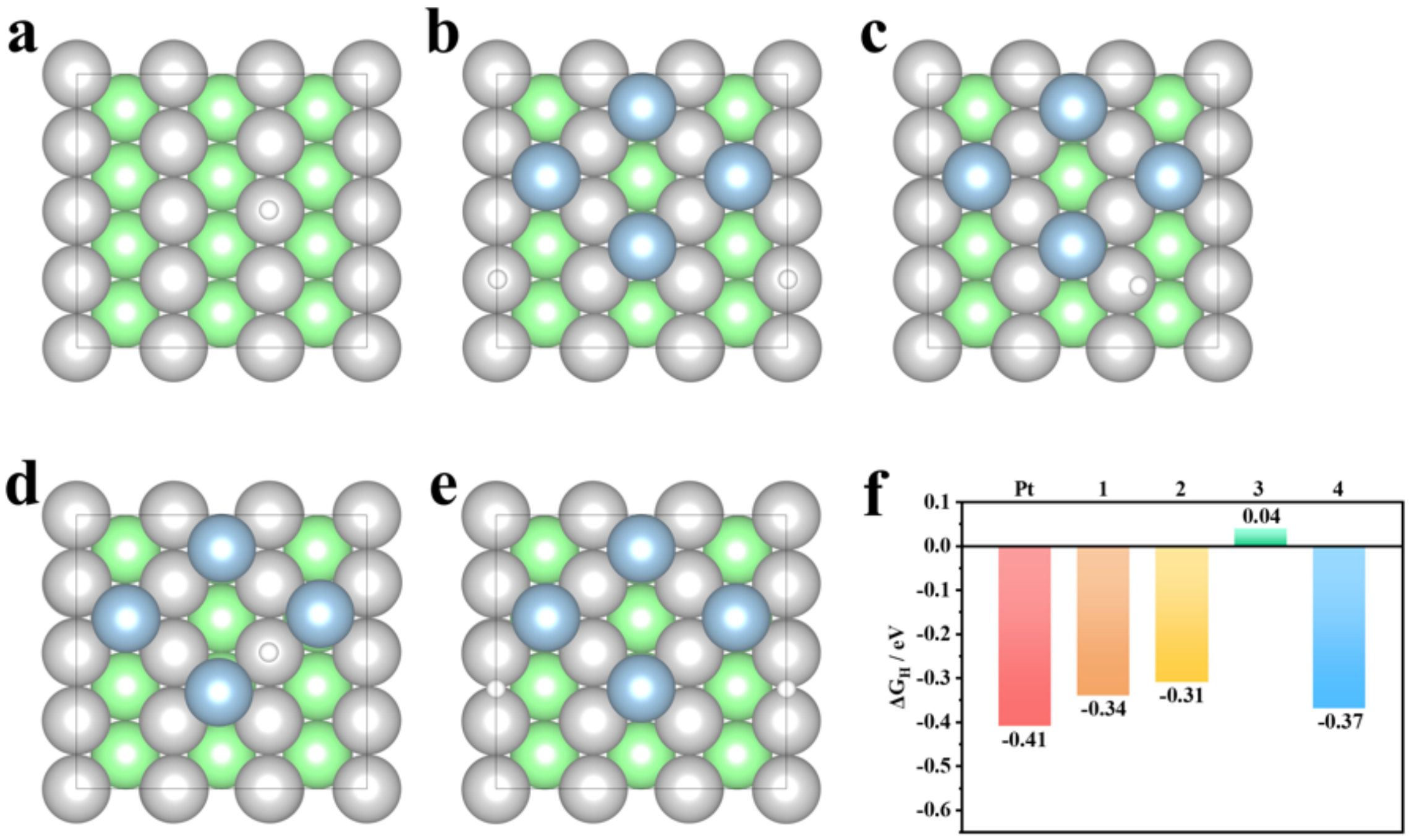

**Figure S16.** Adsorbed H at (A) Pt top site of Pt(110), (B) site 1, (C) site 2, (D) site 3, and (E) site 4 of Bi-Pt(110). (F) The corresponding HBE comparison. Color code: outermost layer Pt, gray; subsurface Pt, green; Bi, blue; and H, white.

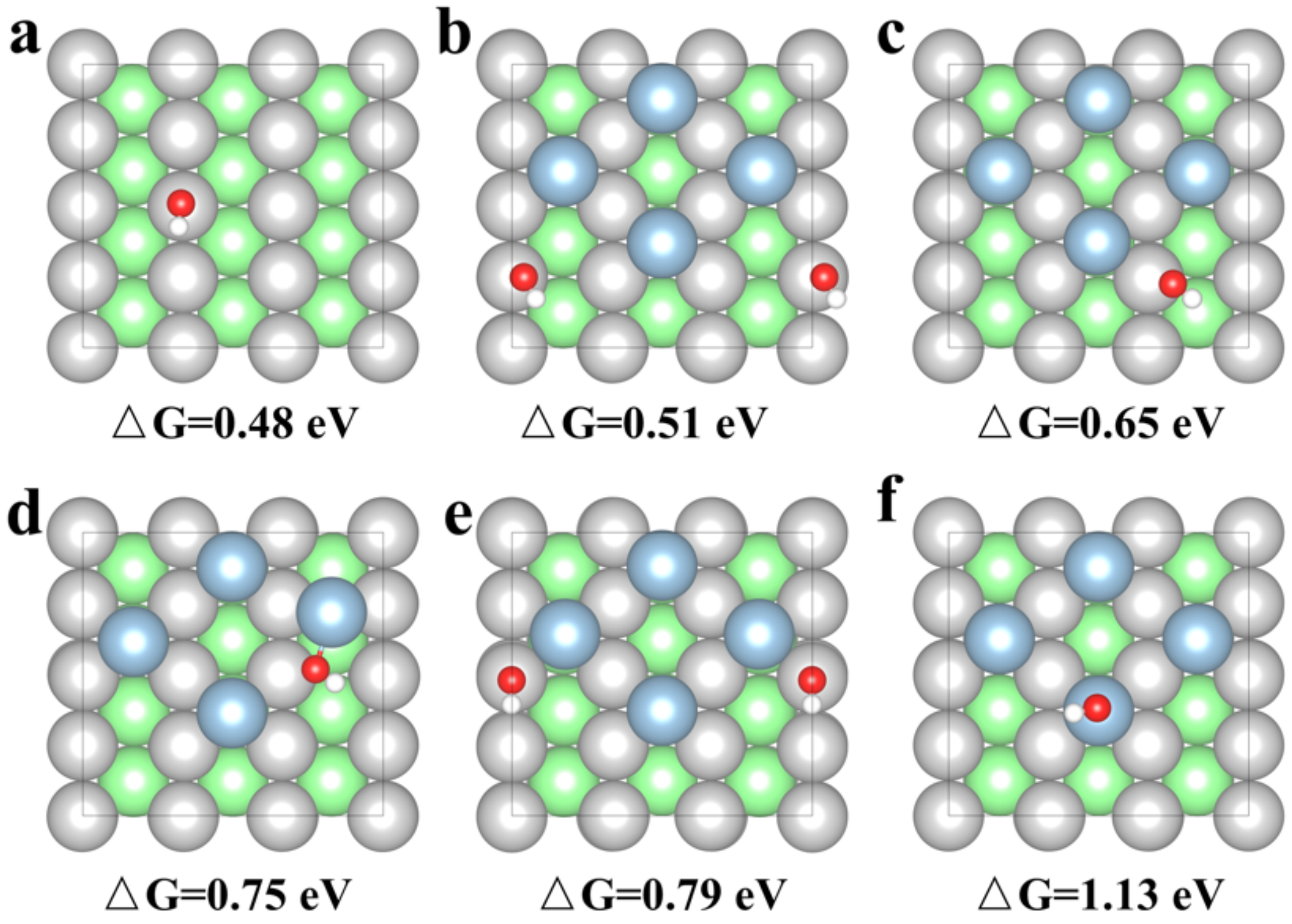

**Figure S17.** Adsorbed OH at (A) Pt top site of Pt(110), (B) site 1, (C) site 2, (D) site 3, (E) site 4, and (F) Bi site of Bi-Pt(110). Color code: outermost layer Pt, gray; subsurface Pt, green; Bi, blue; H, white; and O, red.

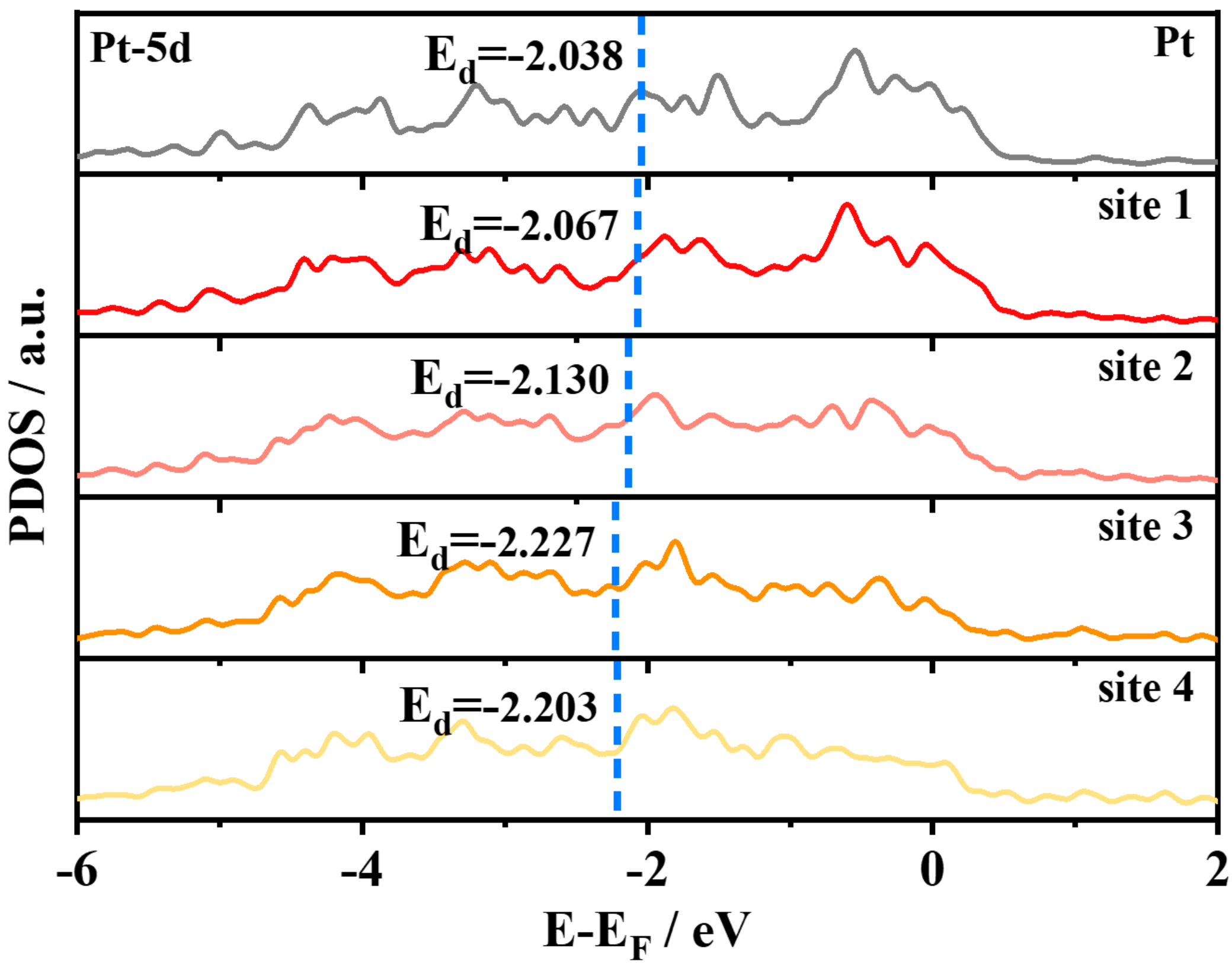

**Figure S18.** PDOS plots and d band center of different sites of Bi-Pt(110).

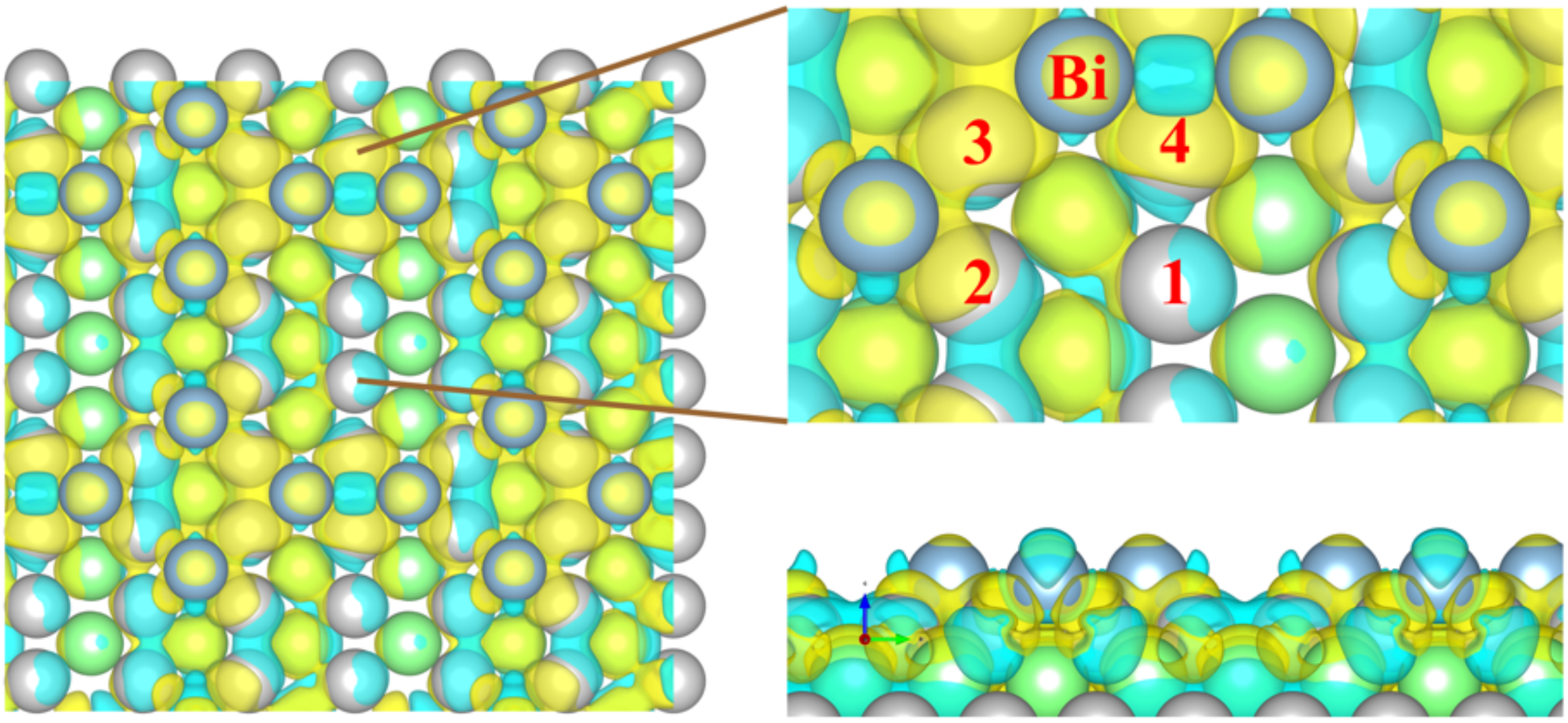

**Figure S19.** Differential charge density images of Bi-Pt(110). The yellow and blue regions present charge accumulation and decrease, respectively. The isosurface value is 0.002 $e^{-}/bohr^{3}$.

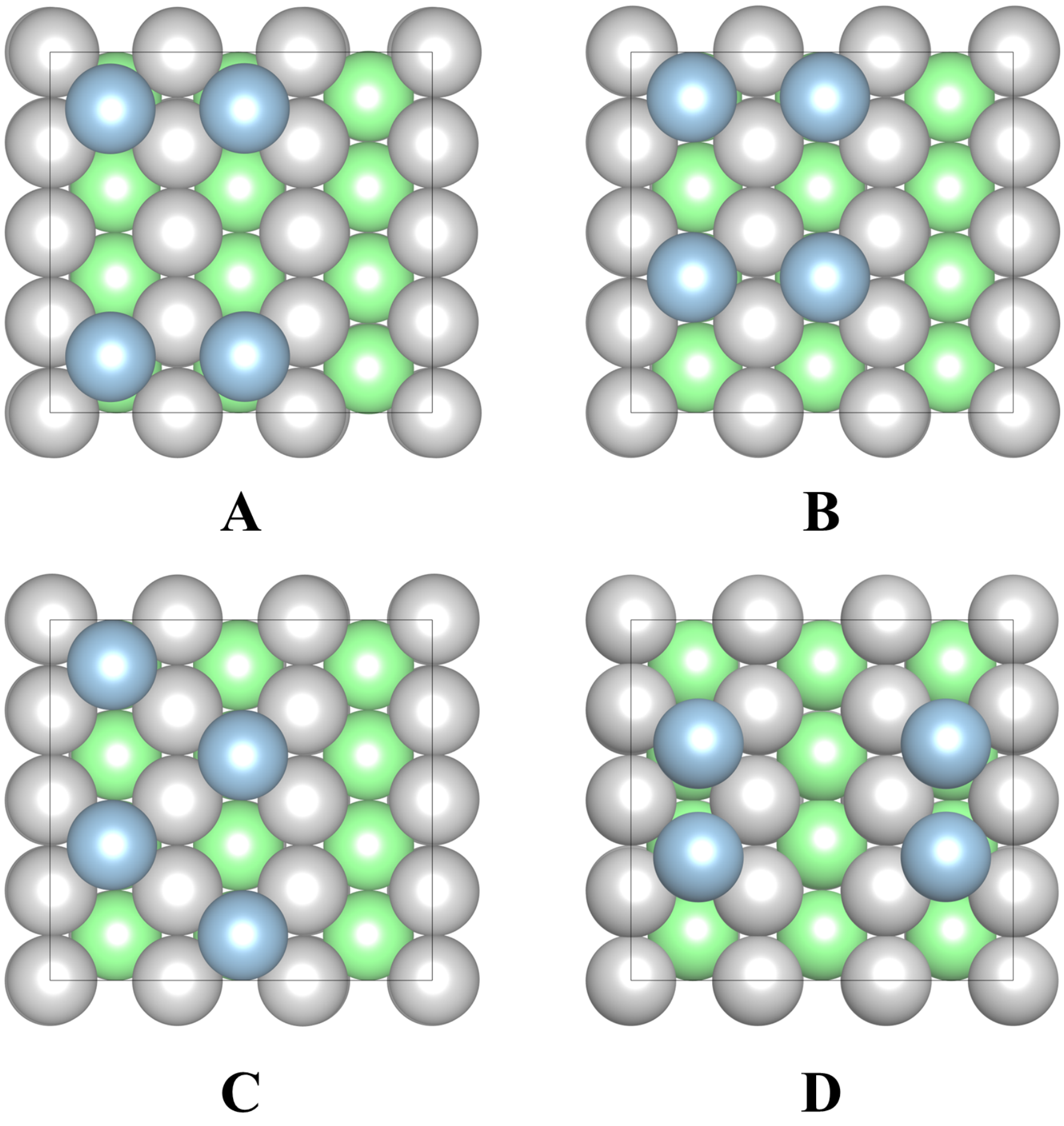

**Figure S20.** Other possible structural models of Bi-Pt(110) (denoted A, B, C, and D). Color code: outermost layer Pt, gray; subsurface Pt, green; and Bi, blue.

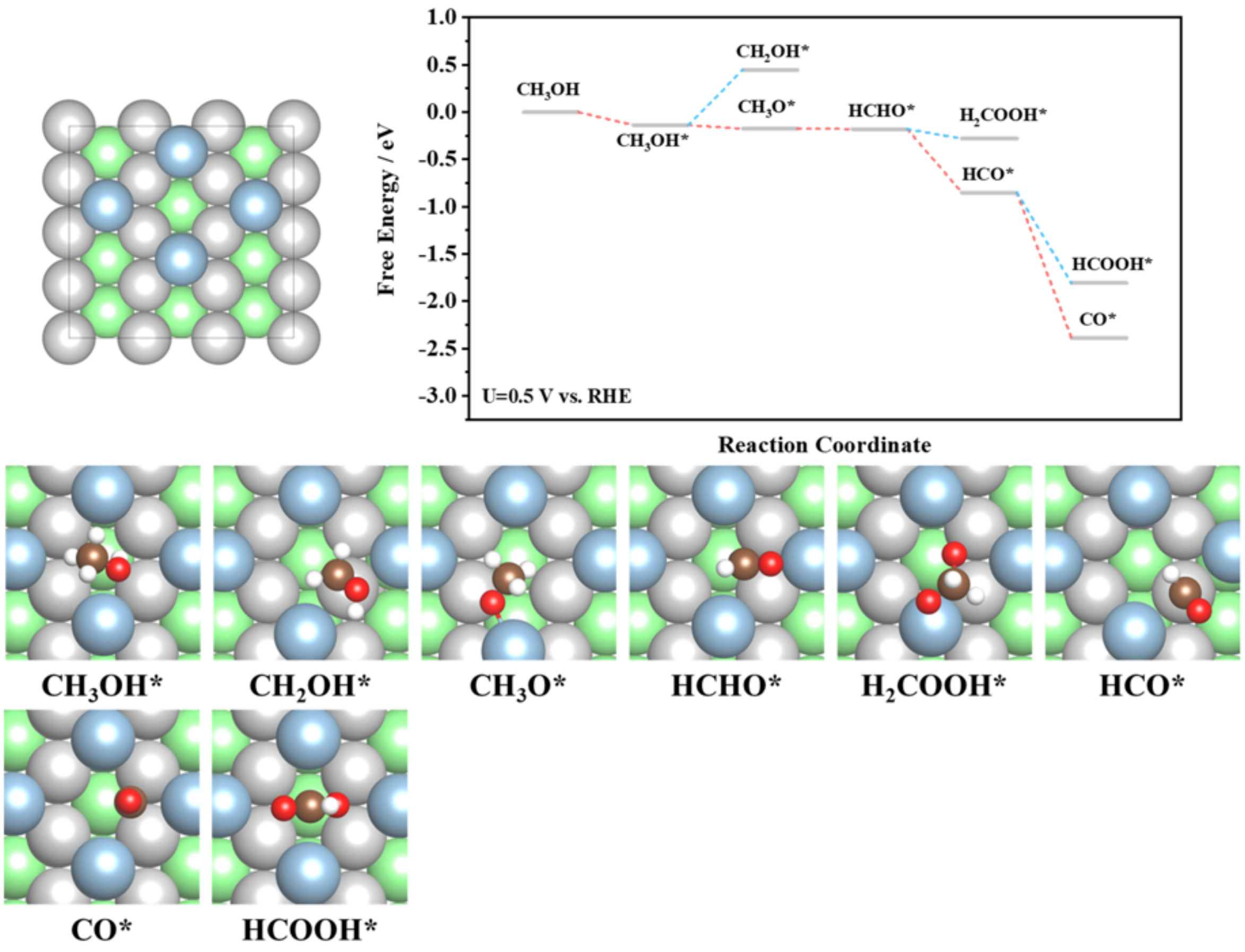

**Figure S21.** Gibbs free energy diagram of methanol oxidation reaction on site 3 of Bi-Pt(110) and corresponding structures of intermediates. Color code: outermost layer Pt, gray; subsurface Pt, green; Bi, blue; H, white; C, brown; and O, red.

Note for Figure S21: MOR at site 3 of Bi-Pt(110) undergoes a CO pathway, i.e., $CH_3OH \rightarrow CH_3OH^* \rightarrow CH_3O^* \rightarrow HCHO^* \rightarrow HCO^* \rightarrow CO^*$, implying that COBE or CO tolerance is not the key intrinsic factor in determining a CO-free pathway.

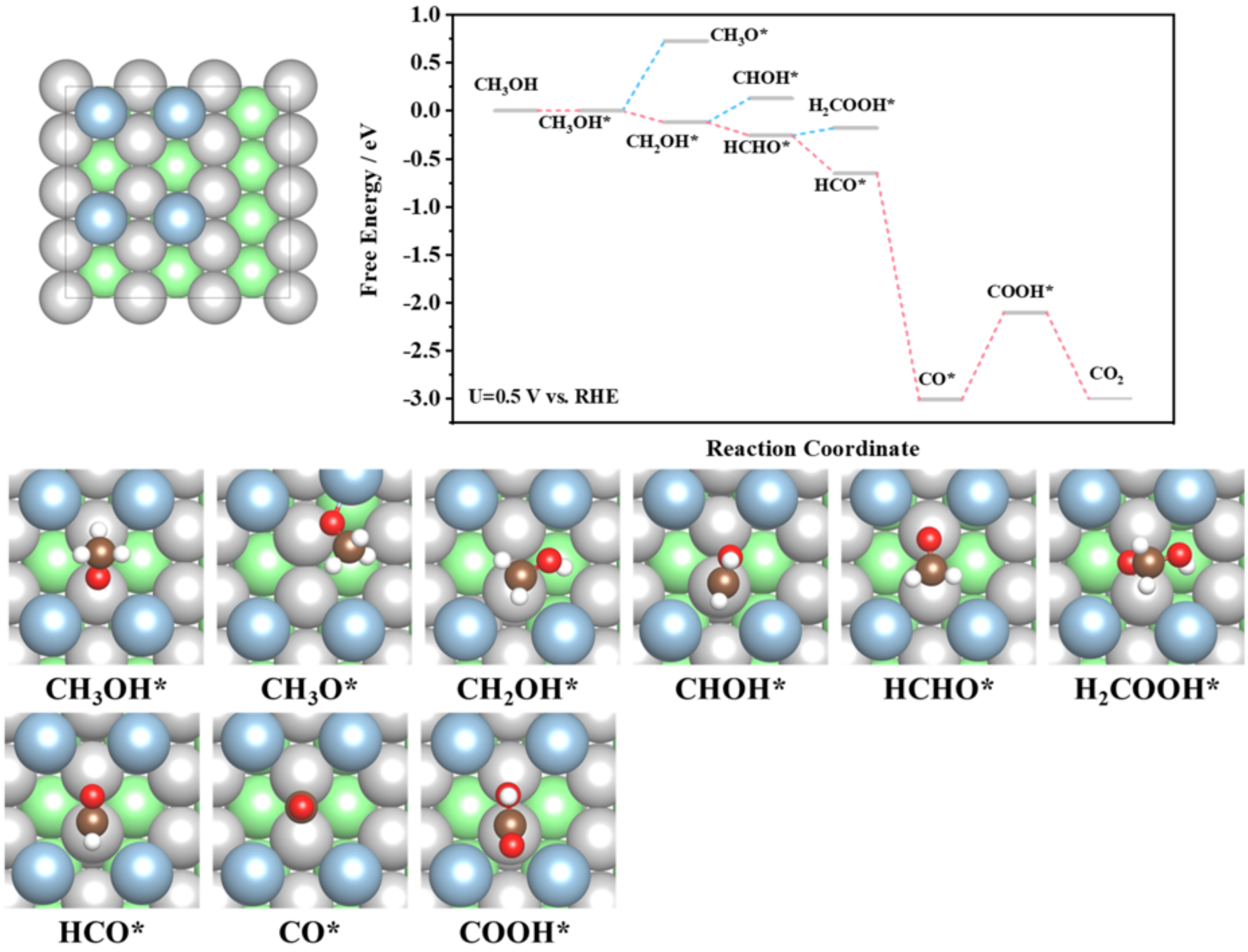

**Figure S22.** Gibbs free energy diagram of methanol oxidation reaction at site 4 of model A and corresponding structures of intermediates. Color code: outermost layer Pt, gray; subsurface Pt, green; Bi, blue; H, white; C, brown; and O, red.

Note for Figure S22: MOR at site 4 of model A undergoes a CO pathway, i.e., $CH_3OH \rightarrow CH_3OH^* \rightarrow CH_3O^* \rightarrow HCHO^* \rightarrow HCO^* \rightarrow CO^*$. Additionally, the CO*-to-COOH* conversion step is the rate-determining step (RDS) in the MOR, which is disagree with experimental results.

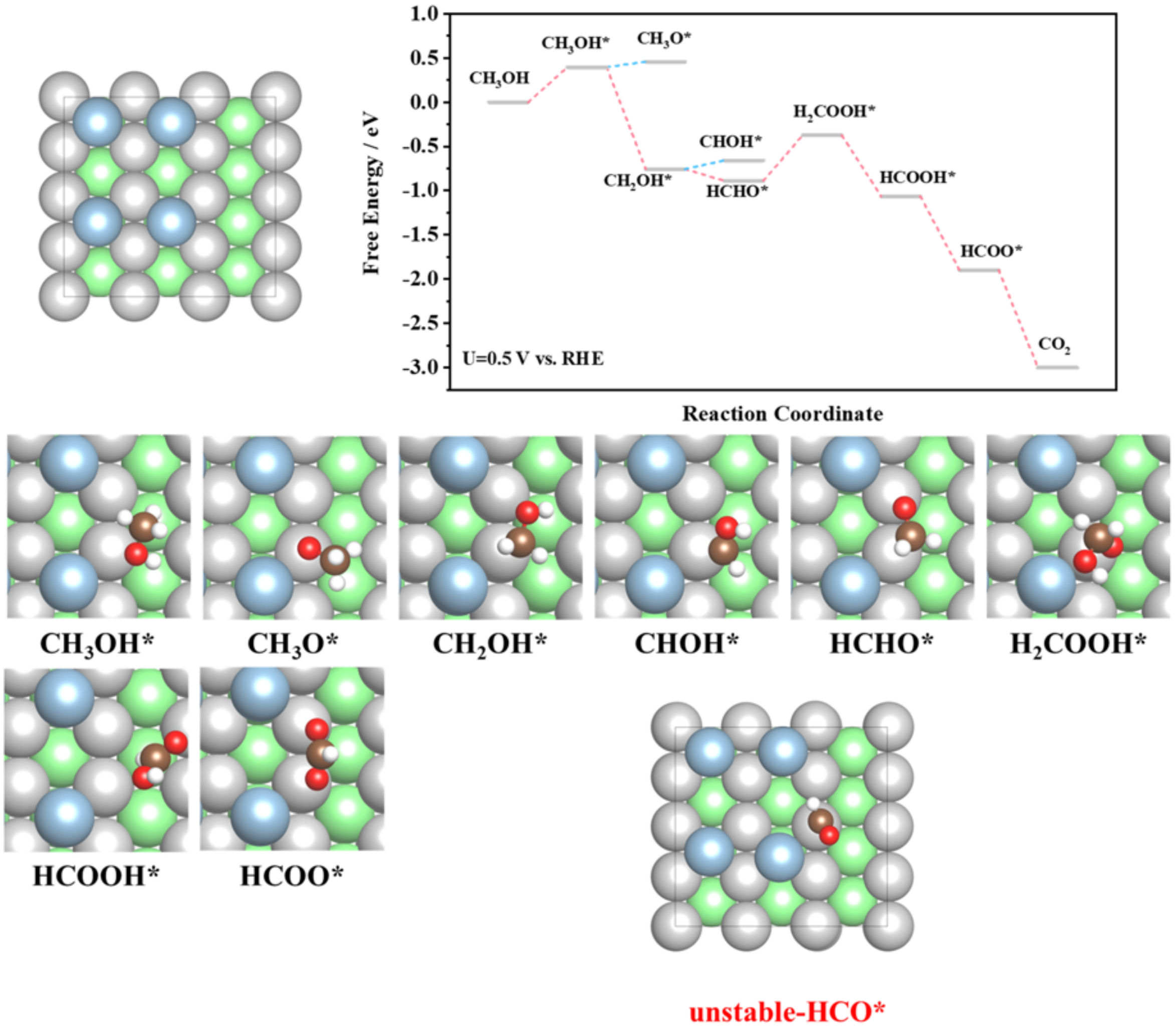

**Figure S23.** Gibbs free energy diagram of methanol oxidation reaction at site 2 of model A and corresponding structures of intermediates. Color code: outermost layer Pt, gray; subsurface Pt, green; Bi, blue; H, white; C, brown; and O, red.

Note for Figure S23: Due to unstable $HCO^*$ at site 2, we cannot obtain the free energy of $HCO^*$. It seems that MOR at site 2 of model A undergo a CO-free pathway, i.e., $CH_3OH \rightarrow CH_3OH^* \rightarrow CH_2OH^* \rightarrow HCHO^* \rightarrow H_2COOH^* \rightarrow HCOOH^* \rightarrow CO_2$. But the $HCHO^*$-to-$H_2COOH^*$ conversion step is the RDS in the MOR, which is disagree with experimental results.

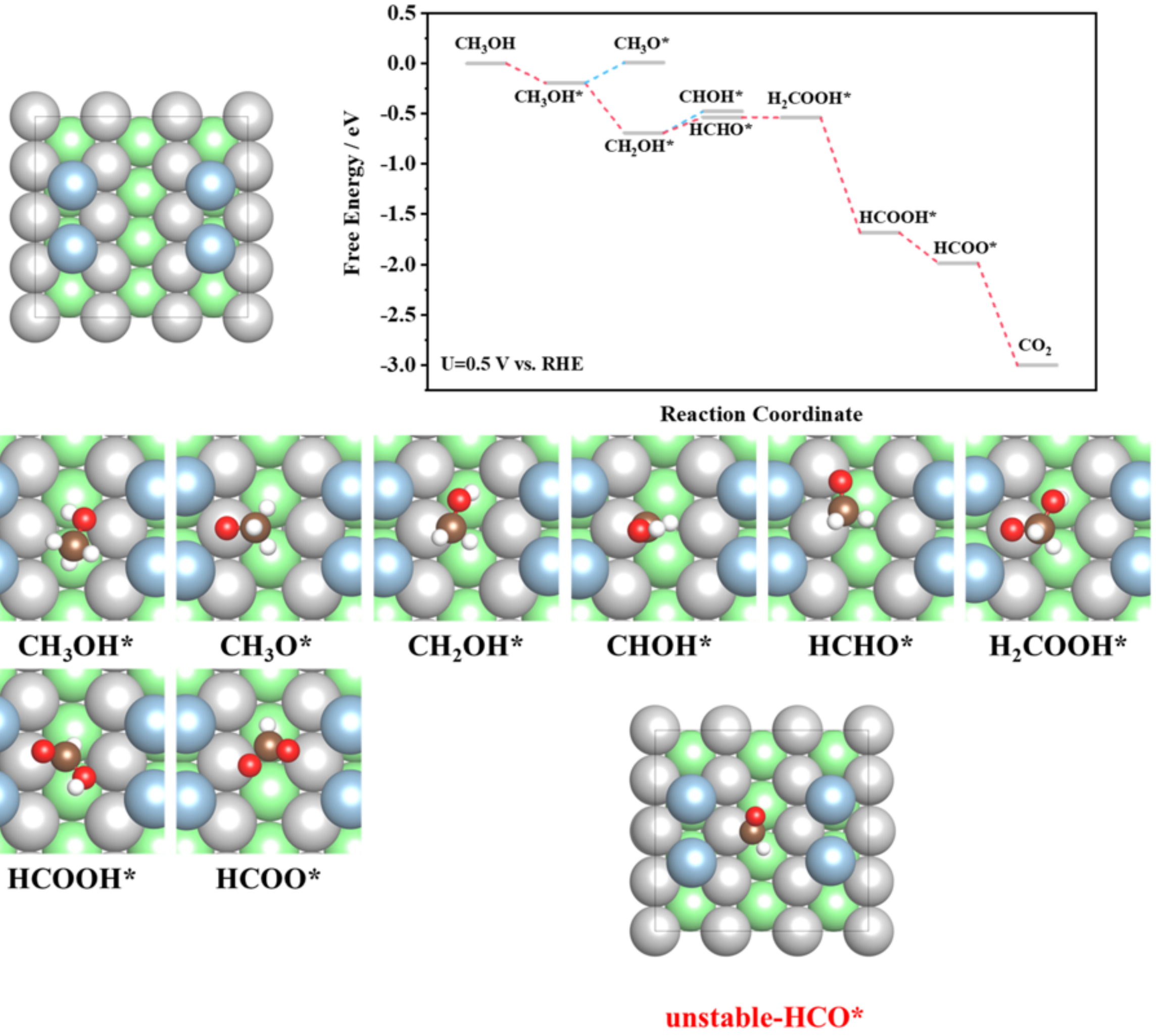

**Figure S24.** Gibbs free energy diagram of methanol oxidation reaction at site 4 of model D and corresponding structures of intermediates. Color code: outermost layer Pt, gray; subsurface Pt, green; Bi, blue; H, white; C, brown; and O, red.

Note for Figure S24: Due to unstable $HCO^*$ at site 4, we cannot obtain the free energy of $HCO^*$. It seems that MOR at site 2 of model D undergo a CO-free pathway, i.e., $CH_3OH \rightarrow CH_3OH^* \rightarrow CH_2OH^* \rightarrow HCHO^* \rightarrow H_2COOH^* \rightarrow HCOOH^* \rightarrow CO_2$.

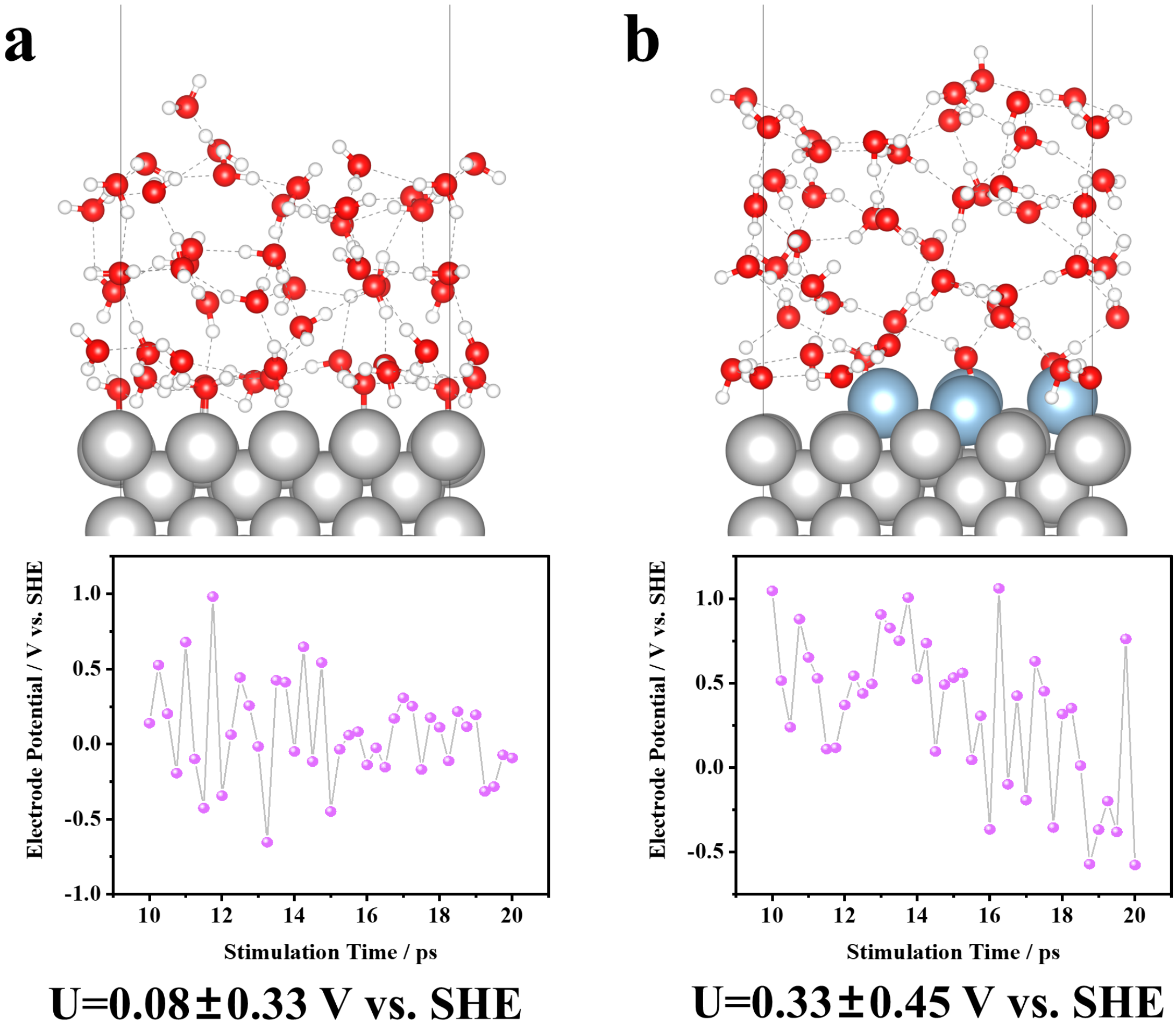

**Figure S25.** Representative snapshots of (A) Pt(110)/water interface and (B) Bi-Pt(110)/water interface and corresponding PZC evaluations. Color code: Pt, gray; Bi, blue; H, white; C, green; and O, red.

Note for Figure S25: Snapshots of one in every 0.25 ps over the final 10 ps AIMD trajectory were collected to calculate the mathematically averaged work function (Φ) of the system. The cutoff method is employed, in which $E_{vac}$ is extracted from the planar-averaged electrostatic potential profile, where the cutoff value of electron density is chosen as $10^{-5}$ e Å$^{-3}$.3[2] Then, electrode potential versus standard hydrogen electrode ($U_{SHE}$) can be calculated by the formula $U_{SHE} = \Phi - \Phi_{SHE}$, where $\Phi_{SHE}$ is the work function of the standard hydrogen electrode and the value of 4.2 V is used in this work.

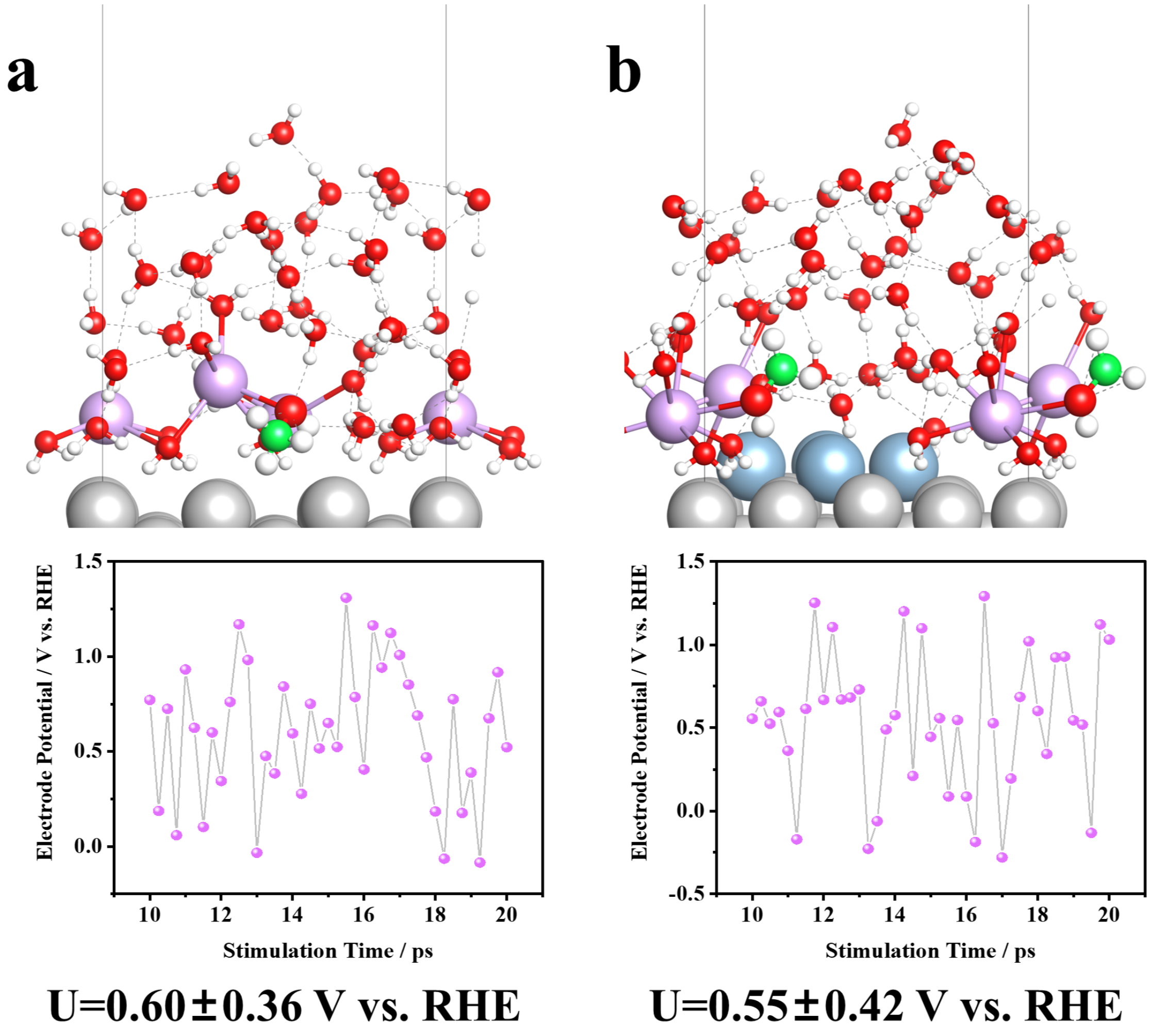

**Figure S26.** Representative snapshots of a $CH_3OH$ on (A) Pt(110) and (B) Bi-Pt(110) at alkaline system and corresponding electrode potential evaluations. Color code: Pt, gray; Bi, blue; H, white; C, green; K, purple; and O, red.

Note for Figure S26: Snapshots of one in every 0.25 ps over the final 10 ps AIMD trajectory are collected to calculate the mathematically averaged work function (Φ) of the system. The cutoff method is employed, in which $E_{vac}$ is extracted from the planar-averaged electrostatic potential profile, where the cutoff value of electron density is chosen as $10^{-5}$ e $Å^{-32}$. Then, electrode potential versus reversible hydrogen electrode ($U_{RHE}$) can be calculated by the formula $U_{RHE} = (\Phi - \Phi_{SHE})/e + 0.059 \times pH$, where pH is set to 14 and $\Phi_{SHE}$ is the work function of the standard hydrogen electrode and the value of 4.2 V is used in this work.

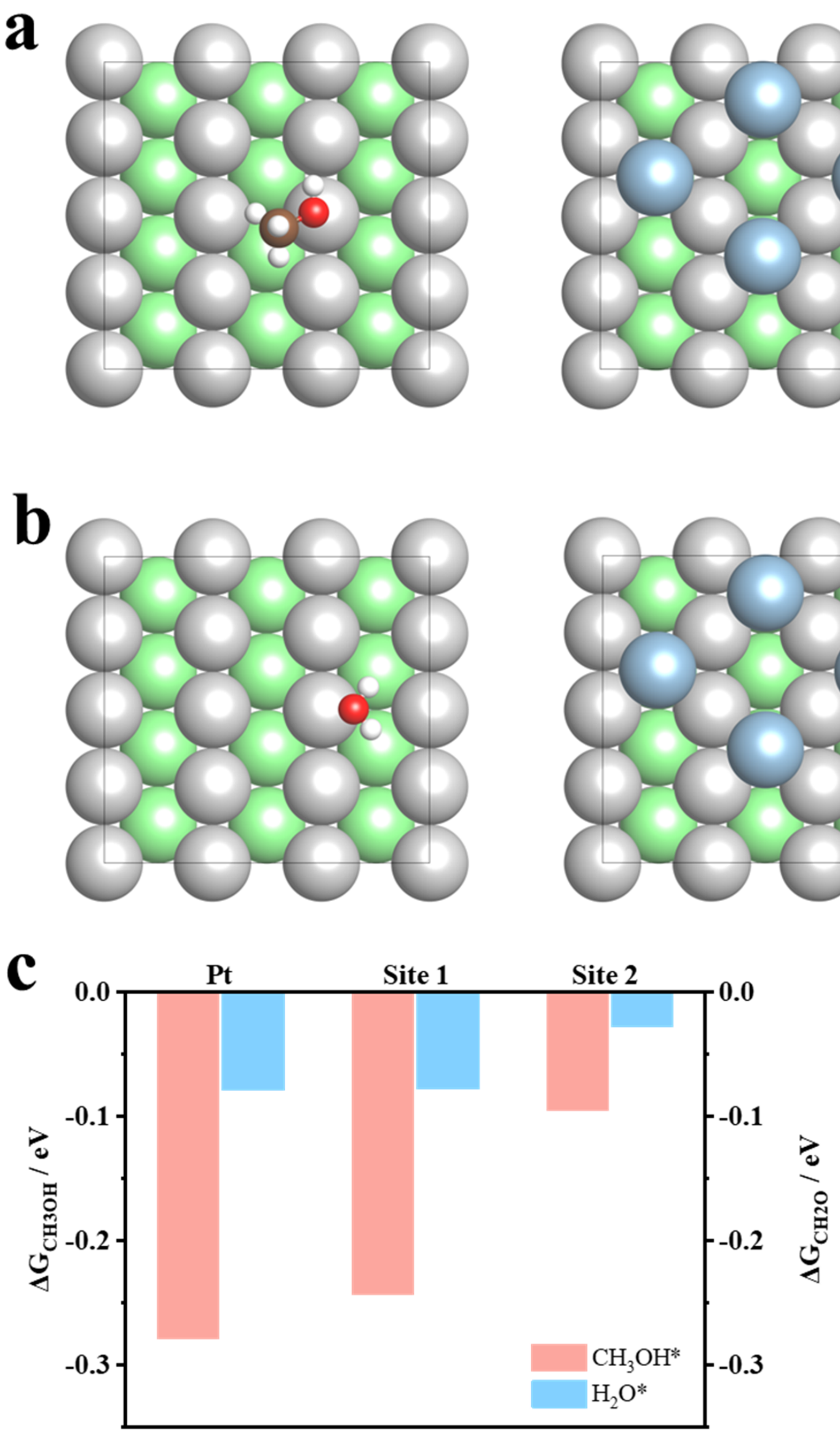

**Figure S27.** (A) Adsorbed $CH_3OH$ and (B) $H_2O$ at Pt top site of Pt(110), site 1 and site 2 of Bi-Pt(110), respectively. (C) The corresponding binding energy of $CH_3OH$ and $H_2O$ comparison. Color code: outermost layer Pt, gray; subsurface Pt, green; Bi, blue; H, white; C, brown; and O, red.

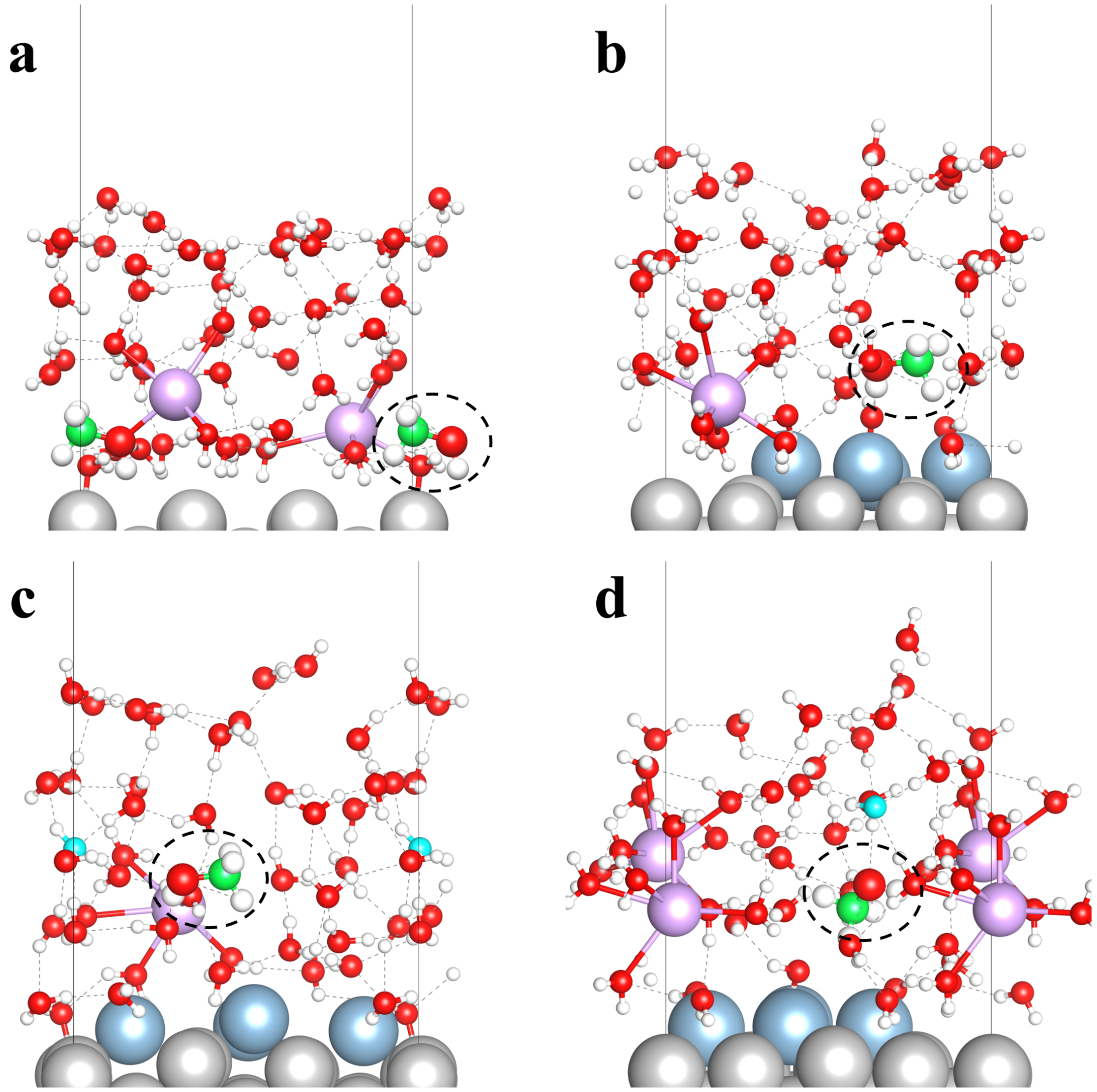

**Figure S28.** Representative snapshots of (A) $CH_3OH$ on Pt(110)-2$K^+$; (B) $CH_3OH$ on Bi-Pt(110)- $K^+$; (C) $CH_3OH$ on Bi-Pt(110)-$K^+$-$OH^-$ and (D) $CH_3OH$ on Bi-Pt(110)- 2$K^+$-$OH^-$. $CH_3OH$ and $OH^-$ are highlighted in black ellipses. Color code: Pt, gray; Bi, blue; H, white; C, green; K, purple; O, red; and O in $OH^-$, cyan.

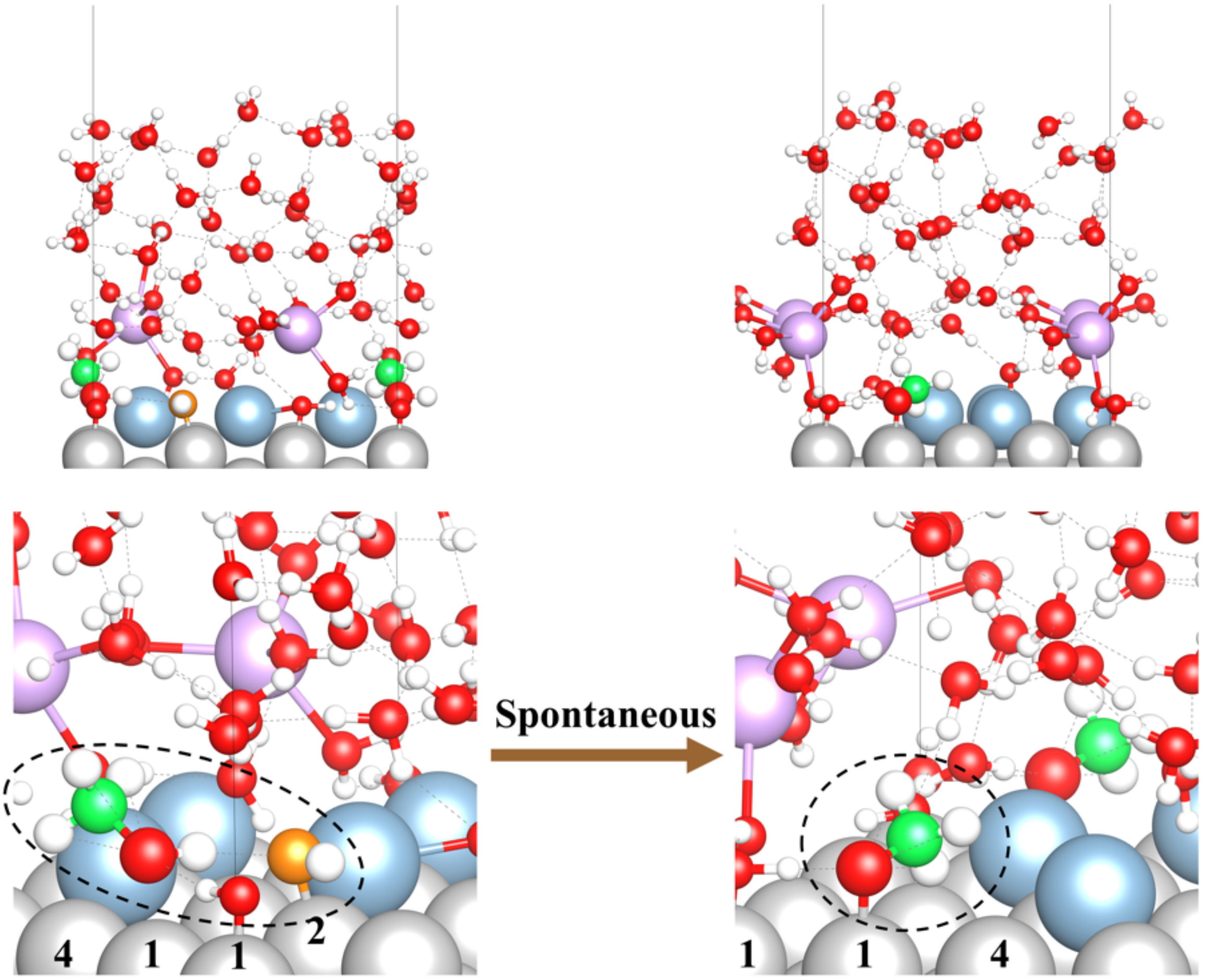

**Figure S29.** Representative snapshots from AIMD trajectory when OH* is located at site 2 and $CH_3OH$ is close to top of adjacent site 1. $CH_3OH$ and OH* are highlighted in black ellipses. Color code: Pt, gray; Bi, blue; H, white; C, green; K, purple; O, red; and O in OH* at site 2, orange.

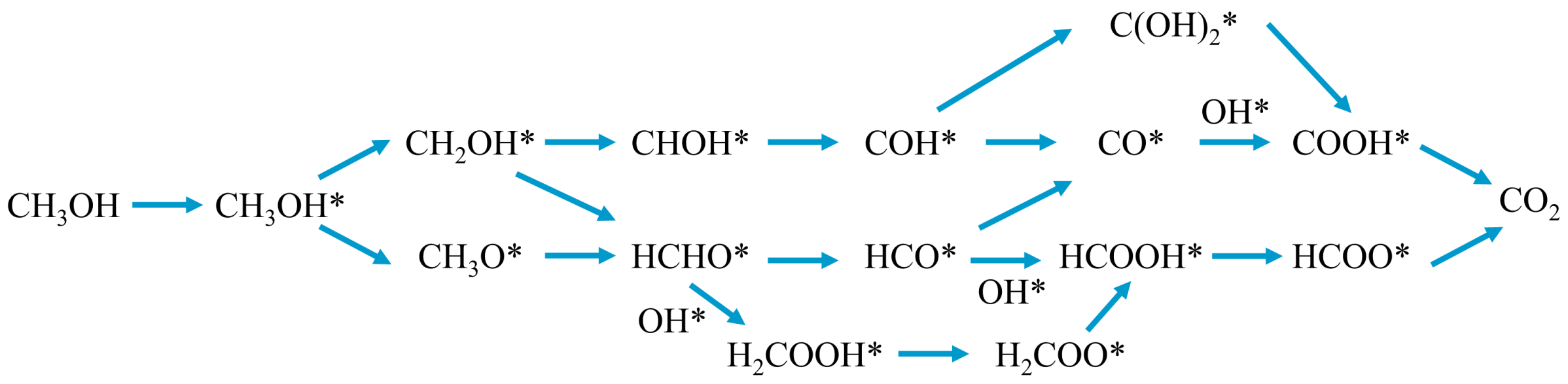

**Figure S30.** Classical reaction network of methanol oxidation reaction. Where * represents an active site on the catalyst surface.

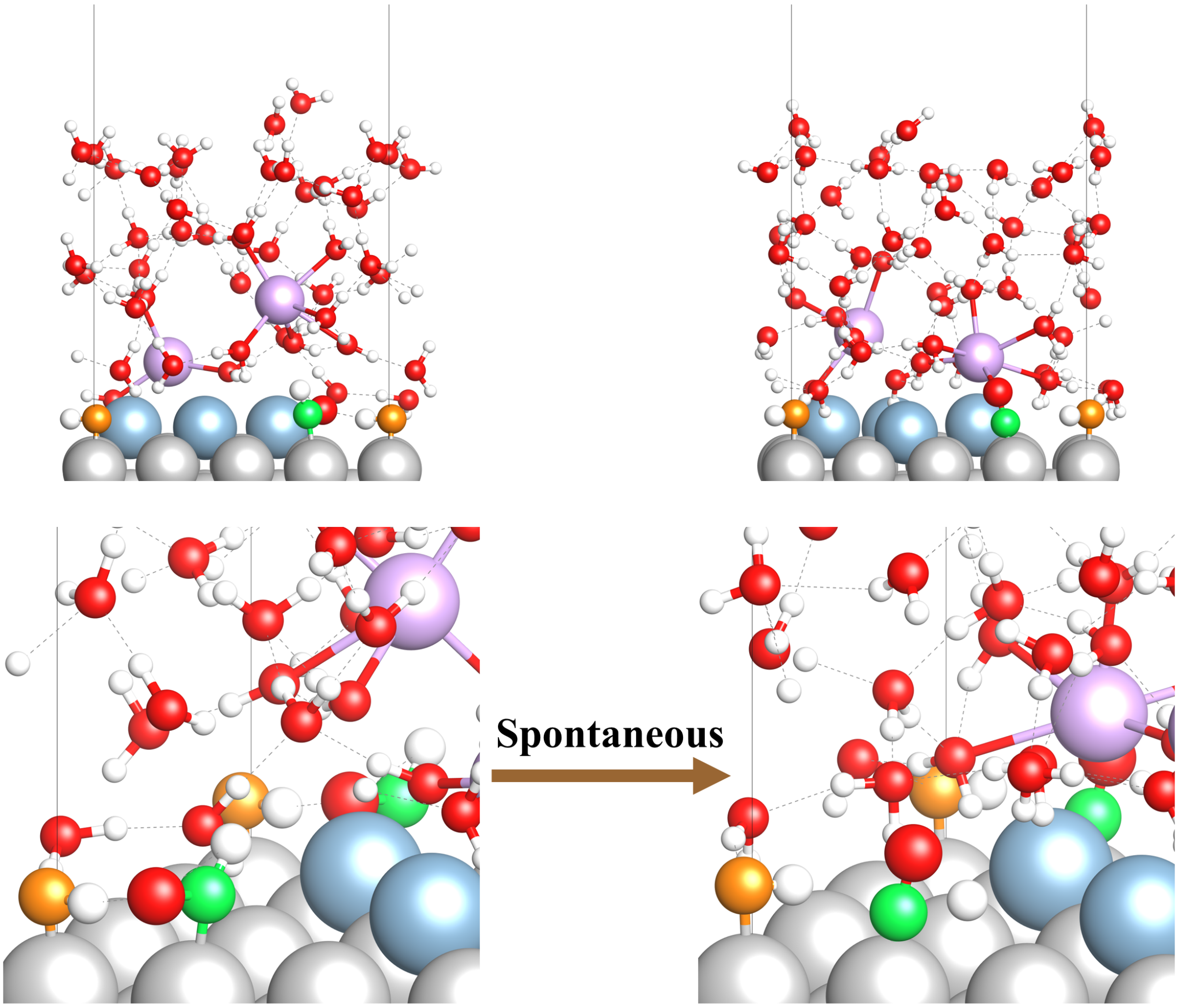

**Figure S31.** Representative snapshots from AIMD trajectory when HCO and OH adsorb at site 1 of Bi-Pt(110), respectively. Color code: Pt, gray; Bi, blue; H, white; C, green; K, purple; O, red; and O in OH*, orange.

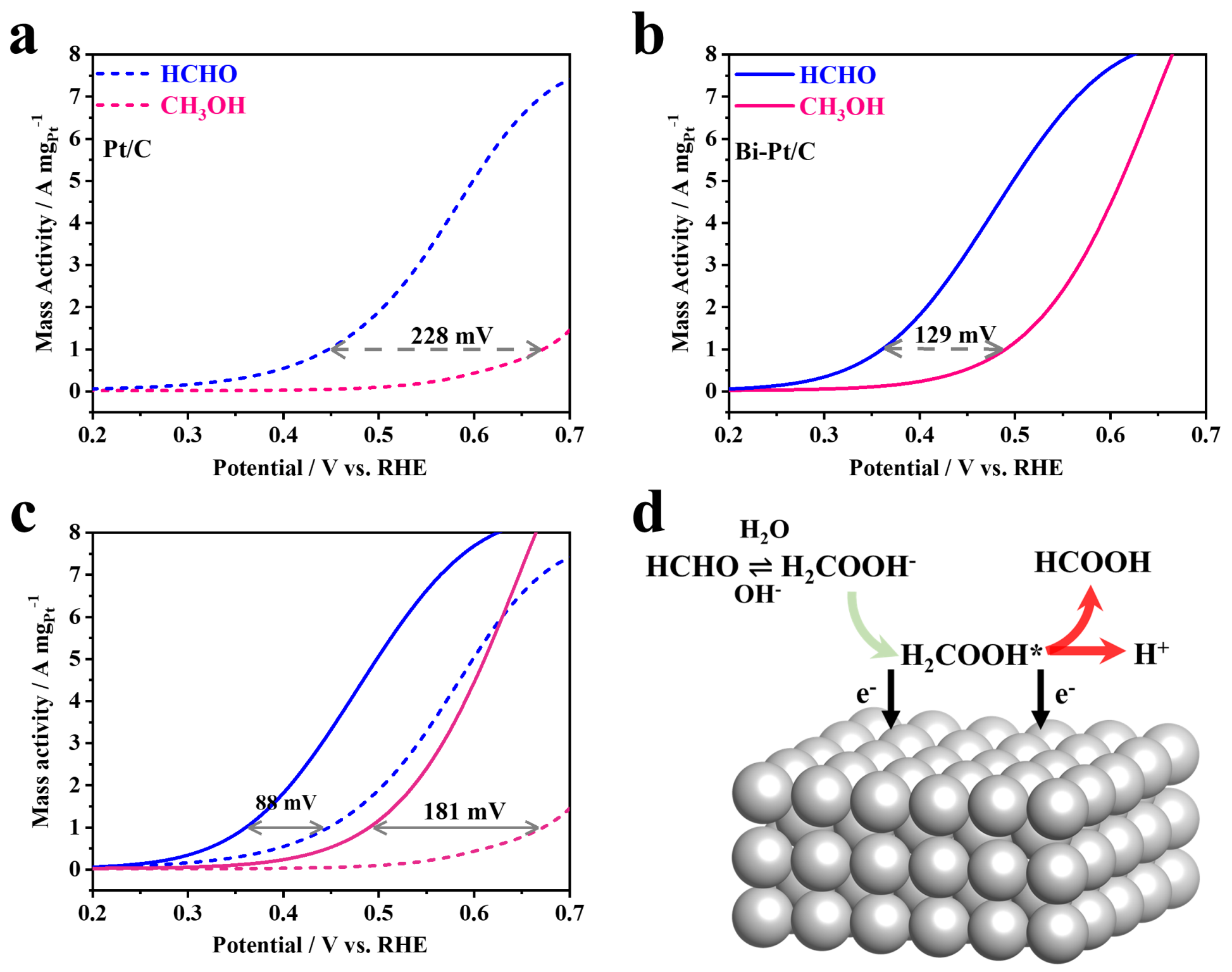

**Figure S32.** A comparison of positive-going polarization curves between in 1M HCHO/1M KOH and in 1M $CH_3OH$/1M KOH for (A) Pt/C and (B) Bi-Pt/C, respectively. (C) Summarized MOR and FOR curves. (D) FOR mechanism on Pt surface under alkaline solution.

Note for Figure S32: Bi-Pt/C exhibits lower onset over-potential (defined as the over-potential required to reach a MA of 0.1 A $mg_{Pt}^{-1}$) than Pt/C, indicating the higher activation ability to HCHO molecules. It is worth noting that the Bi modification can remarkably narrow the activity gap between MOR and FOR by comparing the onset over-potential. As presented in Figures 32A-B, Pt/C shows a substantial onset over-potential gap of 228 mV between MOR and FOR, which can be notably reduced to 129 mV upon Bi modification. Moreover, we also compared the activity and kinetics gap between Bi-Pt/C and Pt/C in MOR and FOR, respectively. As shown in Figure 32C, the onset over-potential gap between Bi-Pt/C and Pt/C in MOR is 181 mV while only 88 mV is observed in FOR, indicating that Bi modification renders a more pronounced improvement in MOR kinetic. The more distinct enhancement in MOR activity and kinetics with Bi-Pt/C compared to Pt/C implies that HCHO intermediate plays an important role in pathway selectivity of MOR.

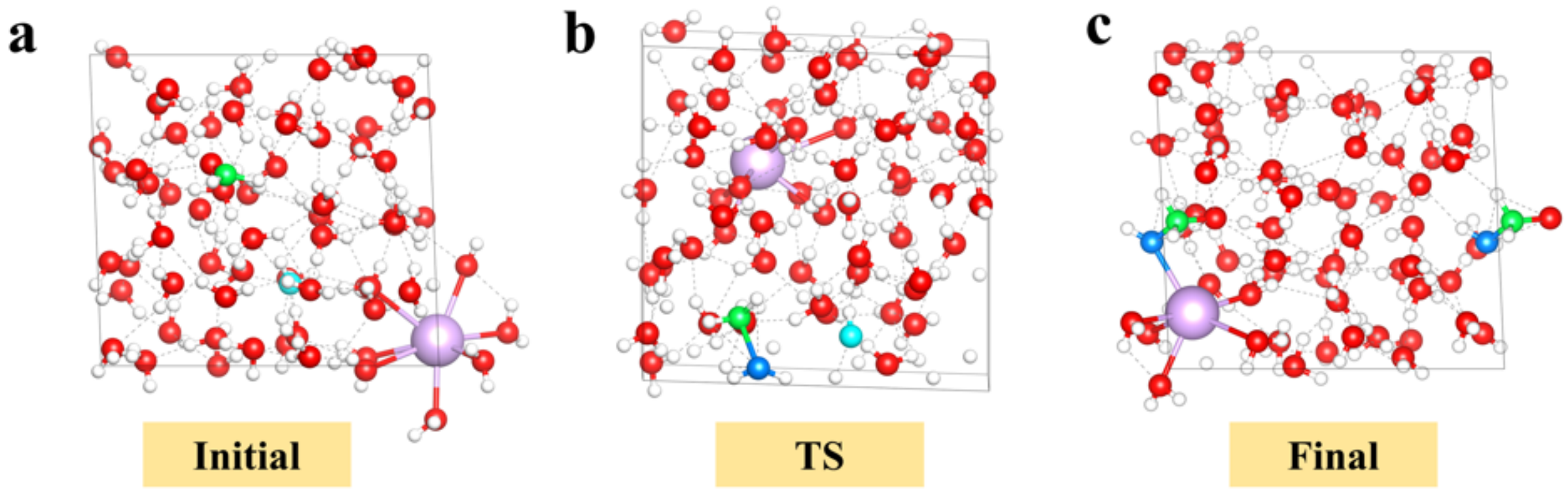

**Figure S33.** Representative snapshots of HCHO(aq)-to-$H_2COOH^-$ anion conversion under alkaline condition including the (A) initial, (B) transition state, and (C) final structures. Color code: H, white; C, green; K, purple; O, red; O in $OH^-$, cyan; and O in $H_2COOH^-$, blue.

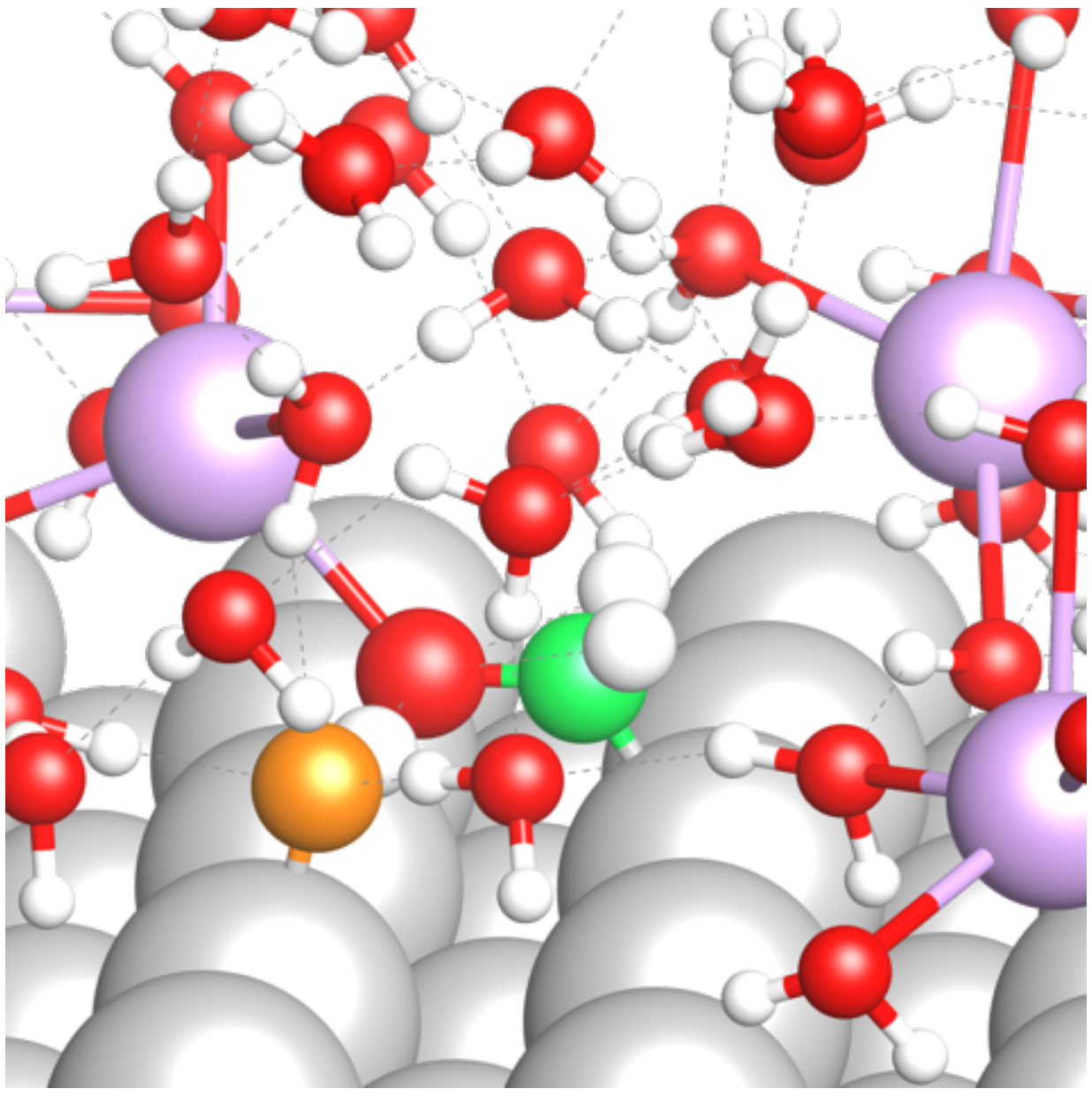

**Figure S34.** Representative snapshots of OH* and HCHO* on Pt (110). Color code: Pt; blue; H; white; C, green; K, purple; O, red; and O in OH*, orange.

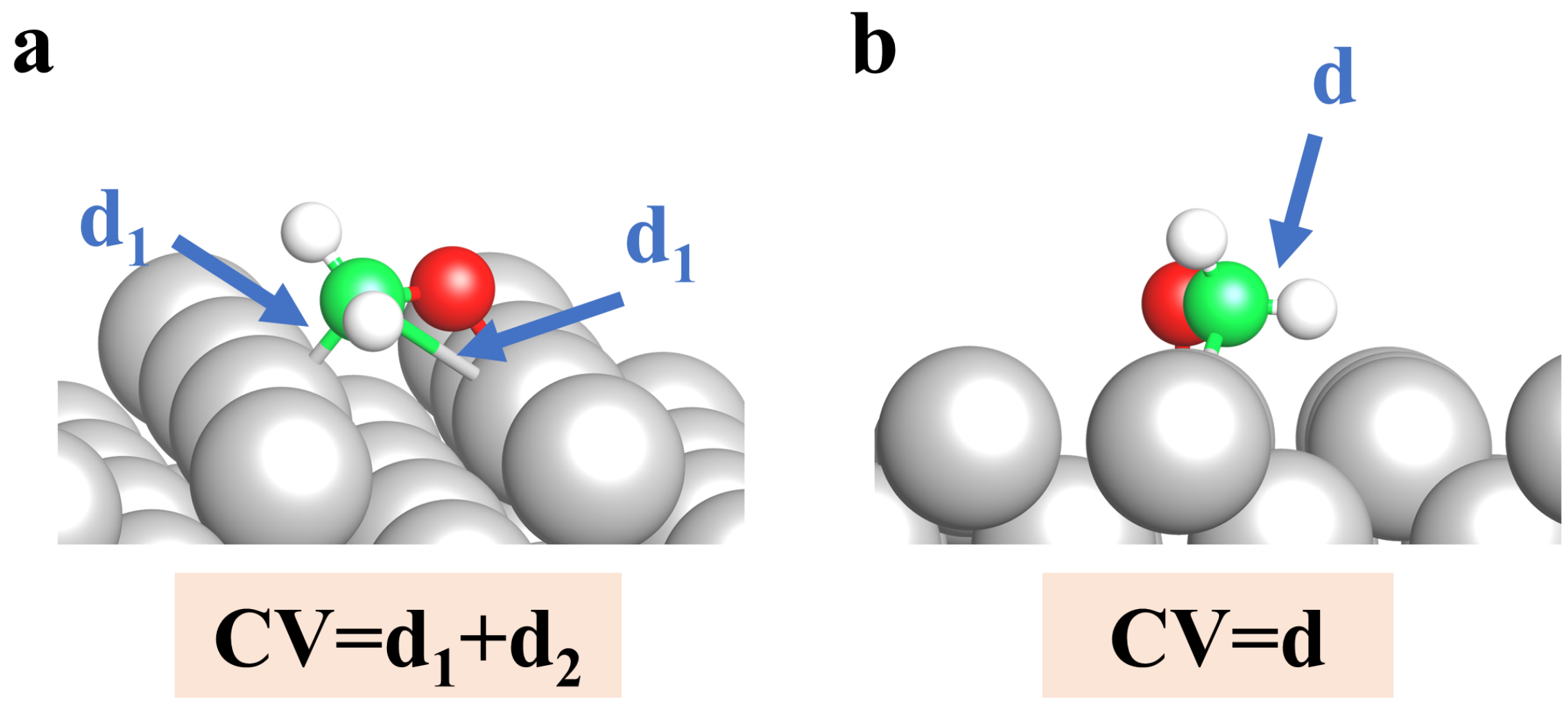

**Figure S35.** Illustration of collective variable of slow-growth approach used in (A) cleavage of Pt-C bond and (B) cleavage of C-H bond. Color code: Pt, gray; Bi, blue; H, white; C, green; and O, red.

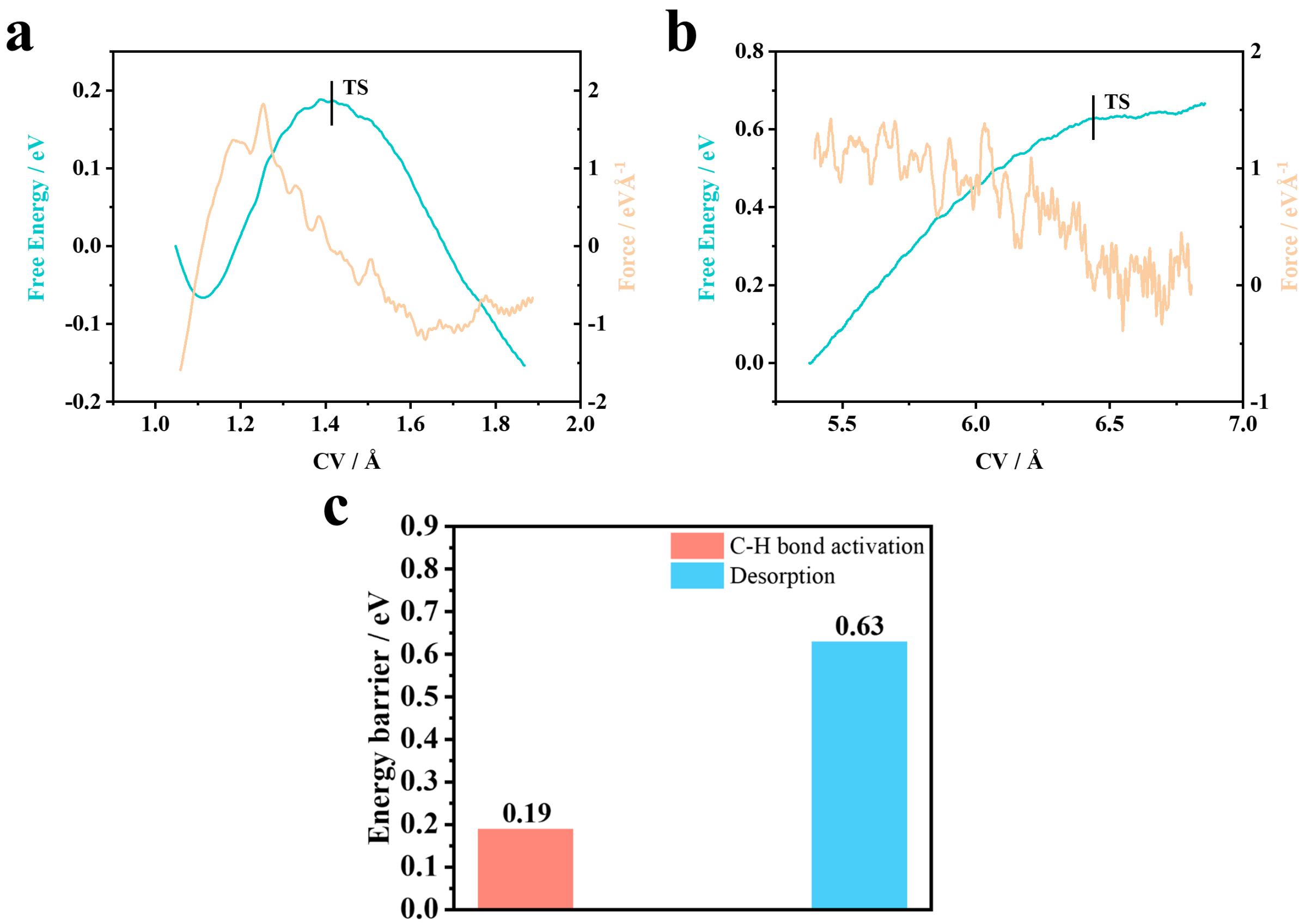

**Figure S36.** Representative potential of mean forces and corresponding free energy changes for (A) HCHO* dehydrogenation and (B) HCHO* desorption on Pt(110). (C) A comparison of desorption barrier of HCHO* and its dehydrogenation barrier on Pt(110) at 0.5 $V_{RHE}$.

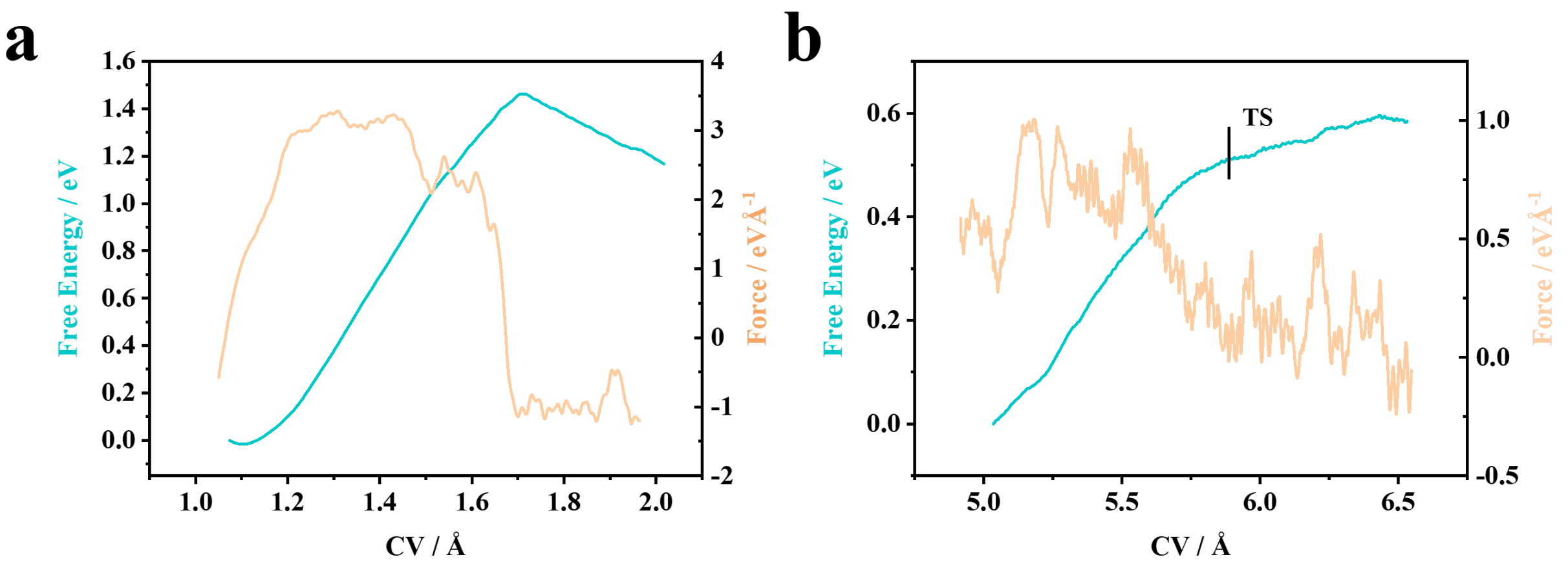

**Figure S37.** Representative potential of mean forces and corresponding free energy changes for (A) HCHO* dehydrogenation and (B) HCHO* desorption on Bi-Pt(110).

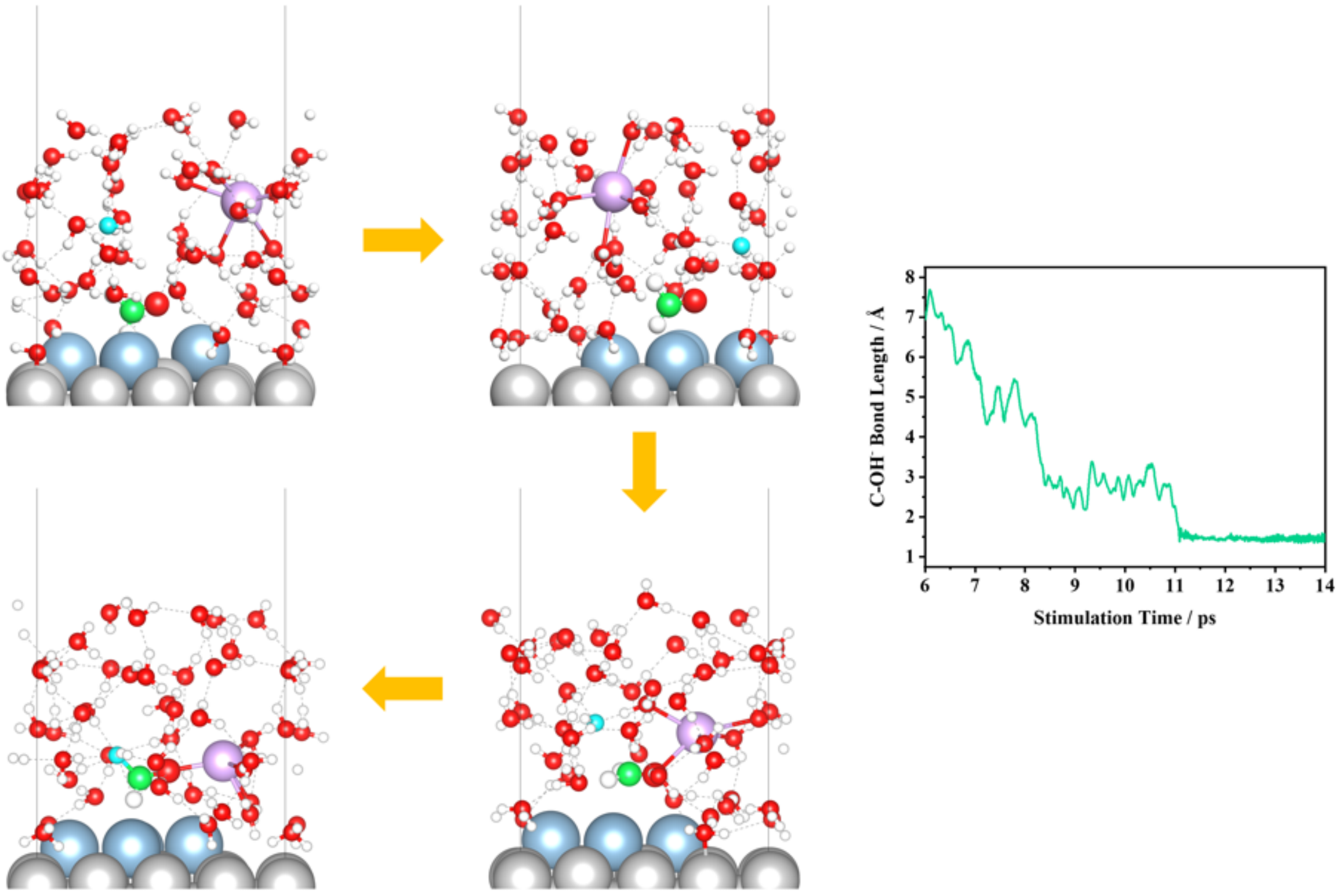

**Figure S38.** Representative snapshots of HCHO(aq)-to-$H_2COOH^-$ conversion in Bi-Pt(110)-$K^+$-$OH^-$ and distance change of C – OH during conversion. Color code: Pt, gray; Bi, blue; H, white; C, green; K, purple; O, red; and O in $OH^-$, cyan.

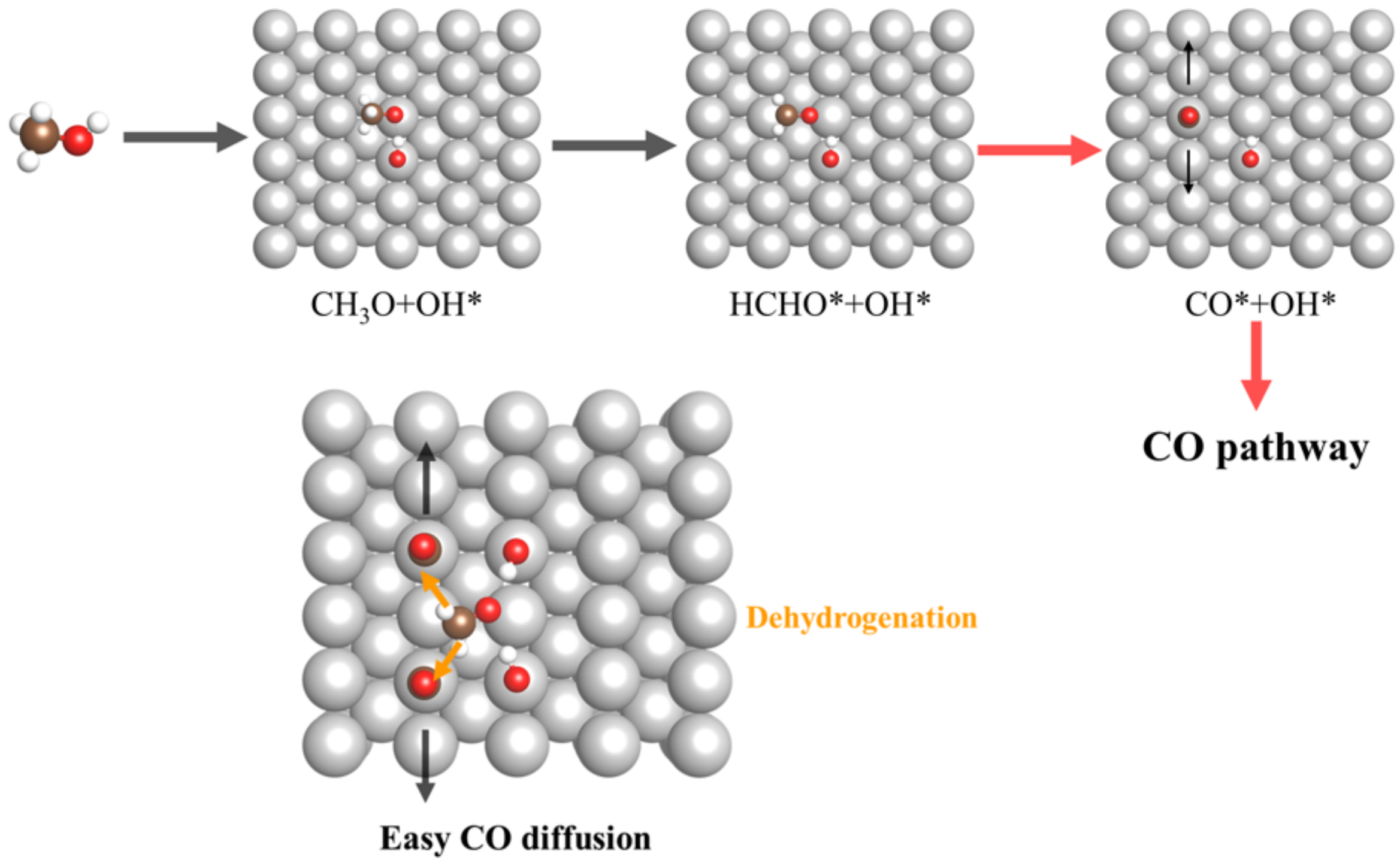

**Figure S39.** Schematic diagram of origin of CO dominated pathway mechanism on Pt surface. Color code: Pt, gray; H, white; C, brown; and O, red.

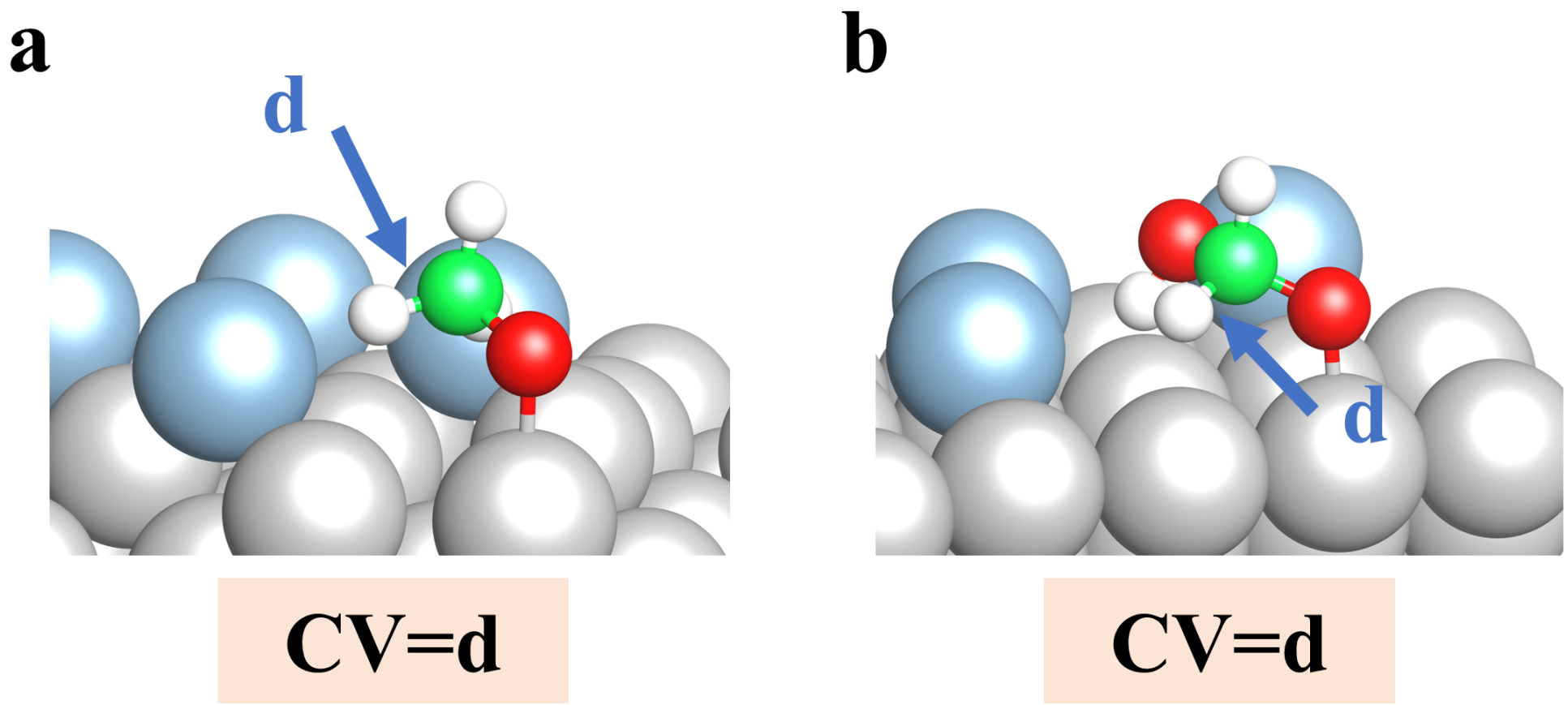

**Figure S40.** Illustration of collective variable of slow-growth approach used in cleavage of Pt-C bond in (a) $CH_3O$* and (b) $H_2COOH$*.

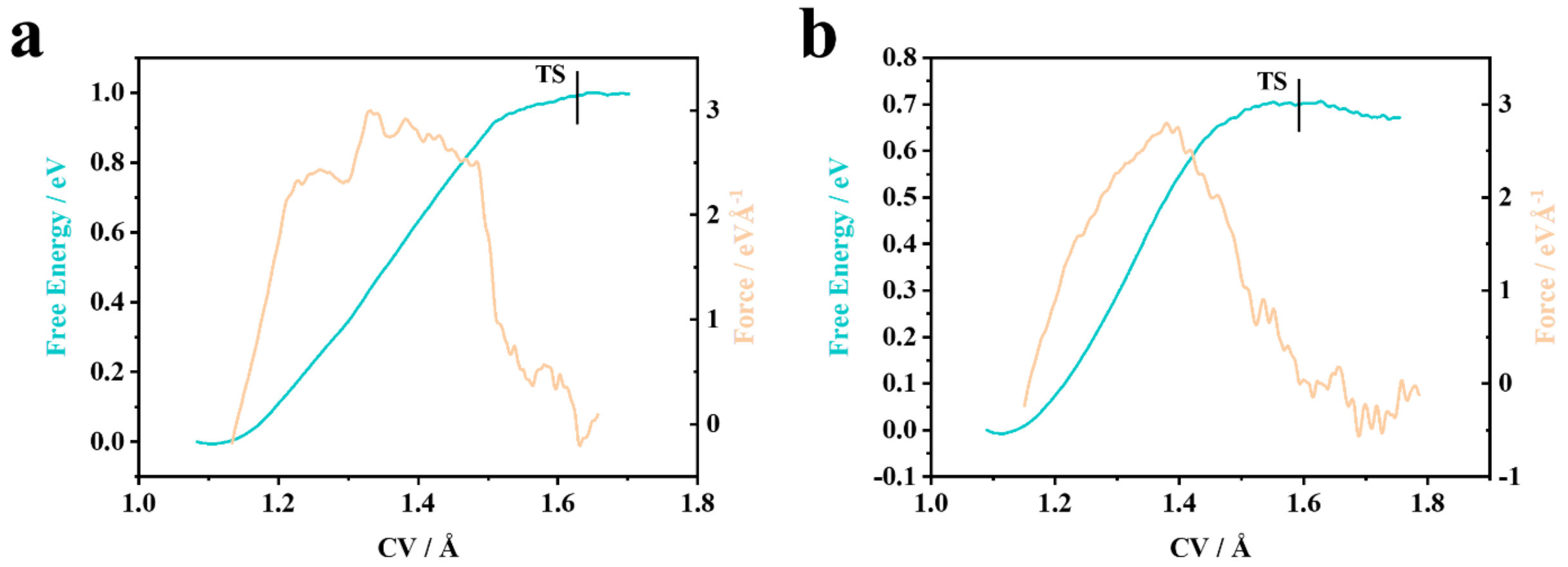

**Figure S41.** Representative potential of mean forces and corresponding free energy changes for (A) $CH_3O^*$ dehydrogenation and (B) $H_2COOH^*$ dehydrogenation on Bi-Pt(110).

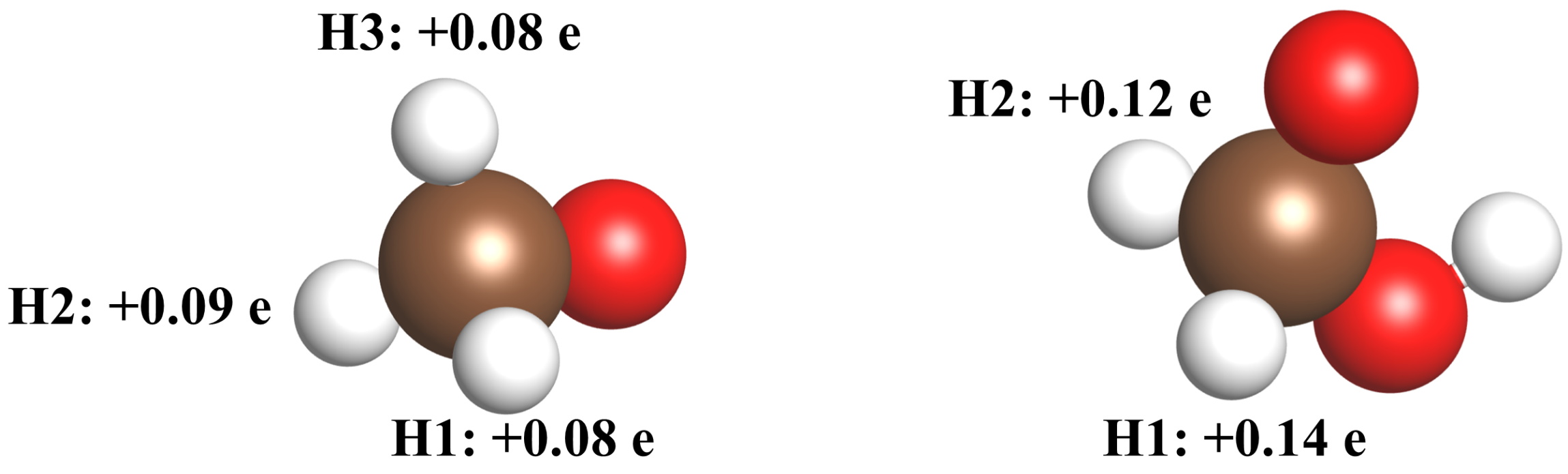

**Figure S42.** Bader charge analysis of $CH_3O$ and $H_2COOH$ group.

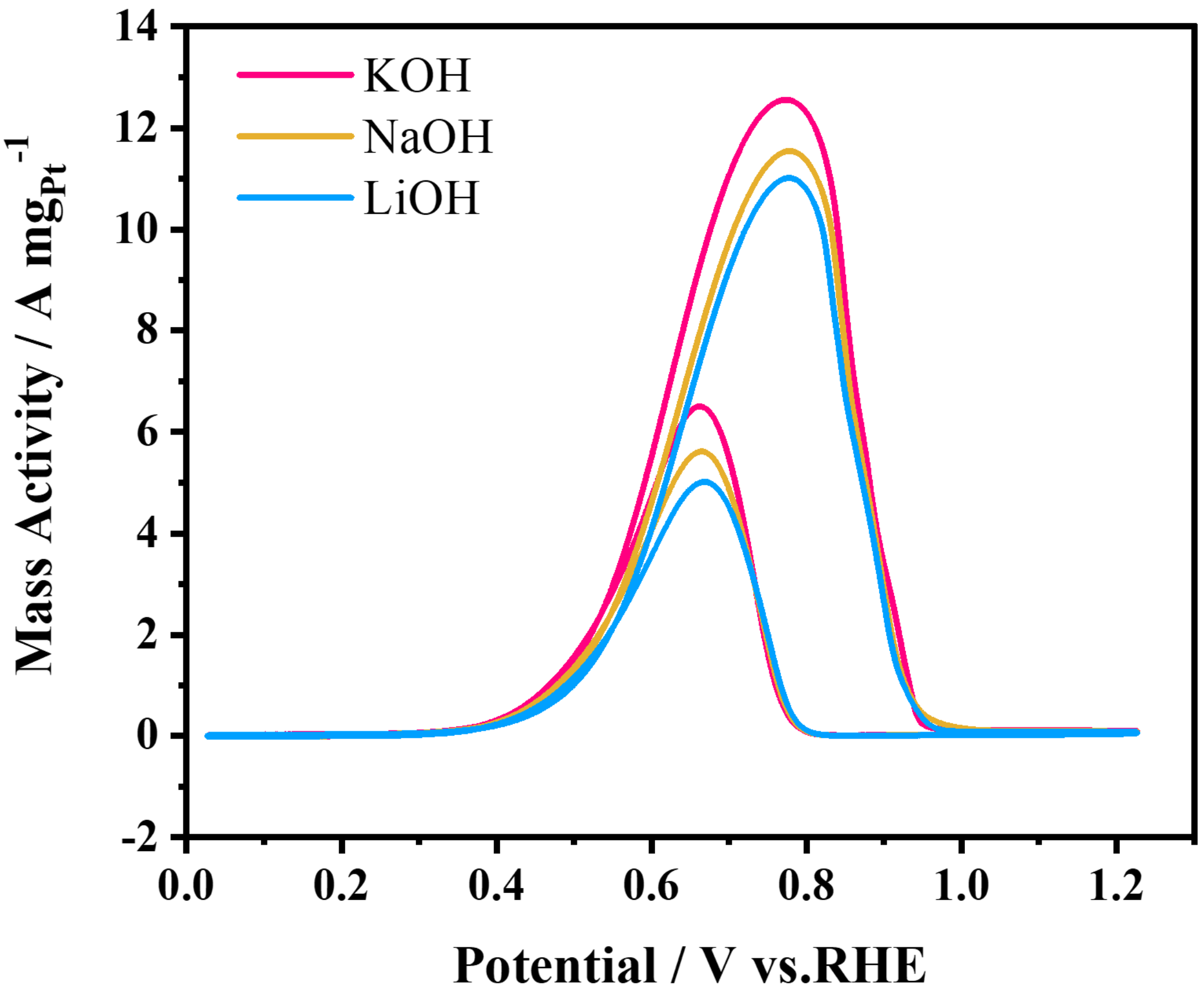

**Figure S43.** CV curves of Bi-Pt/C recorded in 1M $CH_3OH$ containing different cation electrolytes, respectively.

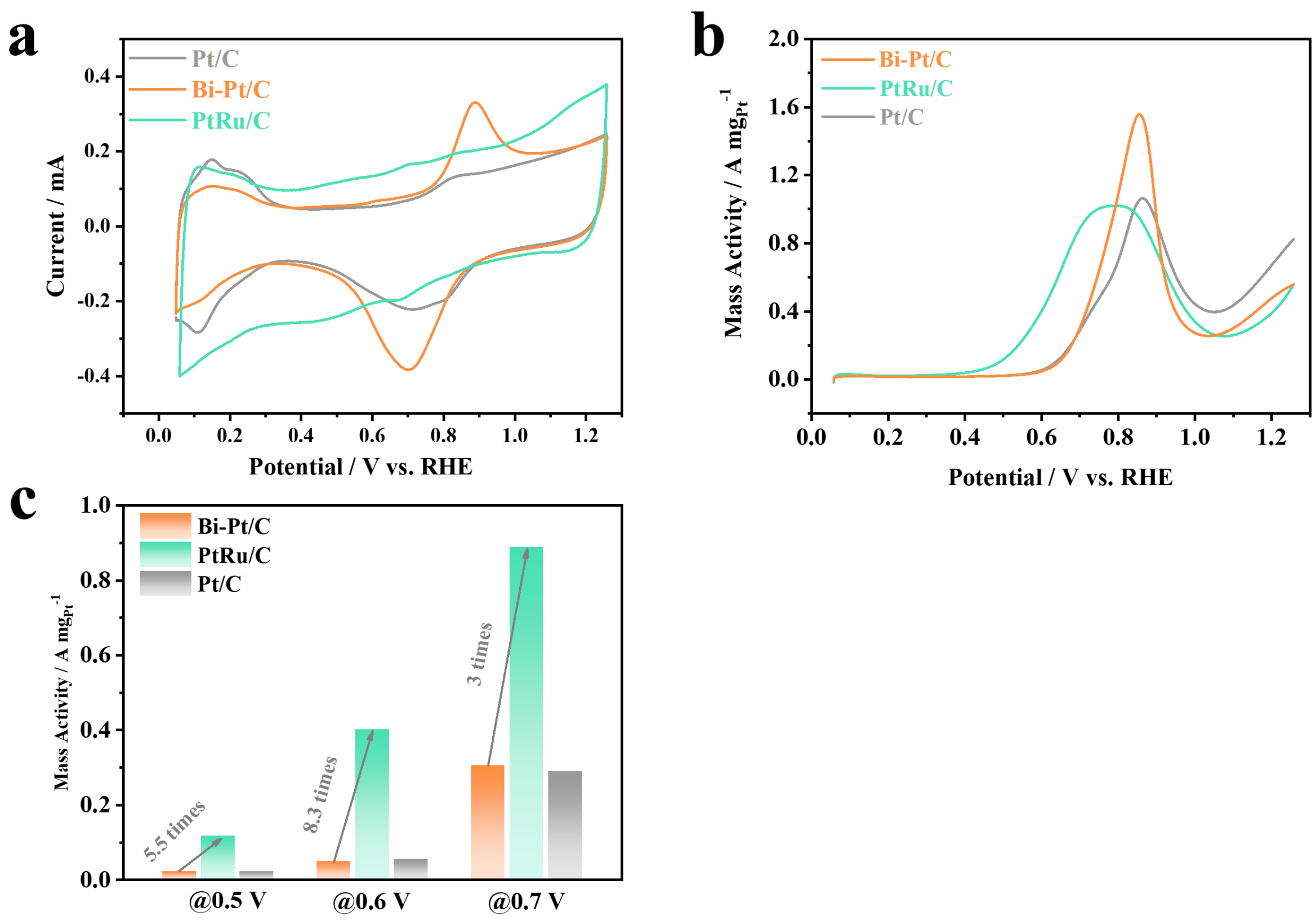

**Figure S44.** (A) CV curves of different catalysts in 0.1M $HClO_4$. (B) Positive-going polarization curves of different catalysts recorded at a scan rate of 50 mV $s^{-1}$ in 0.5M $CH_3OH$/0.1M $HClO_4$. (C) A comparison of the mass activity of different catalysts at different potentials.

Note for Figure S44: Before conducting the acidic MOR test, these catalysts were activated and cleaned in $N_2$-saturated 0.1M $HClO_4$ until stable cyclic voltammograms curves were obtained. As shown in Figure 47a, the voltammetric behavior of Bi-Pt under acidic condition is similar to that in under alkaline condition. Although Bi-Pt/C exhibits mass activity (MA) of 1.56 A $mg_{Pt}^{-1}$ at peak potential, higher than the values found for commercial catalysts, the MA of Bi-Pt/C is far inferior to that of PtRu/C at low potential region (0.5-0.7 $V_{RHE}$). Particularly, the MA of PtRu/C even reaches 8.3 times higher than that of Bi-Pt/C at 0.6 $V_{RHE}$, which processes faster reaction kinetics at low potentials.

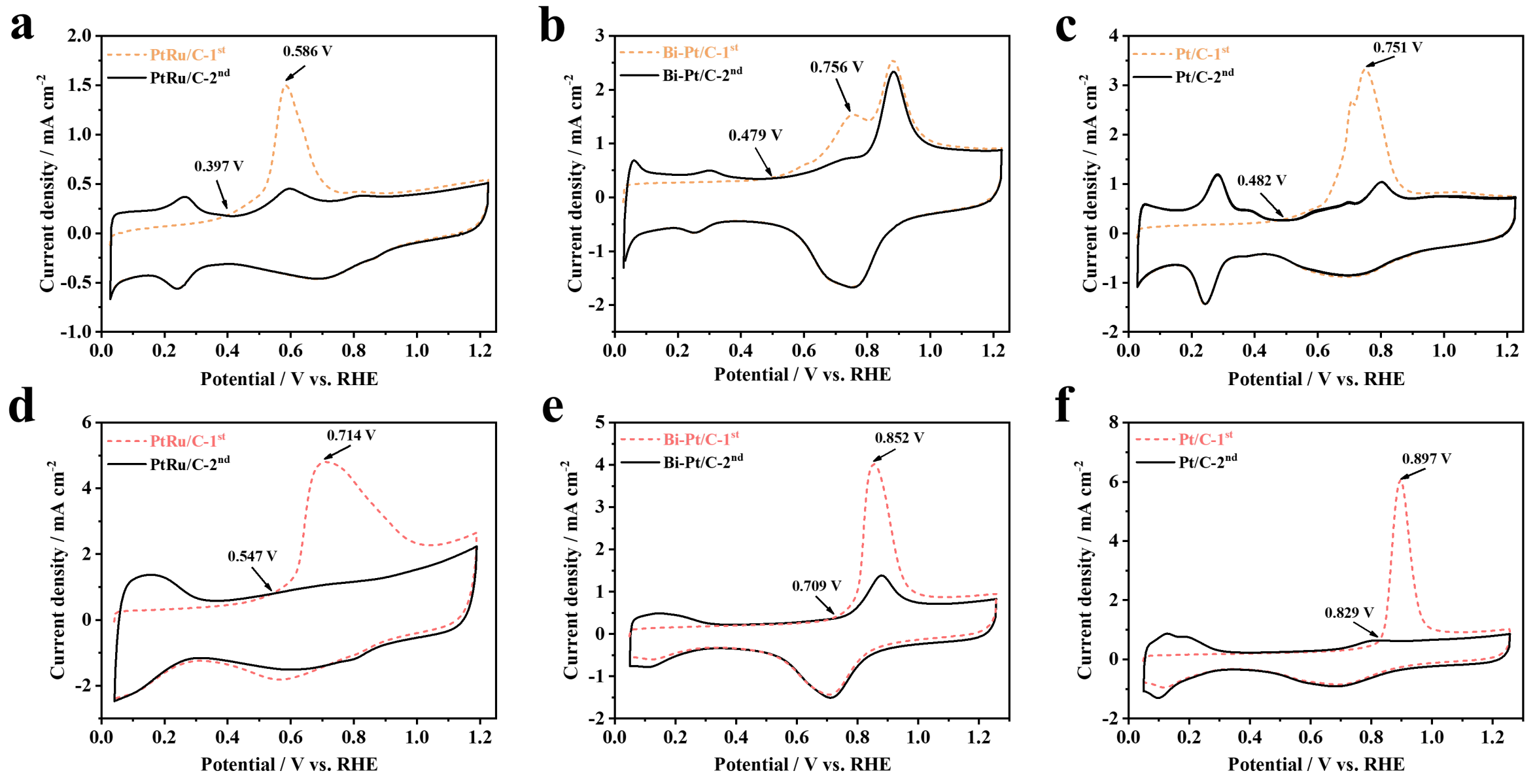

**Figure S45.** CO-stripping curves of PtRu/C, Bi-Pt/C, and Pt/C catalysts in (A-C) 1M KOH solution and in (D-F) 0.1M $HClO_4$ solution, respectively.

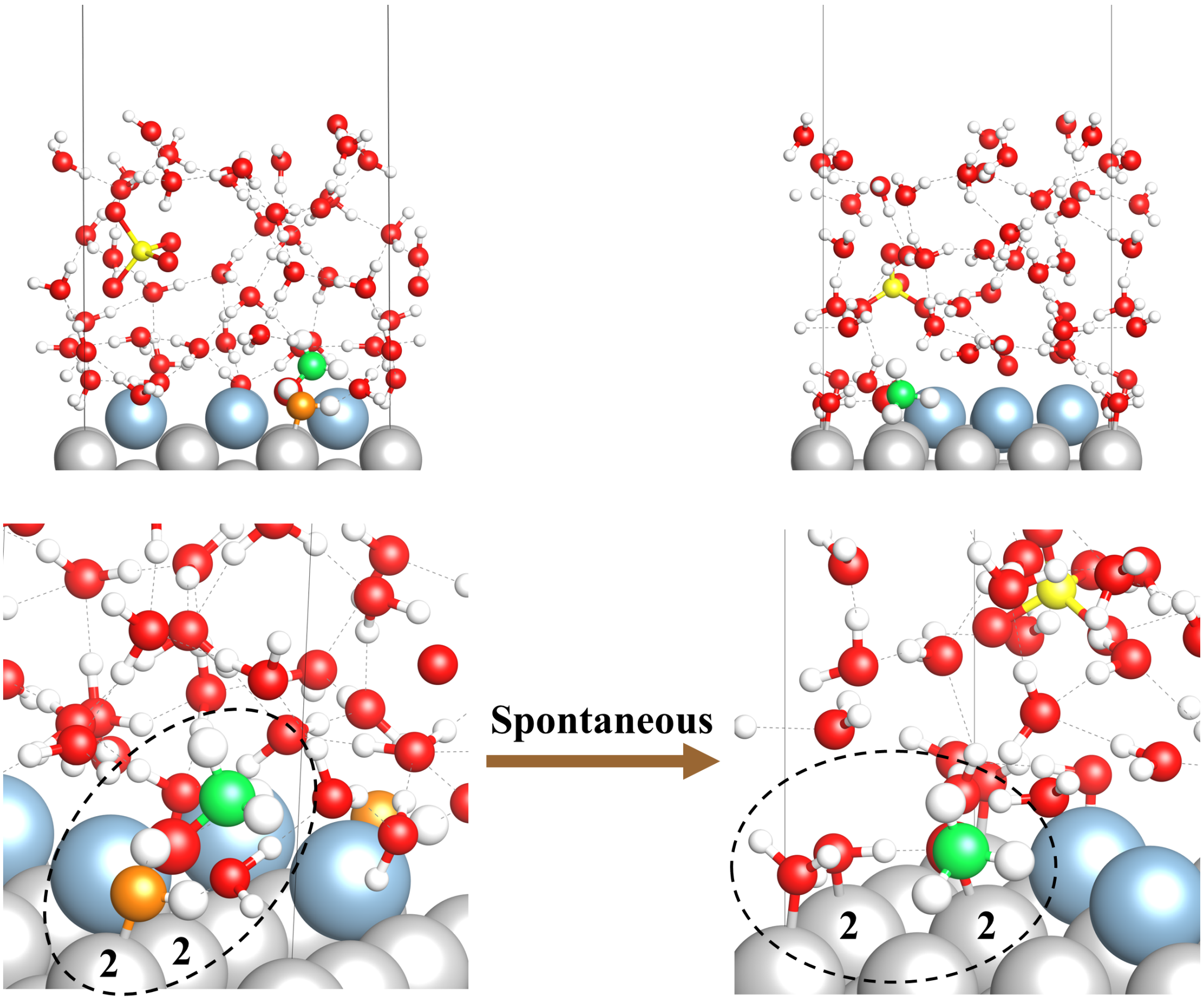

**Figure S46.** Representative snapshots of Bi-Pt(110)-$CH_3OH$-$ClO_4$ from AIMD trajectory when OH* adsorbs at site 2 and $CH_3OH$ is close to top of adjacent site 2 at acidic interface. $CH_3OH$ and OH* are highlighted in black ellipses. Color code: Pt, gray; Bi, blue; H, white; C, green; Cl, yellow; O, red; and O in $OH^*$, orange.

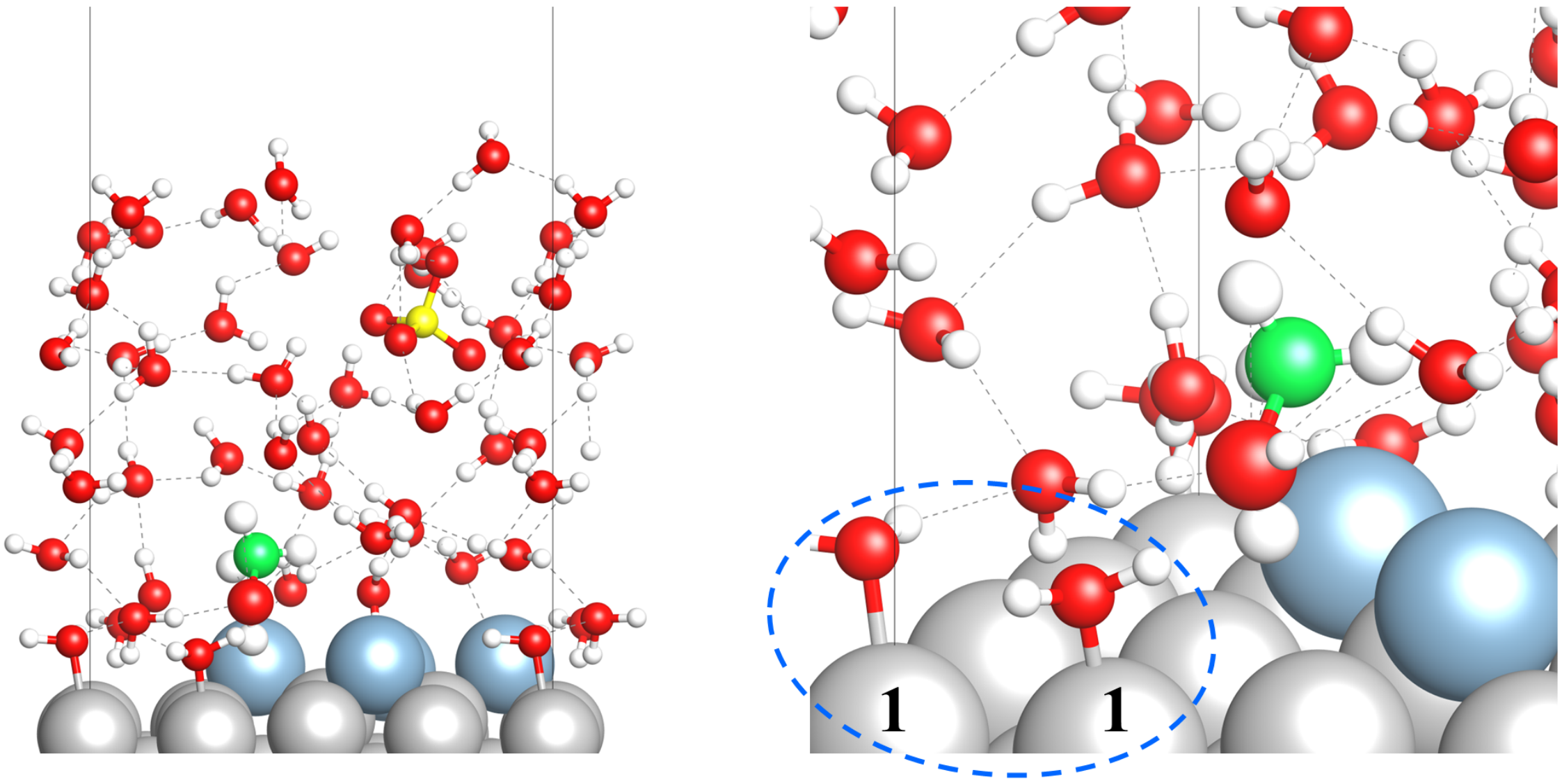

**Figure S47.** Representative snapshots of Bi-Pt(110)-$CH_3OH$-$ClO_4$ at 0.5 $V_{RHE}$ (pH=1) after 10ps by using CP-HS-DM. $H_2O$* are highlighted in blue ellipses. Color code: Pt, gray; Bi, blue; H, white; C, green; Cl, yellow; and O, red.

| Catalysts | Mass Activity / A $mg_{PGM}^{-1}$ | References |
|---|---|---|
| Bi-Pt/C | 13.7 | This work |
| PtRhBiSnSb | 19.529 | 1[3] |
| PtNiFeCoCu HEA NPs | 15.04 | 2[4] |
| PtBi@6.7%Pb | 13.93 | 3[5] |
| HEA-NPs-(14) | 12.6 | 4[6] |
| $Pt_5Ce$ | 9.13 | 5[7] |
| SANi-PtNWs | 7.93 | 6[8] |
| $Ptc/Ti_3C_2T_x$ | 7.32 | 7[9] |
| CS-$Pt_{56}Cu_{28}Ni_{16}$ | 7.0 ± 0.5 | 8[10] |
| Cu–PtBi NFBs | 6.79 | 9[11] |
| $Pt_1/RuO_2$ | 6.766 | 10[12] |

Table S1. A summary of the performances of MOR electrocatalysts in alkaline electrolytes in literature.

| **Time / s** | **Charge / C** | **Produced formate / mmol** | **formate Faradaic efficiency / %** |
|---|---|---|---|
| 3600 | 35.28 | 0.0642 | 90.4 |
| 5400 | 52.92 | 0.094 | 88.1 |
| 7200 | 70.56 | 0.125 | 87.6 |

Table S2. Calculations of formate Faradaic efficiency (%) for the MOR on Bi-Pt/C based on the IC analysis.